\documentclass[letterpaper,11pt]{article}
\pdfoutput=1

\usepackage{jheppub}
\usepackage{multirow}
\usepackage{subfig}
\usepackage{xspace}
\usepackage[countmax]{subfloat}

\newcommand{\al}{\alpha}
\newcommand{\bt}{\beta}
\newcommand{\de}{\delta}

\newcommand{\ga}{\gamma}

\newcommand{\si}{\sigma}
\newcommand{\tr}{\mathrm{tr}}

\newcommand{\cJ}{{\mathcal J}}

\newcommand{\df}{\mathrm{d}}

\newcommand{\be}{\begin{equation}}
\newcommand{\ee}{\end{equation}}
\newcommand{\nn}{\nonumber}

\newcommand{\genang}[2]{{\lambda^{#1}_{#2}}}
\newcommand{\safeang}[1]{{e_{#1}}}
\newcommand{\unsafeang}[1]{{\lambda^{#1}_{0}}}

\newcommand{\nbins}{N_\text{bins}}
\newcommand{\Nev}{N_\text{ev}}

\DeclareRobustCommand{\Sec}[1]{Sec.~\ref{#1}}

\DeclareRobustCommand{\App}[1]{App.~\ref{#1}}

\DeclareRobustCommand{\Fig}[1]{Fig.~\ref{#1}}
\DeclareRobustCommand{\Figs}[2]{Figs.~\ref{#1} and \ref{#2}}
\DeclareRobustCommand{\Eq}[1]{Eq.~(\ref{#1})}
\DeclareRobustCommand{\Eqs}[2]{Eqs.~(\ref{#1}) and (\ref{#2})}
\DeclareRobustCommand{\Ref}[1]{Ref.~\cite{#1}}
\DeclareRobustCommand{\Refs}[1]{Refs.~\cite{#1}}

\newcommand{\pythia}[1]{\textsc{Pythia\xspace #1}}

\newcommand{\fastjet}[1]{\textsc{FastJet\xspace #1}}

\newcommand{\herwigpp}[1]{\textsc{Herwig++\xspace #1}}

\preprint{ 
\begin{flushright}
MIT--CTP 4572 \\
NIKHEF 2014-026
 \end{flushright}}

\title{Gaining (Mutual) Information about \\ Quark/Gluon Discrimination}

\author[a]{Andrew J.~Larkoski,}
\author[a]{Jesse Thaler,}
\author[b,c]{and Wouter J.~Waalewijn}

\affiliation[a]{Center for Theoretical Physics, Massachusetts Institute of Technology, Cambridge, MA 02139, USA}
\affiliation[b]{Nikhef, Theory Group, Science Park 105, 1098 XG, Amsterdam, The Netherlands}
\affiliation[c]{ITFA, University of Amsterdam, Science Park 904, 1018 XE, Amsterdam, The Netherlands}

\emailAdd{larkoski@mit.edu}
\emailAdd{jthaler@mit.edu}
\emailAdd{wouterw@nikhef.nl}

\abstract{Discriminating quark jets from gluon jets is an important but challenging problem in jet substructure.  In this paper, we use the concept of mutual information to illuminate the physics of quark/gluon tagging.  Ideal quark/gluon separation requires only one bit of truth information, so even if two discriminant variables are largely uncorrelated, they can still share the same  ``truth overlap''.   Mutual information can be used to diagnose such situations, and thus determine which discriminant variables are redundant and which can be combined to improve performance.  Using both parton showers and analytic resummation, we study a two-parameter family of generalized angularities, which includes familiar infrared and collinear (IRC) safe observables like thrust and broadening, as well as IRC unsafe variants like $p_T^D$ and hadron multiplicity.  At leading-logarithmic (LL) order, the bulk of these variables exhibit Casimir scaling, such that their truth overlap is a universal function of the color factor ratio $C_A/C_F$.   Only at next-to-leading-logarithmic (NLL) order can one see a difference in quark/gluon performance.   For the IRC safe angularities, we show that the quark/gluon performance can be improved by combining angularities with complementary angular exponents. Interestingly, LL order, NLL order, \pythia{8}, and \herwigpp\ all exhibit similar correlations between observables, but there are significant differences in the predicted quark/gluon discrimination power.  For the IRC unsafe angularities, we show that the mutual information can be calculated analytically with the help of a nonperturbative ``weighted-energy function'', providing evidence for the complementarity of safe and unsafe observables for quark/gluon discrimination.
}

\begin{document} 
\maketitle

%%%%%%%%%%%%%%%%%%%%%%%%%%%%%%%%%%%%%%%%%%%%%%%%%%%%%%%%%%%%%%%%%%%%%%%%%%%%%%%%
\section{Introduction}
%%%%%%%%%%%%%%%%%%%%%%%%%%%%%%%%%%%%%%%%%%%%%%%%%%%%%%%%%%%%%%%%%%%%%%%%%%%%%%%%

Jets are collimated sprays of hadrons that act as proxies for short-distance quarks and gluons.  Because quarks and gluons have different color charges, they have different showering and fragmentation patterns, and one can exploit this information to discriminate quark-initiated jets from gluon-initiated jets on a statistical basis.  Quark/gluon discrimination is one of the key goals of the jet substructure community \cite{Abdesselam:2010pt,Altheimer:2012mn,Altheimer:2013yza}.  A number of quark/gluon tagging methods have been pursued \cite{Gallicchio:2011xq,Gallicchio:2012ez,Krohn:2012fg,Chatrchyan:2012sn,Pandolfi:1480598,Larkoski:2013eya}, with corresponding performance studies \cite{CMS-PAS-JME-13-002,CMS-PAS-JME-13-005,Aad:2014gea} at the Large Hadron Collider (LHC).  

Two seemingly conflicting themes have emerged from these quark/gluon discrimination studies (as well as from other tagging studies).  An optimistic theme is that tagging performance can be substantially improved by combining multiple jet substructure observables, as advocated in \Refs{Gallicchio:2011xq,Gallicchio:2012ez}.  A more pessimistic theme is that even if two discriminant variables are largely uncorrelated, their joint performance may not be much better than their individual performance.  These dueling themes can be seen by comparing the results of recent tagging studies of boosted $W$ bosons \cite{TheATLAScollaboration:2013tia,CMS-PAS-JME-13-006}.  Since quark/gluon discrimination has so many potential physics applications, it is essential to understand why both of these themes can be true.

To achieve this goal, we pursue a twofold approach in this paper.  First, we use the concept of ``mutual information'' to illuminate the statistical aspects of quark/gluon tagging.  Mutual information characterizes the correlations between variables by counting the number of shared bits of information.\footnote{ To our knowledge, the only use of mutual information in the particle physics literature is \Ref{Carruthers:1989gu}, though it has been discussed recently in \Ref{Narsky:2014fya} as a robust measure of correlations.  Elsewhere in high energy physics, mutual information is used in the study of (holographic) entanglement entropy (see e.g.~\cite{Casini:2004bw,2008PhRvL.100g0502W,2009JPhA...42X4005C,Headrick:2010zt}).}  Ideal quark/gluon tagging requires just one bit of ``truth'' information (e.g.\ $0 =$ quark and $1 = $ gluon), so even if a variable has many bits of total information, those bits may or may not have much overlap with the truth.  For an observable $a$ and an equal admixture of quarks and gluons, the mutual information with the truth is
%%%
\be
\label{eq:Iintro}
I(T;A) = \int \df a \left( \frac{p_q(a)}{2} \log_2 \frac{p_q(a)}{p_{\rm tot}(a)} + \frac{p_g(a)}{2} \log_2 \frac{p_g(a)}{p_{\rm tot}(a)} \right),
\ee
%%%
where $p_q$ ($p_g$) is the probability distribution for quarks (gluons), and $p_{\rm tot} = (p_q + p_g)/2$.  Since $0 \leq I(T;A) \leq 1$, we will sometimes refer to it as the ``truth overlap''.  In essence, the conflicting themes above can be traced to the difference between the total information (measured by e.g.\ the Shannon entropy) and relevant information for quark/gluon discrimination (measured by $I(T;A)$).

\begin{figure}
\begin{center}
\includegraphics[scale = 0.7]{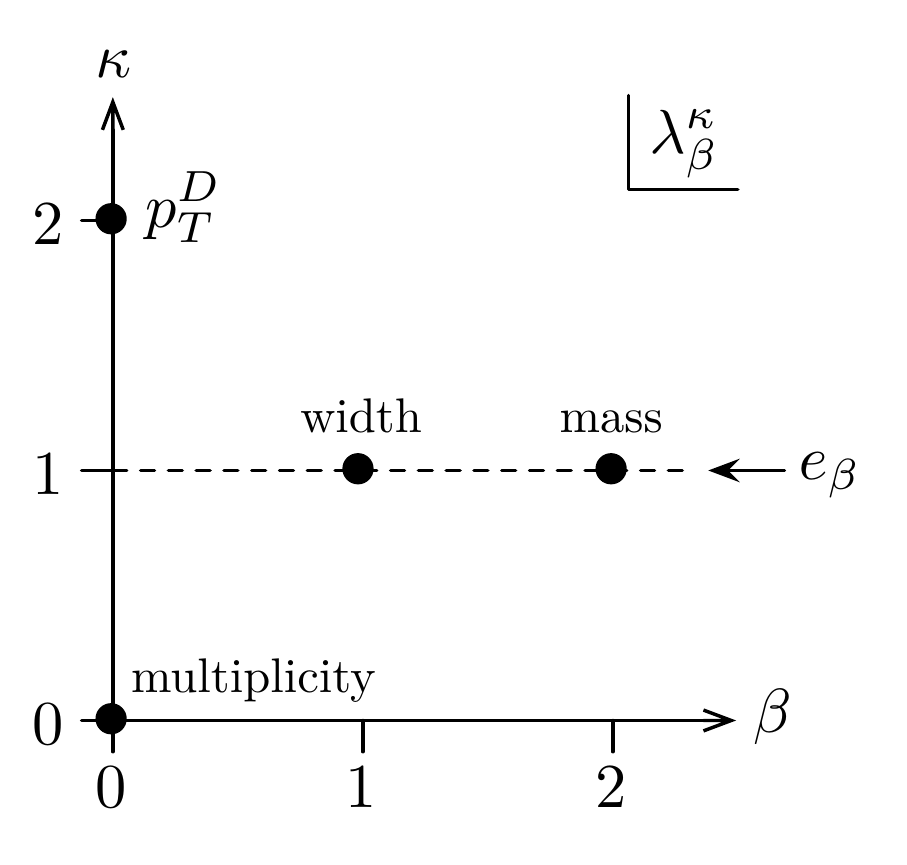}
\end{center}
\caption{Visualization of the space of observables $\genang{\kappa}{\beta}$, which includes several well-known jet observables used in quark/gluon discrimination: the line $\kappa=1$ corresponds to the IRC safe angularities $\safeang{\beta}$, the origin $(\beta,\kappa)=(0,0)$ to multiplicity, and (0,2) to $p_T^D$.  Here,  ``width'' at (1,1) refers also to broadening and girth, and ``mass'' at (2,1) refers to jet-mass-squared divided by energy (i.e.~thrust).
}
\label{fig:lambdaspace}
\end{figure}

Second, we will introduce a two-parameter family of discriminant variables to illuminate the physics aspects of quark/gluon tagging.  We will call them ``generalized angularities'', which depend not only on an angular exponent $\beta \geq 0$, but also on an energy weighting factor $\kappa \geq 0$.  They are defined as 
%%%
\begin{equation}
\label{eq:genang}
\genang{\kappa}{\beta} = \sum_{i \in \text{jet}} z_i^\kappa \left(\frac{R_i}{R_0}\right)^\beta ,
\end{equation}
%%%
where $z_i$ is the momentum fraction of particle $i$, $R_i$ is its rapidity/azimuth angle to a suitable axis,\footnote{To have recoil-free observables, we use the winner-take-all axis \cite{Bertolini:2013iqa,Larkoski:2014uqa,Salambroadening}.  The winner-take-all axis always coincides with one of the particles in the jet, so there is guaranteed to be at least one particle with $R_i = 0$.  For $\beta = 0$, we define $\genang{\kappa}{0} = \sum_{i} z_i^\kappa$.   Though we will not discuss the issue of recoil \cite{Catani:1992jc,Dokshitzer:1998kz,Banfi:2004yd,Larkoski:2013eya,Larkoski:2014uqa,Larkoski:2014bia} in much detail, our analytic results require using recoil-free instead of recoil-sensitive angularities.} and $R_0$ is the jet radius.  These variables are infrared and collinear (IRC) safe only for $\kappa = 1$.  As shown in \Fig{fig:lambdaspace}, certain values of $(\beta,\kappa)$ correspond to well-known observables: $(0,0)$ is particle multiplicity, $(0,2)$ is $p_T^D$ \cite{Chatrchyan:2012sn,Pandolfi:1480598,CMS-PAS-JME-13-005}, and the line $(\beta,1)$ are the (recoil-free) angularities $\safeang{\beta}$ \cite{Berger:2003iw,Almeida:2008yp,Ellis:2010rwa,Larkoski:2014uqa}, including broadening/width/girth at $\beta =1$ \cite{Rakow:1981qn,Ellis:1986ig,Catani:1992jc} and thrust at $\beta =2$ \cite{Farhi:1977sg}.\footnote{The recoil-free angularities are sometimes denoted as $\tau^{(\beta)}$ \cite{Larkoski:2013eya}. The generalized angularities also have the honorific notation of $\phi_\beta^\kappa$.}  We will present analytic calculations and parton shower simulations to understand the quark/gluon discrimination power of the $\genang{\kappa}{\beta}$ variables.

Our analytic calculations build on previous work calculating the substructure of quark and gluon jets \cite{Almeida:2008yp,Ellis:2010rwa,Larkoski:2013eya,Dasgupta:2013ihk,Dasgupta:2013via,Larkoski:2014wba} and well as calculating angularities in $e^+ e^-$ event shapes \cite{Berger:2003iw,Hornig:2009vb,Larkoski:2014uqa}.  For the IRC safe angularities  ($\kappa = 1$), we will be able to analytically study the correlations between two angularities $\safeang{\alpha}$ and $\safeang{\beta}$ up to next-to-leading logarithmic (NLL) accuracy.  For the IRC unsafe angularities ($\kappa \not=1$), we can use the techniques developed in \Refs{Krohn:2012fg,Waalewijn:2012sv,Chang:2013rca,Chang:2013iba} to introduce a new nonperturbative object called the ``weighted-energy function''. 
For $\beta=0$, the discrimination power depends on the details of this object and only the dependence on the jet $p_T$ and radius $R_0$ is calculable. However, for $\kappa\beta \gtrsim 0.5$, just the first (logarithmic) moment of this function enters at NLL.  As long as these moments are sufficiently small, we can predict the quark/gluon tagging performance for an individual $\genang{\kappa}{\beta}$ to NLL accuracy, as well as study correlations between two IRC unsafe angularities to NLL.

Some of the results in this paper are well-known to experts, though perhaps not in the language we use here.  On the statistical side, tagging performance is typically shown in terms of receiver operating characteristic (ROC) curves, which show the background mistag rate for a given signal efficiency.  As discussed in \App{app:rocMI}, ROC curves and mutual information are related to each other, and while ROC curves are perhaps more intuitive, mutual information has a closed form analytic definition and also has a nice visualization in terms of Venn diagrams.  Furthermore, in \App{app:betterroc}, we prove that if one observable has a better ROC curve than another, then it also has a larger truth overlap, showing that $I(T;A)$ is a robust measure of tagging performance.  On the physics side, $\genang{\kappa}{\beta}$ is only a subset of the possible quark/gluon discriminants (see \Ref{Gallicchio:2012ez} for a catalog).  Our goal is not to be exhaustive, but rather explain why different values of $\kappa$ and $\beta$ are sensitive to different properties of quarks and gluons.  To maximize quark/gluon performance when combining variables, clearly one wants to pick variables that are sensitive to different physical effects.

Before presenting our results, we wish to make some general remarks about the definition of ``quark jets'' and ``gluon jets''.   For most jet algorithms, quark and gluon jets are only well-defined at lowest order in $\alpha_s$ (see, however, \Ref{Banfi:2006hf}).  Defining quark and gluon jets to all orders in perturbation theory is a subtle and challenging problem, and becomes even more so when non-perturbative effects are taken into account.
That said, quarks and gluons are well-defined if you are (deep) in the resummation region, i.e.~in the limit of energetic, narrow, well-separated jets. In this regime quark/gluon radiation patterns are universal (including non-perturbative effects \cite{Stewart:2014nna}), although at NNLL soft interference effects start playing a role, introducing a dependence on the color structure of the whole event.  For the analytic predictions in this paper, we will adopt a pragmatic definition: a quark (gluon) jet is what results from the showering of a quark (gluon) parton.  This was also the strategy used in \Ref{Larkoski:2013eya}, and is sufficient to NLL accuracy.  An alternative approach is to avoid trying to directly tag quarks and gluons, and instead use event categories as a well-defined proxy for jet flavor.  For example, at the LHC, one can achieve a quark-enriched sample by looking at $\gamma/W/Z$ plus jet events and a gluon-enriched sample from dijet events, with further enrichment possible through judicious kinematic selections~\cite{Gallicchio:2011xc}.  As long as one accounts for the corresponding dilution factor, the jet-based mutual information techniques in this paper will work equally well on event categories.

In \Sec{sec:mutual}, we review the definition of mutual information, and use it to emphasize why the joint tagging performance of two observables is a separate concept from the correlations between two observables.  In \Sec{sec:casimir}, we show that for observables with ``Casimir scaling'' behavior, the quark/gluon truth overlap is a universal function of the color factor ratio $C_A/C_F$.  In \Sec{sec:angdef}, we discuss the general features of the $\genang{\kappa}{\beta}$ variables and show parton shower results for their mutual information.  We then turn to analytic calculations, treating the IRC safe case of the angularities $\safeang{\beta}$ in \Sec{sec:ircsafeang} and the more general IRC unsafe case of $\genang{\kappa}{\beta}$ in \Sec{sec:unsafeang}.  We conclude in \Sec{sec:conclude}.

%%%%%%%%%%%%%%%%%%%%%%%%%%%%%%%%%%%%%%%%%%%%%%%%%%%%%%%%%%%%%%%%%%%%%%%%%%%%%%%%
\section{Mutual Information}
\label{sec:mutual}
%%%%%%%%%%%%%%%%%%%%%%%%%%%%%%%%%%%%%%%%%%%%%%%%%%%%%%%%%%%%%%%%%%%%%%%%%%%%%%%%

%~~~~~~~~~~~~~~~~~~~~~~~~~~~~~~~~~~~~~~~~~~~~~~~~~~~~~~~~~~~~~~~~~~~~~~~~~~~~~~~
\subsection{Definition}
%~~~~~~~~~~~~~~~~~~~~~~~~~~~~~~~~~~~~~~~~~~~~~~~~~~~~~~~~~~~~~~~~~~~~~~~~~~~~~~~

Mutual information is a measure of the shared information content of two observables.  For continuous distributions of variables $a$ and $b$, the mutual information is (see e.g.~\cite{nielsen00})
%%%
\be
\label{eq:Idef}
I(A;B) = \int \df a \, \df b \, p(a,b) \log_2 \frac{p(a,b)}{p(a) p(b)},
\ee
%%%
where $p(a,b)$ is the joint probability distribution (normalized to have unit integral), and $p(a) \equiv \int db \, p(a,b)$ and $p(b) \equiv \int da \, p(a,b)$  are the marginal probability distributions.\footnote{The typical notation in the information theory literature is lower case symbols ($a$) to denote the observable and upper case symbols ($A$) to indicate the set of possible values, such that $a \in A$.  We will keep this notation for a generic observable, but switch to all lower case in \Sec{sec:angdef} when we specialize to $\genang{\kappa}{\beta}$.}  Here, we are using base 2 logarithms such that $I(A;B) = 1$ corresponds to one binary bit of shared information.  

\begin{figure}
\begin{center}
\subfloat[]{\label{fig:Venn1}
\includegraphics[scale = 0.8]{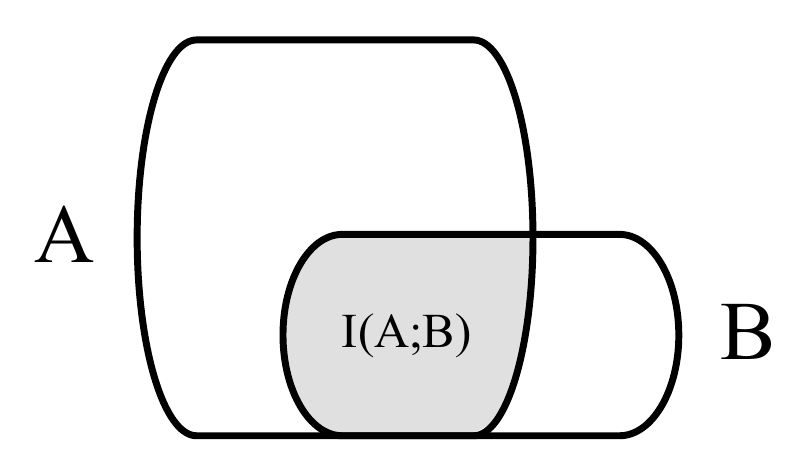}
}
$\qquad$
\subfloat[]{\label{fig:Venn2}
\includegraphics[scale = 0.8]{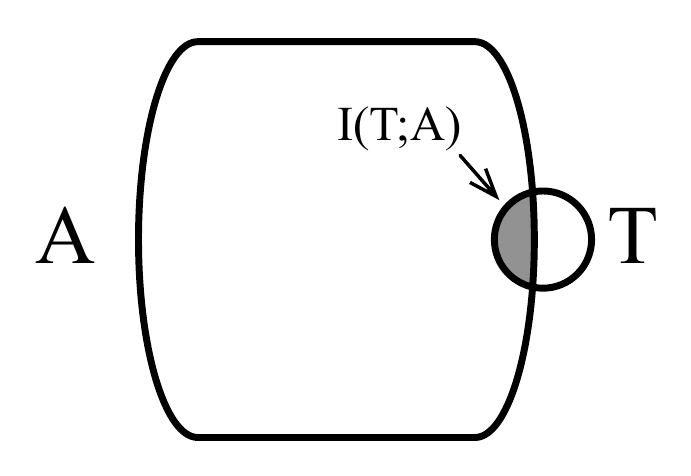}
}
\end{center}
\caption{Left:  The mutual information $I(A;B)$ between observables $A$ and $B$ is visualized as the area of the shaded overlap region in information space.  In keeping with the set-theoretic relation in \Eq{eq:Hinequality}, the region labelled $A$ has area $H(A)$, the region labelled $B$ has area $H(B)$, the union $A \cup B$ has area $H(A,B)$, and the intersection $A \cap B$ has area $I(A;B)$.  Right:  As a special case, we can consider the mutual information $I(T;A)$ between observable $A$ and the truth $T$ (i.e.~the truth overlap).}
\end{figure}

In order to visualize mutual information, it is helpful to rewrite $I(A;B)$ in terms of Shannon entropies $H$:
%%%
\be
\label{eq:altIdef}
I(A;B) = H(A) + H(B) - H(A,B).
\ee
%%%
Strictly speaking, the entropy (unlike the mutual information) is not well-defined for continuous observables, though it can be made sensible by binning the distributions.  For discrete-valued observables, we have
%%%
\be
H(A) = - \sum_{a \in A} \, p(a) \log_2 p(a), \qquad H(A,B) = - \sum_{a \in A} \sum_{b \in B} \, p(a,b) \log_2 p(a,b), 
\ee
%%%
such that $H$ ``counts'' the number of bits of information carried by the corresponding variables.  The entropies satisfy the same inequalities familiar from set theory\footnote{To derive these inequalities, consider any binned distributions $p(a)$ and $p(a,b)$.  The relation $H(A) \geq 0 $ follows from $0 \leq p(a) \leq 1$.  The relation $H(A,B) \geq H(A)$ can be derived by noting that $p(a,b) \leq p(a)$ for any $b$, and therefore
\be
H(A,B) = - \sum_{a,b} \, p(a,b) \log_2 p(a,b) \geq - \sum_{a,b} \, p(a,b) \log_2 p(a) = - \sum_{a} \, p(a) \log_2 p(a) = H(A).
\ee
The relation $H(A) + H(B) \geq H(A,B)$ follows from $-\log_2 x \geq (1-x)/\ln 2$, and therefore
\be
H(A) + H(B) - H(A,B) = - \sum_{a,b} p(a,b) \log_2 \frac{p(a) p(b)}{p(a,b)} \geq \frac{1}{\ln 2} \sum_{a,b} p(a,b) \left(1 - \frac{p(a) p(b)}{p(a,b)} \right) = 0.
\ee
}
%%%
\be
\label{eq:Hinequality}
0 \le H(A) \le H(A,B) \le H(A) + H(B).
\ee
%%%
Thus, mutual information falls in the range
%%%
\be
\label{eq:Iinequality}
0 \le I(A;B) \le \min\{H(A), H(B)\}.
\ee
%%%
As shown in \Fig{fig:Venn1}, $I(A;B)$ can be interpreted as the ``area'' of the intersection $A \cap B$ in information space, and it useful for quantifying the degree of correlation between two variables, with low values corresponding less correlated variables.

%~~~~~~~~~~~~~~~~~~~~~~~~~~~~~~~~~~~~~~~~~~~~~~~~~~~~~~~~~~~~~~~~~~~~~~~~~~~~~~~
\subsection{Single Variable Discrimination}
%~~~~~~~~~~~~~~~~~~~~~~~~~~~~~~~~~~~~~~~~~~~~~~~~~~~~~~~~~~~~~~~~~~~~~~~~~~~~~~~

For a single variable $a$, we can quantify how well it performs as a signal/background discriminant by calculating how much mutual information it shares with the truth $T$.   Consider an event sample with signal fraction $f$ and background fraction $(1-f)$, and let $t = 0$ for signal events and $t = 1$ for background events.  Because $t$ is a discrete variable, it has a well-defined Shannon entropy
%%%
\be
H(T) = - f \log_2 f - (1-f) \log_2 (1-f),
\ee
%%%
which is the number of available ``truth bits''.  The most intuitive choice is $f = 1/2$ which yields $H(T) = 1$, corresponding to one bit of truth (i.e. signal $=0$ vs.\ background $=1$).  

Without knowing the truth information, the measured $a$ distribution would be
%%%
\be
p_{\rm tot}(a) = f \, p_0(a) + (1-f) \, p_1(a),
\ee
%%%
where $p_0$ ($p_1$) is the normalized $a$ probability distribution for signal (background) events.  With the addition of truth information, the joint probability distribution is
%%%
\be
p(t, a) = \delta_{t0} \, f \, p_0(a) + \delta_{t1} \, (1-f)\,  p_1(a),
\ee
%%%
so the mutual information between $A$ and the truth $T$ (i.e.~the truth overlap) is
%%%
\begin{align}
\label{eq:ITA}
I(T;A) &= H(T) + H(A) - H(T,A) \nn \\
&= \int \df a \left(f \, p_0(a) \log_2 \frac{p_0(a)}{p_{\rm tot}(a)} + (1-f) \, p_1(a) \log_2 \frac{p_1(a)}{p_{\rm tot}(a)}   \right).
\end{align}
%%%
By \Eq{eq:Iinequality} and shown in \Fig{fig:Venn2},
%%%
\be
0 \le I(T;A) \le H(T).
\ee
%%%
The mutual information $I(T;A)$ quantifies how well the variable $a$ can separate signal and background, with $I(T;A) = 0$ corresponding to no discrimination power and $I(T;A) = H(T)$ corresponding to full discrimination power. 

In this paper, we will focus almost exclusively on $f = 1/2$  (see \Eq{eq:Iintro}), such that $0 \le I(T;A) \le 1$. As explained in \App{app:rocMI}, by varying $f$, $I(T;A)$ can be related to the ROC curves typically used to quantify discrimination power.  Unlike ROC curves, there is a closed-form expression for the mutual information, which makes it better suited for analytical studies.  To justify that $I(T;A)$ is robust measure of tagging performance, we show  in \App{app:betterroc} that if one observable is Pareto optimal with respect to another (i.e.\ its ROC curve is everywhere improved), then the corresponding truth overlap is strictly larger.  When \Eq{eq:ITA} is applied on a sample of events generated by a Monte Carlo program, the finite sample size requires one to replace the integral by a sum over bins. Care is needed to avoid biasing the mutual information from binning, which is discussed in detail in \App{app:entropystats}.

%~~~~~~~~~~~~~~~~~~~~~~~~~~~~~~~~~~~~~~~~~~~~~~~~~~~~~~~~~~~~~~~~~~~~~~~~~~~~~~~
\subsection{Pairwise Correlations in Discrimination}
%~~~~~~~~~~~~~~~~~~~~~~~~~~~~~~~~~~~~~~~~~~~~~~~~~~~~~~~~~~~~~~~~~~~~~~~~~~~~~~~

Given two variables $a$ and $b$, $I(T;A)$ and $I(T;B)$ quantify how well each performs individually as signal/background discriminants.  Similarly, we can assess how well $a$ and $b$ perform as joint discriminant variables by calculating
%%%
\be
I(T;A,B) = H(T) + H(A,B) - H(T,A,B).
\ee
%%%
Unlike $I(A;B)$, which only tests whether or not $a$ and $b$ or correlated, $I(T;A,B)$ tests whether the (lack of) correlations between $a$ and $b$ is useful for signal/background discrimination. Note that
%%%
\be
\max \{I(T,A),I(T;B)\} \le I(T;A,B),
\ee
%%%
such that $a$ and $b$ always have the same or better joint discrimination than either variable individually.  If $I(T;A,B) = \max \{I(T,A),I(T;B)\}$, then there is no gain in quark/gluon discrimination in considering both $a$ and $b$, and one of the two variables is redundant, at least for this purpose.

\begin{figure}
\begin{center}
\subfloat[]{\label{fig:Venn3}
\includegraphics[scale = 0.8]{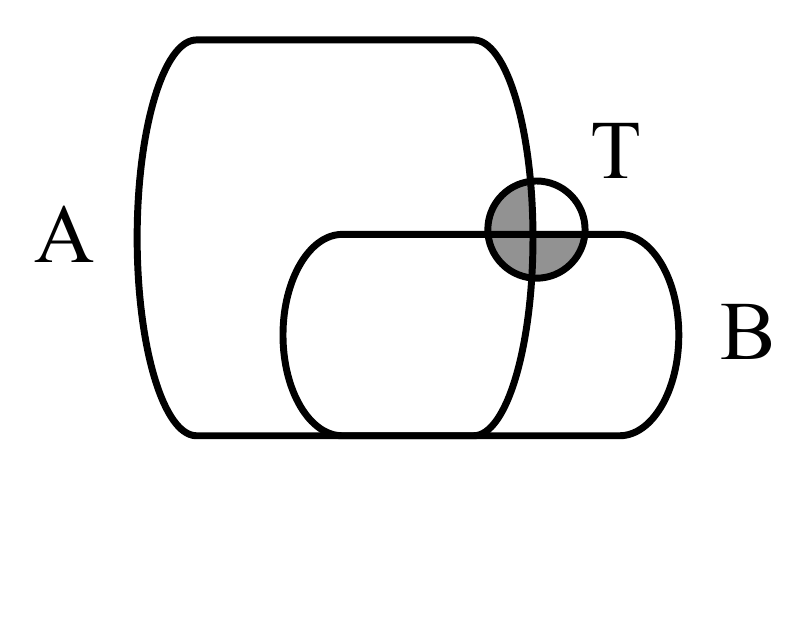}
}
$\qquad$
\subfloat[]{\label{fig:Venn4}
\includegraphics[scale = 0.8]{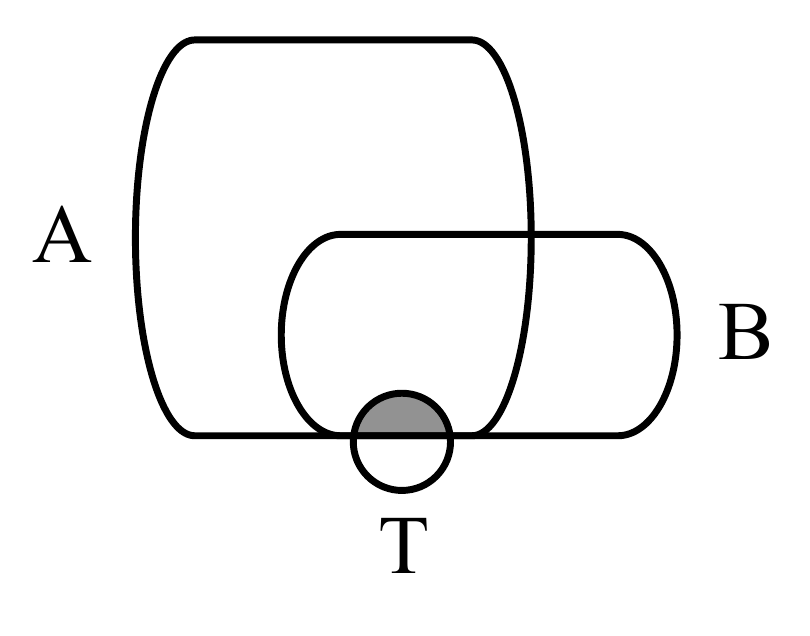}
}\end{center}
\caption{The mutual information $I(T;A,B)$ between observables $A$, $B$, and the truth $T$ are shown as the area of their respective intersection. Though the mutual information between $A$ and $B$ (i.e.~their correlation) is the same for both figures, their overlap with the truth differs. In the left figure, the mutual information with the truth (shaded) is complementary, such that combining $A$ and $B$ increases the truth overlap. This is not so in the right figure.}
\label{fig:Venn34}
\end{figure}

To highlight the difference between $I(A;B)$ and $I(T;A,B)$, consider \Fig{fig:Venn34} which shows two variables with a low degree of correlation (i.e.\ $I(A;B)$ is relatively small).  In the left example, $a$ and $b$ are both decent discriminant variables individually (i.e.\ $A$ and $B$ both have overlap with the truth $T$), but they have considerably improved joint discrimination power (i.e.\ $I(T;A,B)$ is larger than both $I(T;A)$ and $I(T;B)$).  In the right example, despite the fact that $a$ and $b$ are largely uncorrelated (as measured by $I(A;B)$), there is no gain in discrimination power by considering $a$ and $b$ jointly (i.e.\ $I(T;A,B) = I(T;A) = I(T;B)$.)

In the jet substructure literature, there are known examples of both situations in \Fig{fig:Venn34}.  The left example is the ideal case (see e.g.~\cite{TheATLAScollaboration:2013tia}), where two variables $a$ and $b$ give complementary information for discrimination.  The right example is the more puzzling case (see e.g.~\cite{CMS-PAS-JME-13-006}), where two variables exhibit comparable discrimination power, a low degree of correlation, yet little gain in performance when considered jointly.  Using mutual information, it is straightforward to diagnose and visualize this situation, helping to identify redundant variables.  (Of course, redundant but largely uncorrelated variables can still be helpful for other purposes, such as calibration.)

%%%%%%%%%%%%%%%%%%%%%%%%%%%%%%%%%%%%%%%%%%%%%%%%%%%%%%%%%%%%%%%%%%%%%%%%%%%%%%%%
\section{Quark/Gluon Discrimination from Casimir Scaling}
\label{sec:casimir}
%%%%%%%%%%%%%%%%%%%%%%%%%%%%%%%%%%%%%%%%%%%%%%%%%%%%%%%%%%%%%%%%%%%%%%%%%%%%%%%%

As a simple example of using mutual information, consider an observable $a$ that satisfies the property of ``Casimir scaling''.  For such an observable, the normalized cumulative distribution for quarks ($\Sigma_q$) and gluons ($\Sigma_g$) can be written as
%%%
\begin{equation}
\label{eq:defcasimirscaling}
\Sigma_q(a) =e^{-C_F \, r(a)}, \qquad \Sigma_g(a) =e^{-C_A \, r(a)},
\end{equation}
%%%
where $r(a)$ is any monotonically decreasing function of $a$, $C_F = 4/3$ is the color factor for quarks, and $C_A = 3$ is the color factor for gluons.  As we will see in \Sec{sec:ircsafeang}, the angularities $\safeang{\beta}$ obey Casimir scaling at leading-logarithmic (LL) order.  In \Sec{subsec:betaposregime}, we will even find that the generalized angularities $\genang{\kappa}{\beta}$ with $\kappa \beta \gtrsim 0.5$ obey Casimir scaling at LL as well. 

As discussed in \Ref{Larkoski:2013eya}, any observable that exhibits Casimir scaling has a ROC curve of
%%%
\begin{equation}
\text{ROC} = x^{C_A/C_F} = x^{9/4},
\end{equation}
%%%
where $x$ is the quark jet efficiency and $x^{C_A/C_F}$ is the gluon jet mistag rate.  This result follows from making a cut $a < a_{\rm cut}$, which keeps a fraction $\Sigma_q(a_{\rm cut})$ of quarks and a fraction $\Sigma_g(a_{\rm cut}) = (\Sigma_q(a_{\rm cut}))^{C_A/C_F}$ of gluons.  Since (approximate) Casimir scaling is so ubiquitous among quark/gluon discriminants, this explains why so many discriminant variables have such similar performance.  To improve performance, one has to probe the jet beyond just its overall color charge $C_i$.

We can understand this same feature from the point of view of mutual information by showing that the truth overlap $I(T;A)$ is a universal function of $C_A/C_F$ for observables that exhibit Casimir scaling.\footnote{This result can also be derived from the relationship between the ROC curve and mutual information presented in \App{app:rocMI}.}  The probability distribution for $a$ is just the derivative of the cumulative distribution with respect to $a$:  
%%%
\begin{align}
p_q(a) &= \frac{\df }{\df  a}\,\Sigma_q(a) = -C_F \, r'(a) \, e^{-C_F r(a)},\\
p_g(a) &= \frac{\df }{\df  a}\,\Sigma_g(a) = -C_A \,r'(a) \, e^{-C_A r(a)},
\end{align}
%%%
If $f$ is the fraction of quark jets in the sample and $(1-f)$ is the fraction of gluon jets, the total probability distribution is
%%%
\be
p_{\rm tot}(a) = f \, p_q(a)+(1-f) \, p_g(a),
\ee
%%%
and the truth overlap is
%%%
\begin{align}
I(T;A) &=  f \int\! \df a\, p_q(a) \log_2 \frac{p_q(a)}{p_{\rm tot}(a)}
+ (1-f )\int\! \df a \, p_g(a) \log_2 \frac{p_g(a)}{p_{\rm tot}(a)} \nonumber \\
&= f\int\! \df a\, \left(C_F \,r'(a) \,e^{-C_F r(a)} \right) \log_2\left(f +(1-f)\frac{C_A}{C_F}e^{-(C_A-C_F) r(a)}\right) \nonumber \\
&\quad+(1-f) \int \df a \, \left(C_A \, r'(a) \, e^{-C_A r(a)} \right)\log_2\left(f \frac{C_F}{C_A} e^{-(C_F-C_A) r(a)}+(1-f)\right).
\end{align}
%%%
By making the change of variables 
%%%
\begin{equation}
u \equiv e^{-C_F \, r(a)},
\end{equation}
%%%
all dependence on the distribution $r(a)$ can be removed and the integrals can be evaluated exactly.  We find
%%%
\begin{align}
I(T;A) &= \frac{1}{\ln 2} \bigg[ f\frac{\left(C_A-C_F\right)^2}{C_F C_A} \bigg( 1- {}_2F_1\Big(1,\frac{C_F}{C_A-C_F};\frac{C_A}{C_A-C_F};\frac{(f-1) C_A}{f
   C_F}\Big) \bigg)
   \nonumber \\
   &\qquad \qquad -f \ln \Big(\frac{C_A}{C_F}-f\frac{ C_A-C_F}{C_F}\Big)-(1-f) \ln \Big(1-f\frac{C_A- C_F}{C_A}\Big)\bigg], \label{eq:mutualinfocasiiarscaling}
\end{align}
%%%
where $_2F_1(a,b;c;z)$ is the hypergeometric function and $\ln$ is the natural logarithm. As advertised, for any observable exhibiting Casimir scaling, $I(T;A)$ is a universal function of $C_A/C_F$.

Setting the quark fraction $f$ equal to $1/2$ and $C_A/C_F = 9/4$ for QCD, the mutual information for quark/gluon discrimination is
%%%
\begin{equation}\label{eq:LLang_base}
I(T;A)_{f=1/2} \simeq 0.103.
\end{equation}
%%%
This will be the baseline value to which all observables will be compared.  Note that $I(T,A)$ is quite far from $1$ (i.e.~a full truth bit), demonstrating the inherent challenge of quark/gluon tagging.

%%%%%%%%%%%%%%%%%%%%%%%%%%%%%%%%%%%%%%%%%%%%%%%%%%%%%%%%%%%%%%%%%%%%%%%%%%%%%%%%
\section{Generalized Angularities}
\label{sec:angdef}
%%%%%%%%%%%%%%%%%%%%%%%%%%%%%%%%%%%%%%%%%%%%%%%%%%%%%%%%%%%%%%%%%%%%%%%%%%%%%%%%

Our analytic studies of quark/gluon separation will focus on the generalized angularities $\genang{\kappa}{\beta}$ defined in \Eq{eq:genang}, repeated for convenience:
%%%
\begin{equation}
\genang{\kappa}{\beta} = \sum_{i \in \text{jet}} z_i^\kappa \theta_i^\beta
\,.\end{equation}
%%%
Here $z_i$ is the energy fraction and $\theta_i = R_i/R_0$ the angular fraction with respect to the jet radius $R_0$, such that $0 \leq z_i, \theta_i \leq 1$.   We measure the angles $R_i$ with respect to the recoil-free winner-take-all axis~\cite{Bertolini:2013iqa,Larkoski:2014uqa,Salambroadening} and we use a jet algorithm that centers the jet on the winner-take-all axis, such that $\theta_i \leq 1$ is strictly enforced.  For the IRC safe angularities $\safeang{\beta}$, it is known that a recoil-free axis improves quark/gluon discrimination power \cite{Larkoski:2013eya}.  For the generalized angularities $\genang{\kappa}{\beta}$, a recoil-free axis is crucial for the calculations with $\beta \lesssim \kappa$, since it ensures that $\genang{\kappa}{\beta}$ measures the radiation pattern around the initiating hard quark or gluon and not the displacement (i.e.~recoil) of the hard parton away from the jet axis.

These variables are effective quark/gluon discriminants because they probe the angular and energetic structure of jets, both of which are sensitive to the differing color factors between quarks and gluons, among other effects.  Large $\beta$ emphasizes wide-angle radiation whereas small $\beta$ emphasizes collinear radiation.  Large $\kappa$ emphasizes harder hadrons, whereas small $\kappa$ emphasizes softer hadrons.  For reference, we highlight the $\kappa = 1$ and $\beta = 0$ cases:
%%%
\begin{align}
 \safeang{\beta} \equiv \genang{1}{\beta} &= \sum_{i \in \text{jet}} z_i \theta_i^\beta,\\
\unsafeang{\kappa} & = \sum_{i \in \text{jet}} z_i^\kappa.
\end{align}
%%%
While $\unsafeang{1} = 1$ is a trivial observable, we can expand around $\kappa=1$ to find
%%%
\be
\label{eq:kappatoone}
\lim_{\kappa \to 1} \unsafeang{\kappa} = 1 + \sum_{i \in \text{jet}} (\kappa - 1) z_i \ln z_i,
\ee
%%%
so when we present studies for $\unsafeang{1}$, we really mean $\lim_{\kappa \to 1} \unsafeang{\kappa}$, which is effectively the same as the observable $\sum_{i \in \text{jet}} z_i \ln z_i$.

\begin{figure}
\begin{center}
\subfloat[]{\label{fig:pythiasinglelambda}
\includegraphics[width=7.5cm]{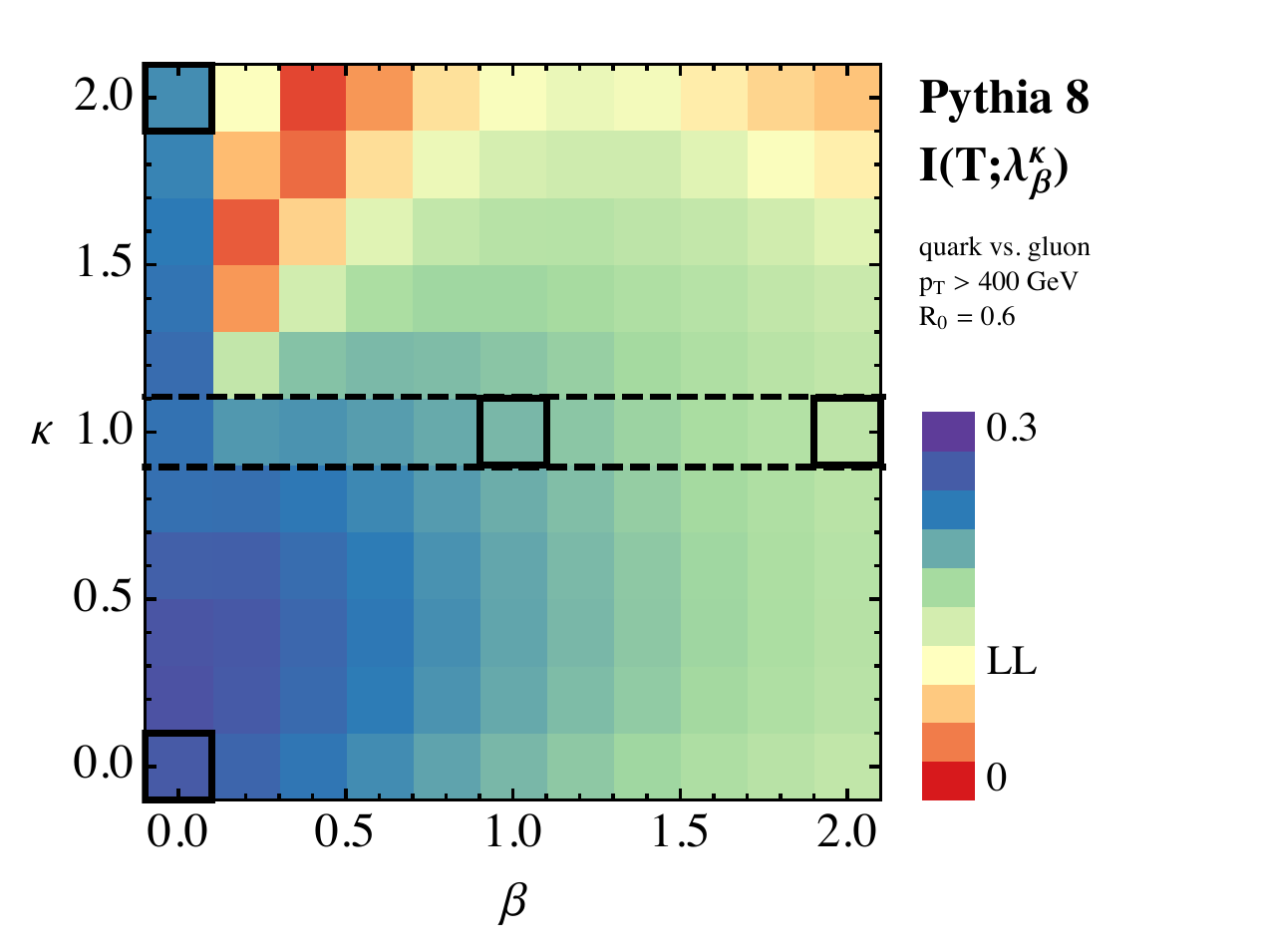}
}
\subfloat[]{\label{fig:herwigsinglelambda}
\includegraphics[width=7.5cm]{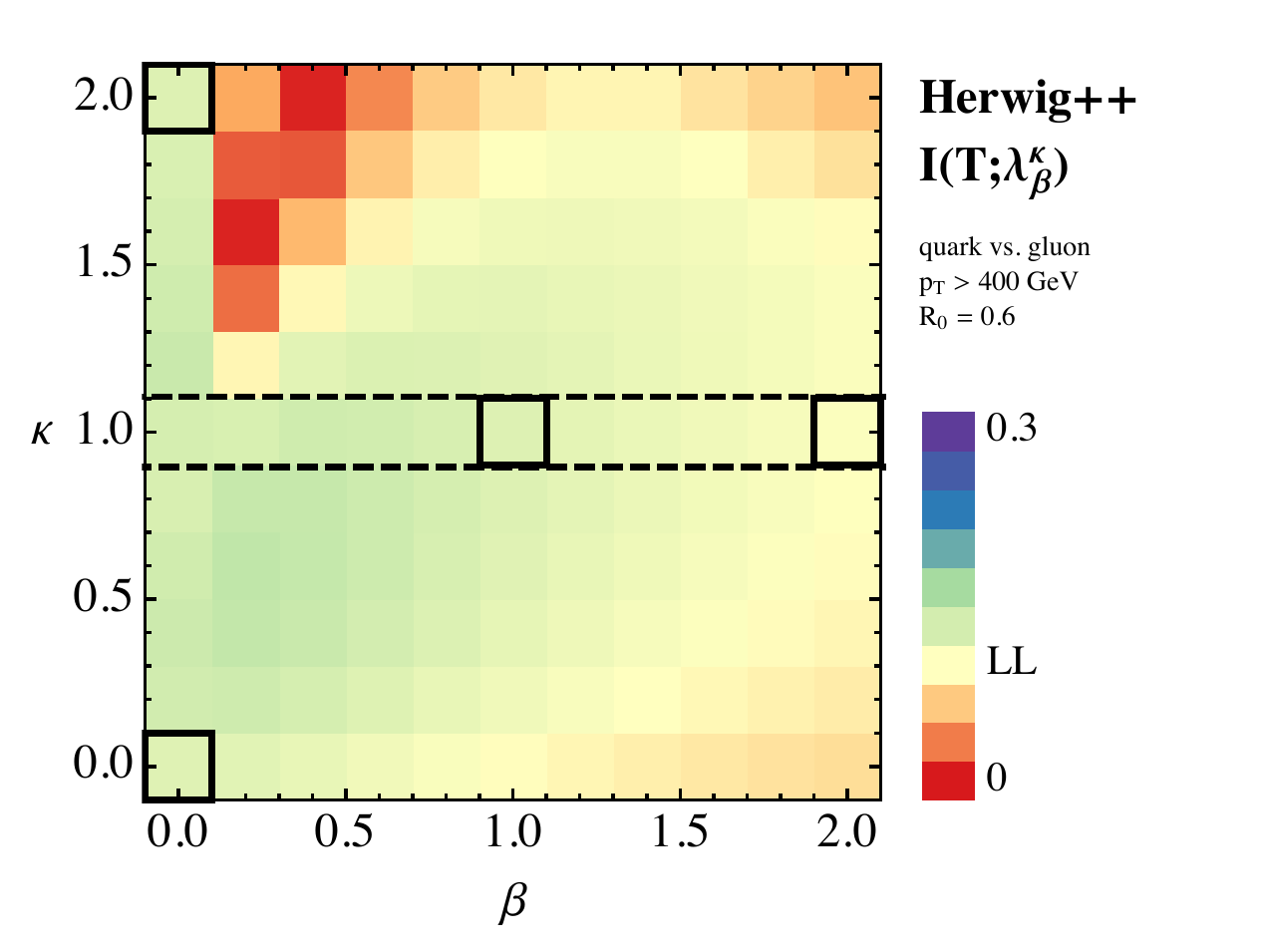}
}
\\
\subfloat[]{\label{fig:pythiazerozeropluslambda}
\includegraphics[width=7.5cm]{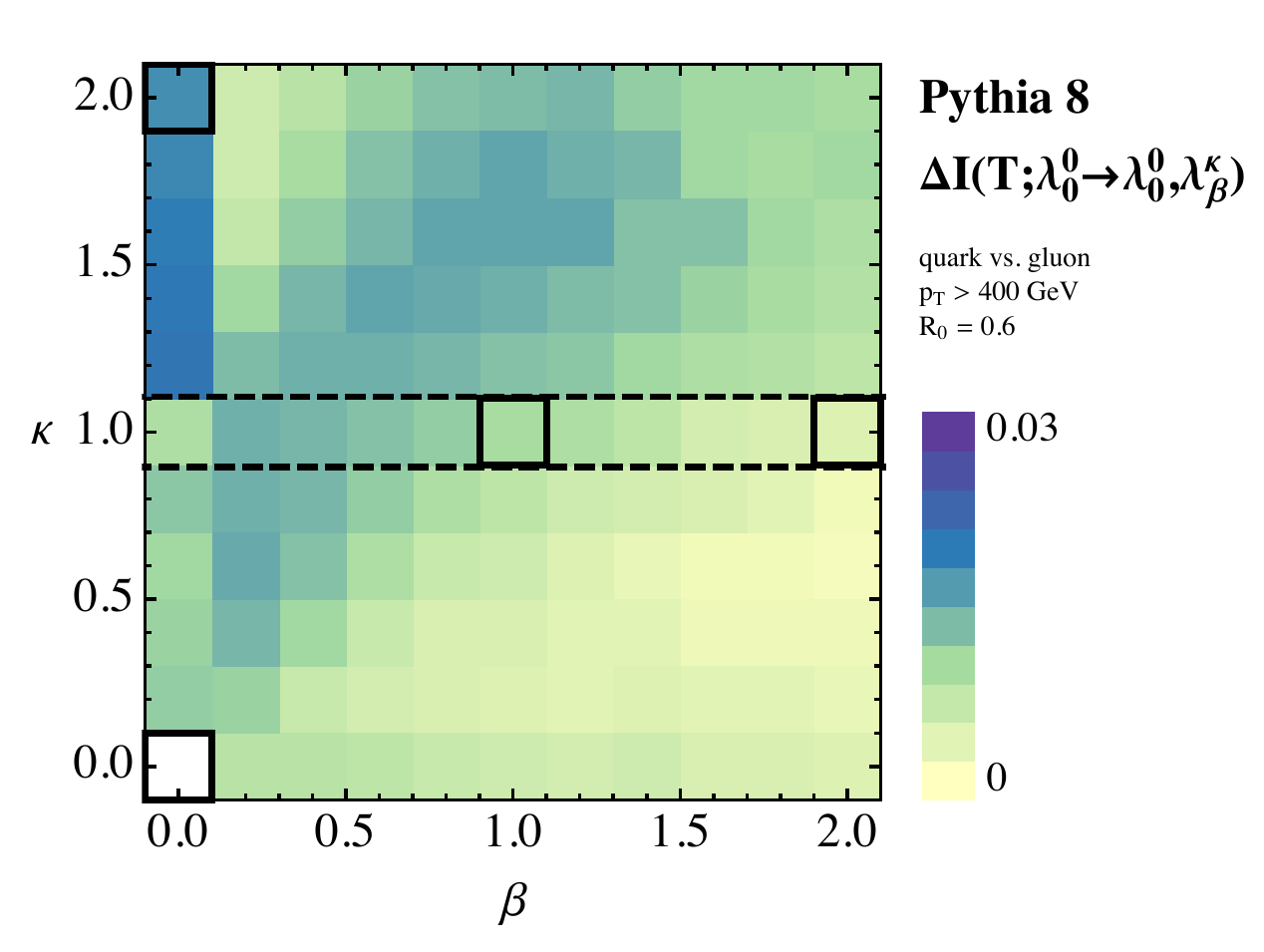}
}
\subfloat[]{\label{fig:herwigzerozeropluslambda}
\includegraphics[width=7.5cm]{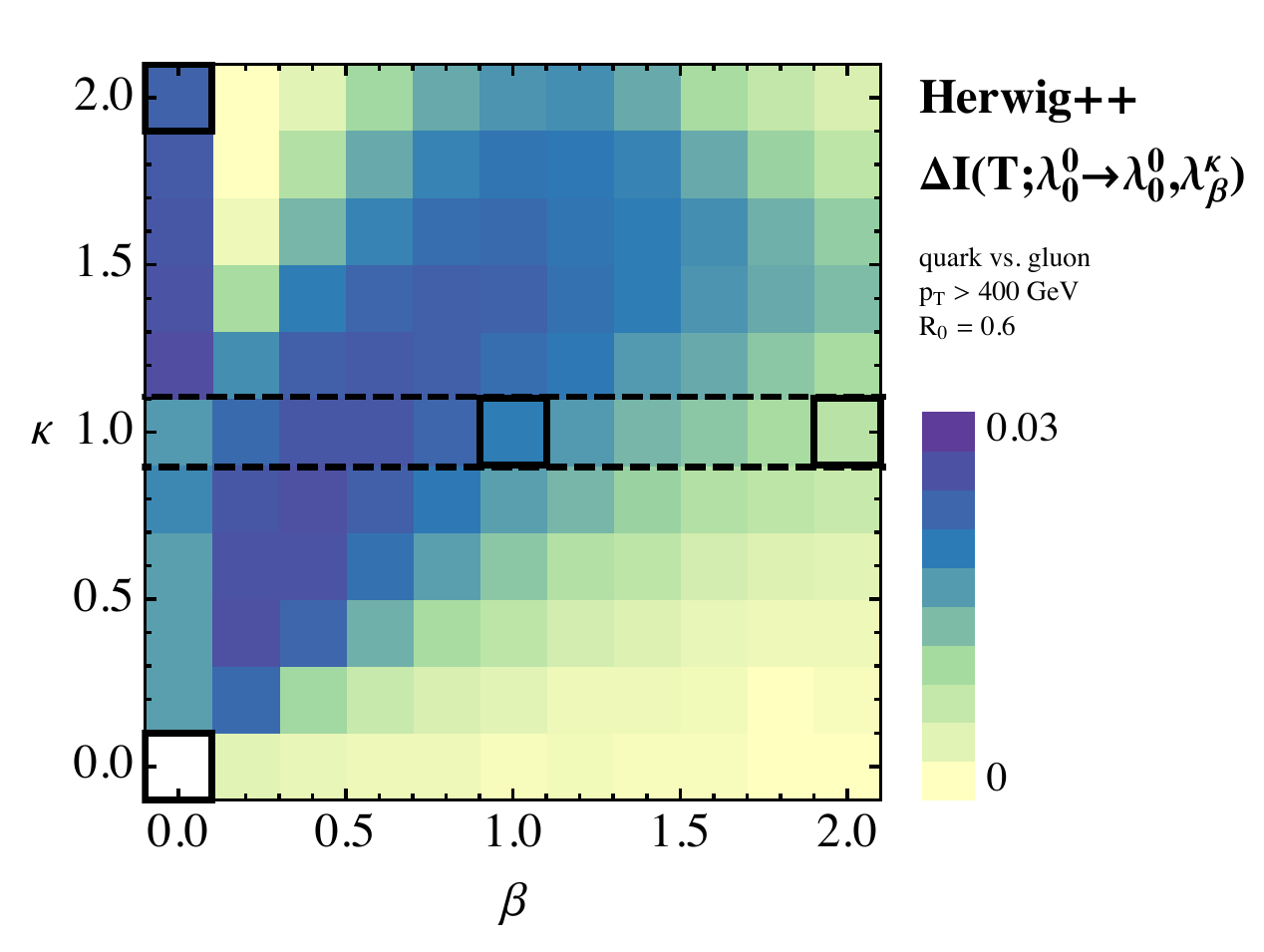}
}
\end{center}
\caption{Parton shower study of quark/gluon discrimination for \pythia{8} (left) and \herwigpp\ (right).  Top:  quark/gluon discrimination power of $\genang{\kappa}{\beta}$ as characterized by the truth overlap $I(T;\genang{\kappa}{\beta})$.  Bottom: improvement in discrimination power from supplementing multiplicity with $\genang{\kappa}{\beta}$, $\Delta I(T;\genang{0}{0} \to \genang{0}{0},\genang{\kappa}{\beta}) \equiv I(T;\genang{0}{0},\genang{\kappa}{\beta}) - I(T;\genang{0}{0})$.  The small solid boxes correspond to the dots indicated in \Fig{fig:lambdaspace}, the wide dashed box indicates the IRC safe angularities $e_\beta$, and ``LL'' in light yellow indicates the result from Casimir scaling (i.e.~$I(T;\genang{\kappa}{\beta}) \simeq 0.1$ from \Eq{eq:LLang_base}).
}
\label{fig:pythiainitialstudy}
\end{figure}

To get a feel for the performance of the various $\genang{\kappa}{\beta}$, we can use parton shower simulations to estimate their quark/gluon truth overlap.  We generate an equal admixture of quark and gluon jets (i.e.~$f = 1/2$) from the processes $qq\to qq$ and $gg\to gg$ using \pythia{8.183} \cite{Sjostrand:2006za, Sjostrand:2007gs} and \herwigpp{2.6.3} \cite{Bahr:2008pv,Gieseke:2011na} at the 8 TeV LHC.\footnote{The choice of 8 TeV allows us to use the same event sample and event selection as \Ref{Larkoski:2013eya}.  Results at 14 TeV are qualitatively similar.}  The transverse momenta of the jets is required to be $p_T>400$ GeV with a jet radius of $R_0 = 0.6$.  To avoid any effects from recoil \cite{Catani:1992jc,Dokshitzer:1998kz,Banfi:2004yd,Larkoski:2013eya,Larkoski:2014bia}, we identify jets using 1-jettiness~\cite{Stewart:2010tn,Thaler:2010tr} as a jet finder \cite{Thaler:2011gf}, taking the winner-take-all axis \cite{Bertolini:2013iqa,Larkoski:2014uqa,Salambroadening} as the jet center.  This style of jet finding always returns one perfectly circular jet cone, and \fastjet{3.0.3} \cite{Cacciari:2011ma} code is available from the \texttt{Nsubjettiness} package through the \fastjet{} contrib project (\url{http://fastjet.hepforge.org/contrib/}).\footnote{We thank T.J.~Wilkason for providing a beta version of his code.}

In \Figs{fig:pythiasinglelambda}{fig:herwigsinglelambda}, we show the truth overlap $I(T;\genang{\kappa}{\beta})$ from \Eq{eq:ITA} for different choices of $\genang{\kappa}{\beta}$.\footnote{As discussed in \App{app:entropystats}, there is an important subtlety in calculating mutual information for binned samples with finite statistics.  To avoid sample size artifacts, we use the same number of events to estimate $p_q(a)$, $p_g(a)$, and $p_{\rm tot}(a)$.}  Confirming the results of \Refs{Gallicchio:2011xq,Gallicchio:2012ez}, one of the best single discriminant variables is $\unsafeang{0}$ (i.e.~hadron multiplicity).  In \Figs{fig:pythiazerozeropluslambda}{fig:herwigzerozeropluslambda}, we show the truth gain 
%%%
\be
\Delta I(T;\genang{0}{0} \to \genang{0}{0}, \genang{\kappa}{\beta}) \equiv I(T;\genang{0}{0},\genang{\kappa}{\beta}) - I(T;\genang{0}{0}),
\ee
%%%
which is a measure of the information gain by using a second $\genang{\kappa}{\beta}$ in addition to $\genang{0}{0}$.  We see that observables like $\genang{1}{1} \equiv \safeang{1}$ (i.e.~width) and $\unsafeang{2}$ (i.e.~$p_T^D$) do add additional information, in agreement with LHC performance studies \cite{CMS-PAS-JME-13-002,CMS-PAS-JME-13-005,Aad:2014gea}.

Of course, these parton shower results should be taken as just illustrative, especially since it is known that \pythia{8} typically overestimates the quark/gluon separation power \cite{Aad:2014gea}.  The differences between \pythia{8} and \herwigpp\ are quite striking, but the origin of the disagreement is not known at present.  For this reason, we want to calculate $I(T;\genang{\kappa}{\beta})$ from first principles to predict which observable (or combination of observables) has the best discrimination power, which is the subject of the next sections.

%%%%%%%%%%%%%%%%%%%%%%%%%%%%%%%%%%%%%%%%%%%%%%%%%%%%%%%%%%%%%%%%%%%%%%%%%%%%%%%%
\section{IRC Safe Angularities}
\label{sec:ircsafeang}
%%%%%%%%%%%%%%%%%%%%%%%%%%%%%%%%%%%%%%%%%%%%%%%%%%%%%%%%%%%%%%%%%%%%%%%%%%%%%%%%

We start our analytic studies with the IRC safe limit $\kappa = 1$, corresponding to the recoil-free angularities $\safeang{\beta} \equiv \genang{1}{\beta}$.  For all $\beta>0$, these are IRC safe.  To the order of accuracy of our calculations, $\safeang{\beta}$ are identical to the energy correlation functions $C_1^{(\beta)}$ \cite{Larkoski:2013eya}.  The case $\beta=1$ is also known as width (or broadening or girth) and $\beta = 2$ is known as thrust (which is related to mass-squared at a fixed jet energy).

It was observed in \Ref{Larkoski:2013eya} that the recoil-free angularities are good quark/gluon discriminants, with better performance at fixed $\beta$ than the traditional recoil-sensitive angularities (i.e.\ angularities measured with respect to the jet momentum axis).   The discrimination power of $\safeang{\beta}$ increased as the angular exponent $\beta$ decreases towards zero, and we will verify this behavior from the mutual information viewpoint.  In addition, using the double differential cross sections from \Refs{Larkoski:2013paa,Larkoski:2014tva}, we can study the correlations between different angularities $\safeang{\alpha}$ and $\safeang{\beta}$ to show how using additional information can improve tagging performance.

%~~~~~~~~~~~~~~~~~~~~~~~~~~~~~~~~~~~~~~~~~~~~~~~~~~~~~~~~~~~~~~~~~~~~~~~~~~~~~~~
\subsection{Truth Overlap for One Angularity}
\label{subset:oneAng}
%~~~~~~~~~~~~~~~~~~~~~~~~~~~~~~~~~~~~~~~~~~~~~~~~~~~~~~~~~~~~~~~~~~~~~~~~~~~~~~~

The properties of $\safeang{\beta}$ are particularly simple at LL accuracy.\footnote{We define logarithmic accuracy through the cumulative distribution of the observable of interest.  For an observable $e$, the cumulative distribution has the expansion
%%%
\begin{equation}
\label{eq:log_exp}
\ln \Sigma(e) = \alpha_s \ln^2 e+ \alpha_s \ln e +\alpha_s +{\cal O}(\alpha_s^2) \ .
\end{equation}
%%%
We define ``LL'' as keeping the leading terms in this expansion with the scaling $\alpha_s \ln^2 e \sim 1$.}
The normalized cumulative distribution of the angularity $\safeang{\beta}$ was computed in, e.g. \Ref{Larkoski:2013eya}:
%%%
\begin{equation}
\label{eq:LLangcum}
\Sigma_i(\safeang{\beta}) = \exp\Big(-\frac{\alpha_s}{\pi}\,\frac{C_i}{\beta}\,\ln^2 \safeang{\beta}\Big) .
\end{equation}
%%%
Here, $C_i$ is the color of the jet:  $C_F = 4/3$ for quarks and $C_A = 3$ for gluons.  This distribution satisfies the Casimir scaling property of \Eq{eq:defcasimirscaling}, and therefore the truth overlap $I(T;\safeang{\beta})$ is given by the formula in \Eq{eq:mutualinfocasiiarscaling}, independent of $\beta$.

To determine the $\beta$-dependence of $I(T;\safeang{\beta})$, we have to go to next-to-leading logarithmic (NLL) accuracy, as in \Ref{Larkoski:2013eya}.\footnote{We define ``NLL'' as the leading terms in the expansion of \Eq{eq:log_exp} with the scaling $\alpha_s \ln e \sim 1$.}  We use the NLL distributions for the recoil-free angularities computed in \Ref{Larkoski:2014uqa} (which are identical to the NLL resummation of the energy correlation functions from \Ref{Banfi:2004yd}) and compute the mutual information of the angularities with truth.  For $\beta > 1$, our NLL distributions correspond to the calculations for (recoil-sensitive) angularities performed in \Ref{Ellis:2010rwa}. For reference, the cumulative distributions are given in \App{app:IRCsafeNLL}, and we determine $I(T;\safeang{\beta})$ through numeric integration.

\begin{figure}
\begin{center}
\subfloat[]{\label{fig:NLL_mutinf}
\includegraphics[width=7cm]{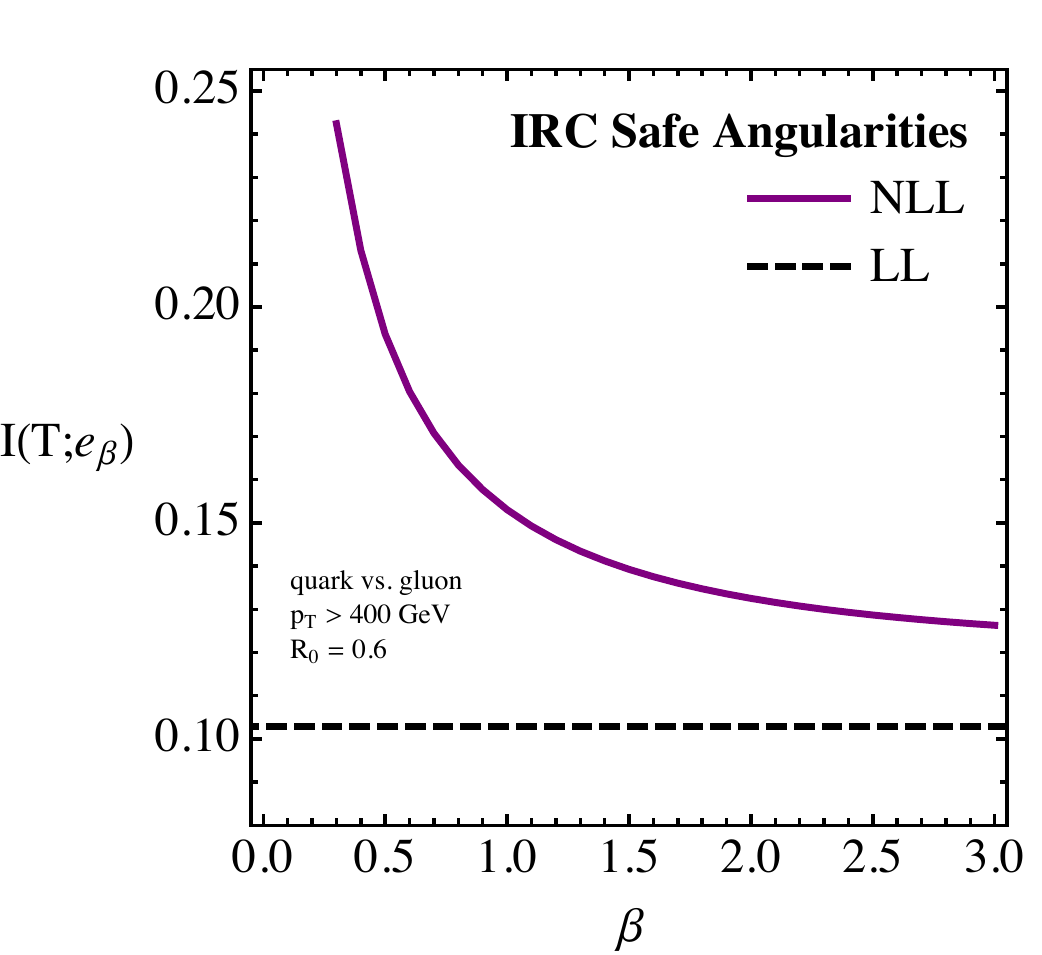}
}
$\qquad$
\subfloat[]{\label{fig:PY_mutinf} 
\includegraphics[width=7cm]{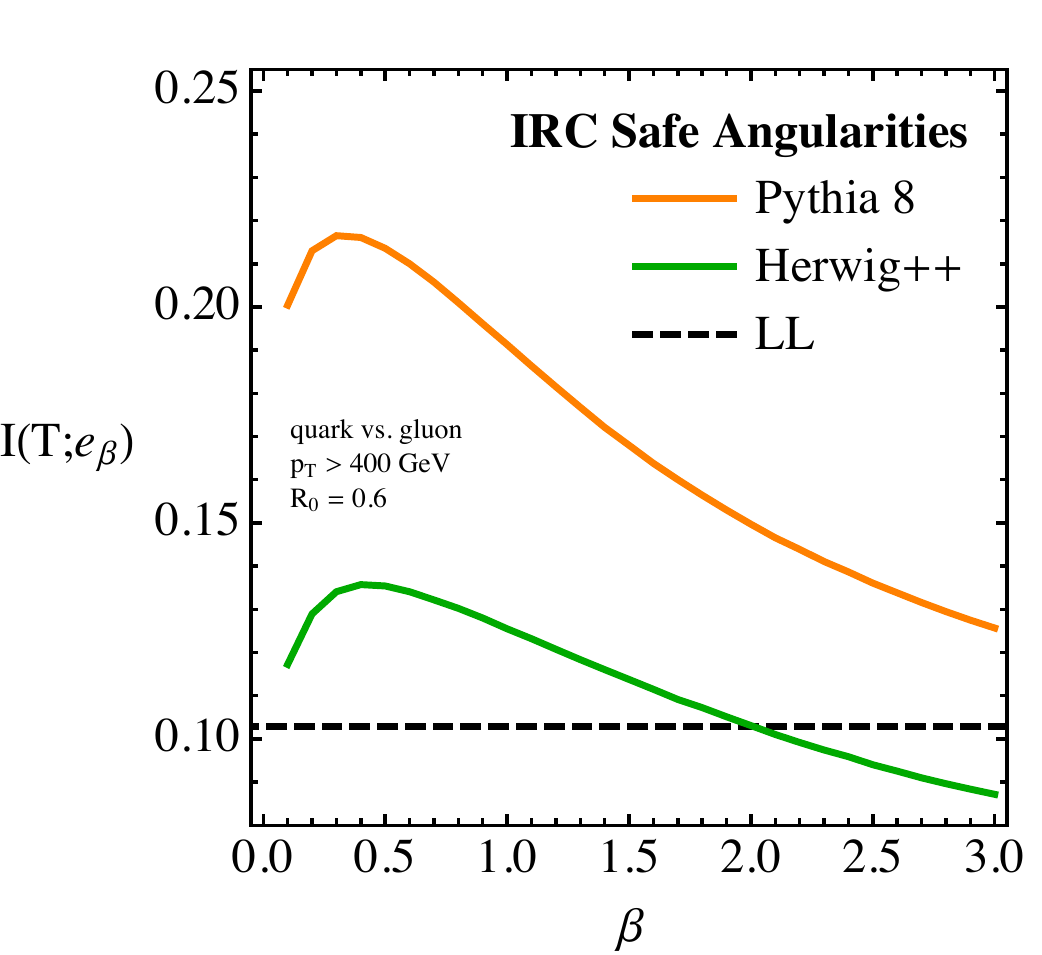}
}
\end{center}
\caption{
The quark/gluon truth overlap for an individual IRC safe angularity $\safeang{\beta}$ as a function of angular exponent $\beta$.  The transverse momentum of the jets is $p_T > 400$ GeV and the jet radius is $0.6$.  Left: comparing the NLL truth overlap to the baseline LL result.  Right:  comparing the  \pythia{8} and \herwigpp\ samples. 
}
\label{fig:mutinf_plot}
\end{figure}

The truth overlap $I(T;\safeang{\beta})$ as a function of the angular exponent $\beta$ is shown in \Fig{fig:mutinf_plot} for $f=1/2$.  The left plot is from the NLL calculation and the right plot shows \pythia{8} and \herwigpp, using the same event generation settings as in \Sec{sec:angdef} (i.e.\ $p_T> 400$ GeV and $R_0 = 0.6$).  The LL result from Casimir scaling is plotted for reference.  We see that $I(T;\safeang{\beta})$ increases significantly as $\beta$ decreases, showing that the quark/gluon discrimination improves. As discussed in \Ref{Larkoski:2013eya}, the qualitative $\beta$-dependence is the same at NLL compared to the two parton shower programs, but there are significant numerical differences.  Part of that is because the NLL result is lacking effects like nonperturbative power corrections which modify  the quark/gluon discrimination power.  The large difference between \pythia{8} and \herwigpp\ has been seen in other contexts \cite{Aad:2014gea}, and the underlying reason is as-of-yet unknown.

%~~~~~~~~~~~~~~~~~~~~~~~~~~~~~~~~~~~~~~~~~~~~~~~~~~~~~~~~~~~~~~~~~~~~~~~~~~~~~~~
\subsection{Truth Overlap for Two Angularities}
\label{sec:twoAng}
%~~~~~~~~~~~~~~~~~~~~~~~~~~~~~~~~~~~~~~~~~~~~~~~~~~~~~~~~~~~~~~~~~~~~~~~~~~~~~~~

We now turn to a study of the quark/gluon discrimination power of two angularities.  This will highlight the analytic benefits of using mutual information (instead of ROC curves) to study correlated observables.  Constructing the ROC curve for more than a single observable is a formidable challenge because contours of constant signal/background significance can be non-trivial functions of the observables.  Typically, the procedure for determining the discrimination power is to use a multivariate analysis (MVA) such as a boosted decision tree.  In contrast, mutual information is defined by simply integrating over the joint probability distribution, so all correlations between observables are automatically taken into account.

At LL accuracy, the double differential cross section of two angularities was computed in \Ref{Larkoski:2013paa}.  For angularities $\safeang{\alpha}$ and $\safeang{\beta}$ with different angular exponents $\al > \beta$, the double cumulative distribution is
%%%
\begin{equation}\label{eq:LLdoubSud}
\Sigma_i(\safeang{\alpha},\safeang{\beta})= \exp\left[{- \frac{\alpha_s}{\pi}C_i\left( \frac{\ln^2 \safeang{\beta}}{\beta}+\frac{\ln^2 \frac{\safeang{\alpha}}{\safeang{\beta}}}{\alpha-\beta}  \right)} \right].
\end{equation}
%%%
At this order, the angularities satisfy the inequalities
\be
\label{eq:safephasespace}
\safeang{\beta}>\safeang{\alpha}, \qquad (\safeang{\alpha})^\beta > (\safeang{\beta})^\alpha.
\ee
While the LL distribution does exhibit Casimir scaling, it does so for a multivariate exponential function, so the analysis of \Sec{sec:casimir} does not apply.  The double differential cross section is defined by differentiating  
%%%
\begin{equation}
\frac{\df^2\sigma_i}{\df \safeang{\alpha}\, \df \safeang{\beta}} = \left( \frac{\partial^2}{\partial \safeang{\alpha} \, \partial \safeang{\beta}}\,\Sigma_i(\safeang{\alpha},\safeang{\beta}) \right) \Theta_0(\safeang{\alpha},\safeang{\beta}),
\end{equation}
%%%
with explicit expressions in \App{app:twoIRCsafeNLL}.  The function $\Theta_0$ enforces the phase space restrictions in \Eq{eq:safephasespace}, and has to be \emph{outside} of the derivatives.

At NLL accuracy, the (conjectured) double differential cross section was determined in \Ref{Larkoski:2014tva} by interpolating between effective theories at the $\safeang{\beta} = \safeang{\alpha}$ and $(\safeang{\alpha})^\beta = (\safeang{\beta})^\alpha$ boundaries of phase space.  The NLL expression is given in \App{app:twoIRCsafeNLL} for reference, and an equivalent derivation using Soft-Collinear Effective Theory (SCET)~\cite{Bauer:2000ew,Bauer:2000yr, Bauer:2001ct, Bauer:2001yt} is given in \App{app:scet_pair_interpolation}.

\begin{figure}
\begin{center}
\subfloat[]{\label{fig:LL_ang_mutinf}
\includegraphics[width=7cm]{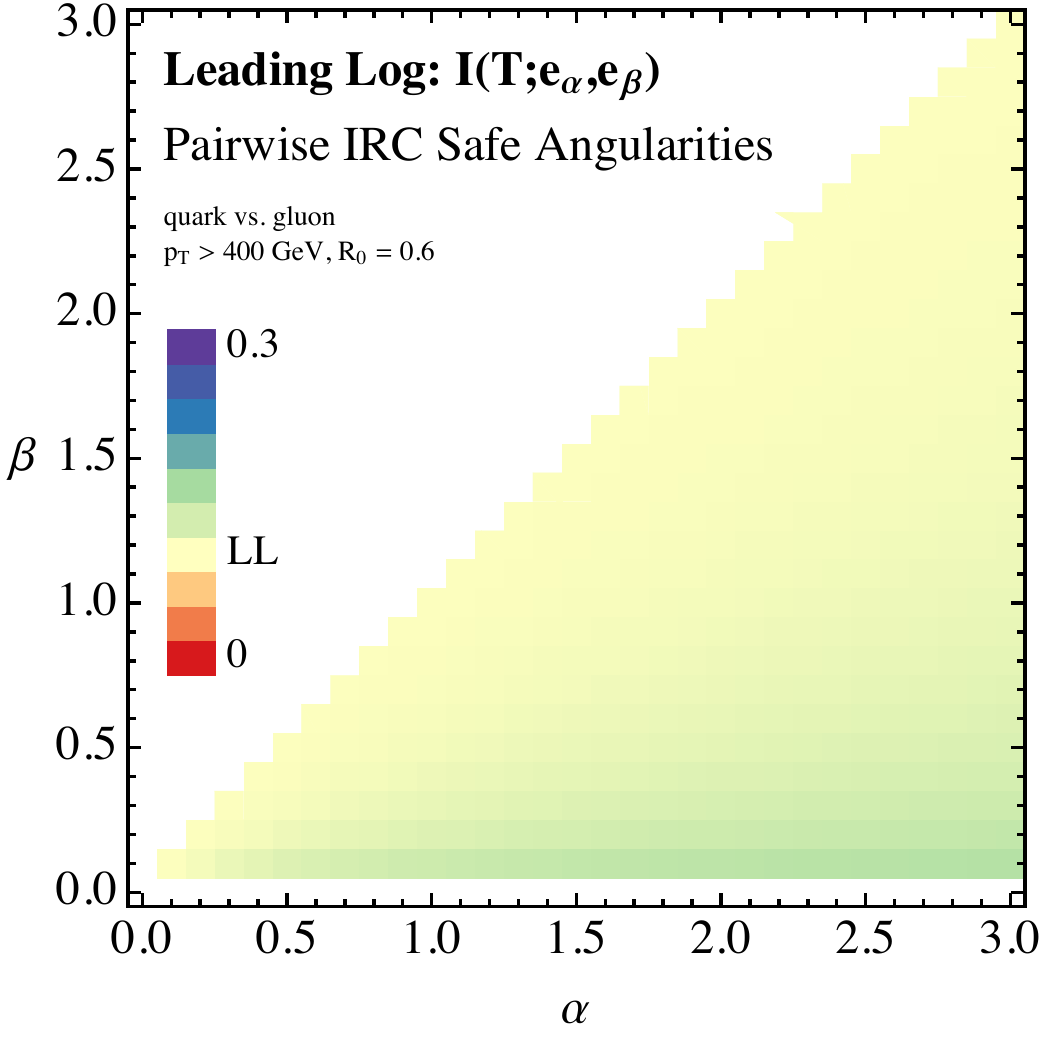}
}
$\qquad$
\subfloat[]{\label{fig:NLL_ang_mutinf} 
\includegraphics[width=7cm]{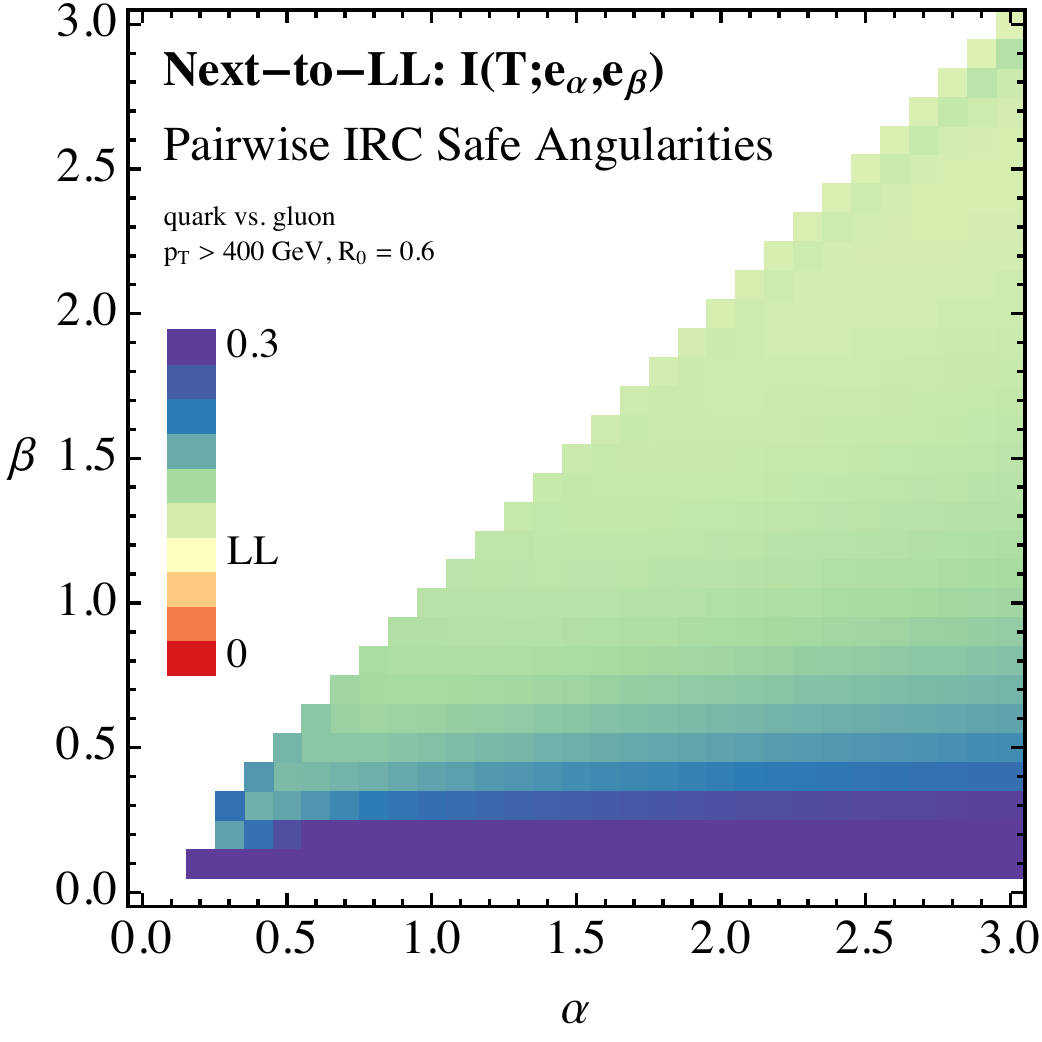}
}\\
\subfloat[]{\label{fig:pythia_ang_mutinf} 
\includegraphics[width=7cm]{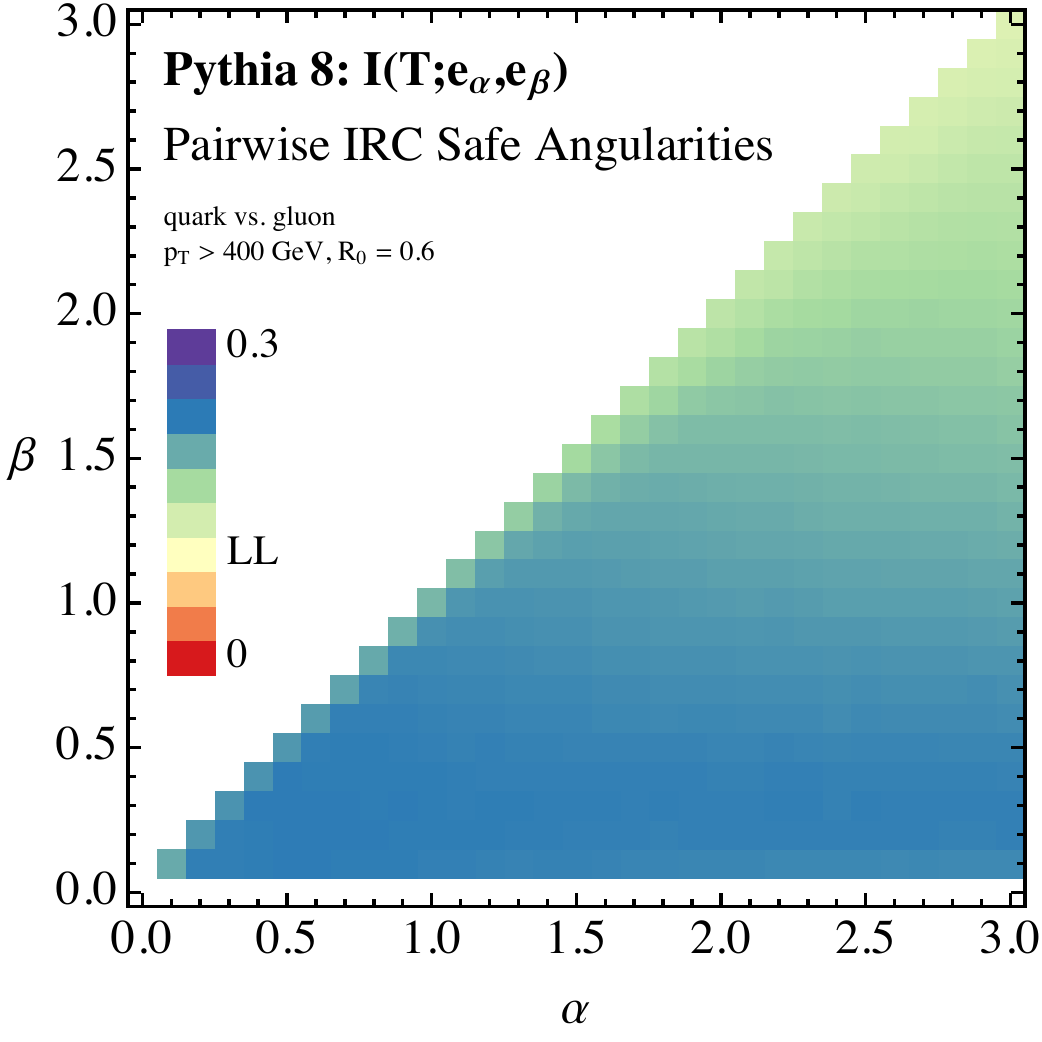}
}$\qquad$
\subfloat[]{\label{fig:herwig_ang_mutinf} 
\includegraphics[width=7cm]{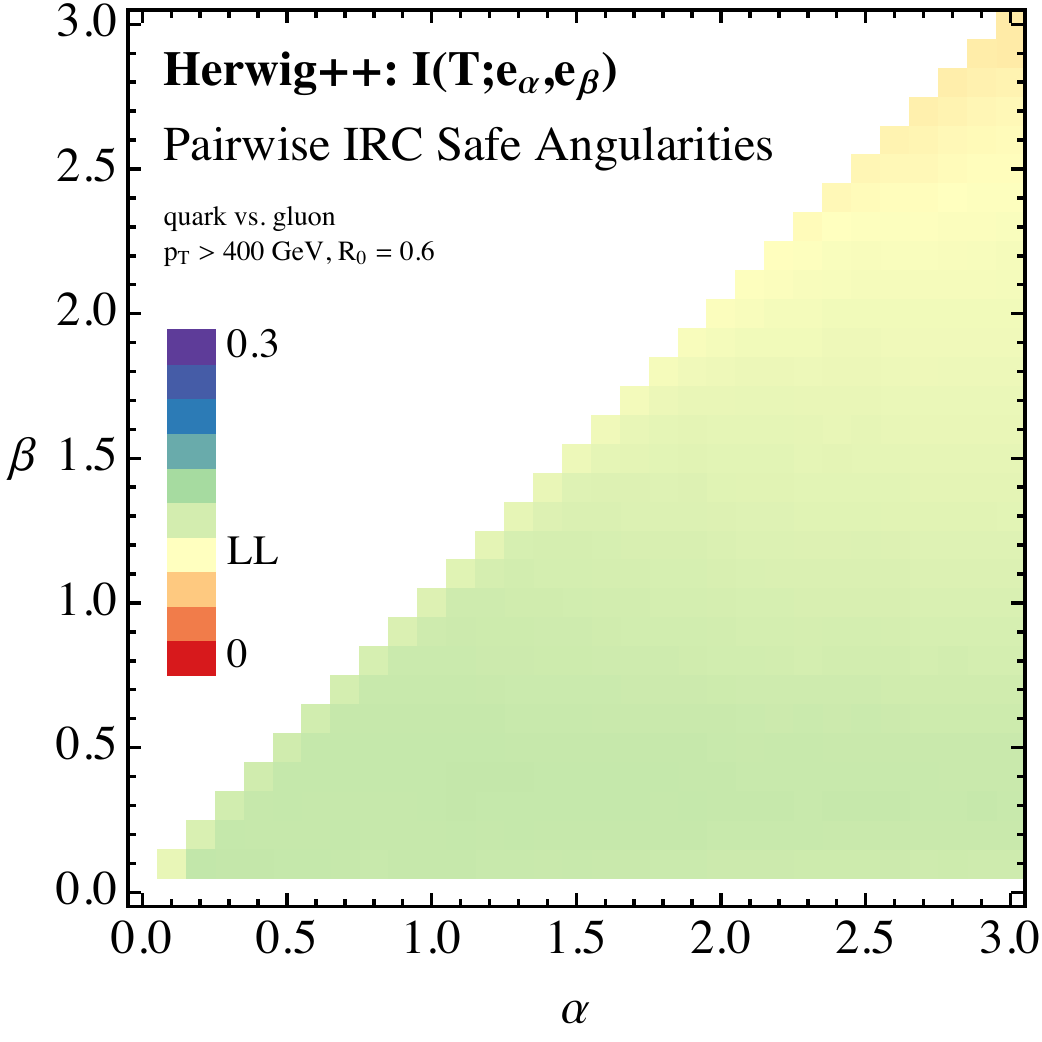}
}
\end{center}
\caption{
The quark/gluon truth overlap for pairs of IRC safe angularities $(\safeang{\alpha},\safeang{\beta})$.  Top: the LL and NLL analytic calculations.  Bottom: the \pythia{8} and \herwigpp\ parton showers.  The single observable LL baseline ($I(T;\safeang{\beta}) \simeq 0.1$) is indicated by light yellow.  Note that the LL and NLL results are only trustable for $\beta \gtrsim 0.5$.  Also, near the $\alpha = \beta$ diagonal, the NLL results suffer from numerical issues due to the small phase space allowed by \Eq{eq:safephasespace}.}
\label{fig:ang_mutinf_truth}
\end{figure}

In \Fig{fig:ang_mutinf_truth} we show the truth overlap $I(T;\safeang{\alpha},\safeang{\beta})$, comparing the LL expression, the NLL expression, \pythia{8}, and \herwigpp.  As before we have set the quark fraction $f=0.5$, and the diagonal entries correspond to the single observable values from \Fig{fig:mutinf_plot}.  From the baseline LL value of a single angularity in \Eq{eq:LLang_base} (i.e.\ $I(T;\safeang{\beta}) \simeq 0.1$) the truth overlap can be increased noticably even at LL.  For example, for angularities $e_2$ and $e_{0.5}$, the joint truth overlap is greater than $0.12$ at LL.  At NLL, the discrimination power uniformly rises, as expected from \Fig{fig:mutinf_plot}.  Because our NLL expressions do not account for the nonperturbative region of phase space, one should be cautious interpreting the results for $\beta \lesssim 0.5$.  Turning to the parton showers, they give quite different prediction for $I(T;\safeang{\alpha},\safeang{\beta})$, with \pythia{8} even more optimistic than the NLL result and \herwigpp\ closer to the LL result.  

The large numerical differences between these methods highlights the considerable theoretical uncertainties present in quark/gluon discrimination.  It is important to note that these large differences are not present when trying to model quarks and gluons individually, and only arise in the context of discrimination.  In \App{app:correlations} we show the mutual information $I(\safeang{\alpha};\safeang{\beta})$, which measures the degree of correlation between two angularities on separate quark and gluon samples.  The four methods (LL, NLL, \pythia{8}, and \herwigpp) show much closer agreement for $I(\safeang{\alpha};\safeang{\beta})$ than for $I(T;\safeang{\alpha},\safeang{\beta})$, suggesting that the truth overlap is more sensitive to subtle (and difficult to predict) differences between quark and gluon jets.

For completeness, in \App{app:correlations} we show the truth gain
%%%
\be
\label{eq:DeltaITemax}
\Delta I (T,e_{\rm max} \to \safeang{\alpha},\safeang{\beta}) \equiv I(T;\safeang{\alpha},\safeang{\beta}) - \max\{I(T;\safeang{\alpha}),I(T;\safeang{\beta})\},
\ee
%%%
which makes it easier to see that there is improved quark/gluon discrimination power from measuring two angularities instead of just one.\footnote{Of course, while mutual information is helpful to characterize the possible gains from combining observables, a multivariate analysis is still needed to realize these gains in practice.  As discussed in \App{app:ROCatLL}, it is challenging to determine the optimal cuts analytically, even at LL.  Alternatively, in the spirit of \Ref{Soper:2011cr}, one could use the ratio of the quark/gluon double differential distributions as a weighting factor.}   Roughly speaking, pairs of angularities with the smallest values of $I(\safeang{\alpha};\safeang{\beta})$ (i.e.~least correlation) lead to the largest increase in $I(T;\safeang{\alpha},\safeang{\beta})$ (i.e.~discrimination power), though there is considerable variability.  Because the four methods have different predictions for which angularities should be combined, care should be taken when using any of these methods to estimate quark/gluon discrimination performance.

%%%%%%%%%%%%%%%%%%%%%%%%%%%%%%%%%%%%%%%%%%%%%%%%%%%%%%%%%%%%%%%%%%%%%%%%%%%%%%%%
\section{IRC Unsafe Angularities}
\label{sec:unsafeang}
%%%%%%%%%%%%%%%%%%%%%%%%%%%%%%%%%%%%%%%%%%%%%%%%%%%%%%%%%%%%%%%%%%%%%%%%%%%%%%%%

We now turn to the more interesting case of the IRC unsafe angularities with $\kappa \not= 1$.  As seen in \Sec{sec:angdef} and known in the literature, hadron multiplicity ($\unsafeang{0}$) and $p_T^D$ ($\unsafeang{2}$) are effective quark/gluon discriminants.  But much of the rest of the $(\kappa,\beta)$ plane is still unexplored (apart from the angularity line at $\kappa = 1$).

One challenge to gaining an analytic understanding of the $\kappa\neq1$ case is that $\genang{\kappa}{\beta}$ is collinear unsafe (unlike $\safeang{\beta}$ studied above).  This introduces an intrinsic sensitivity to nonperturbative physics that describes how the emitted radiation is split into hadrons, prohibiting a purely perturbative calculation.  That said, using the techniques developed in~\Refs{Krohn:2012fg,Waalewijn:2012sv,Chang:2013rca,Chang:2013iba}, we can encode the nonperturbative information into a ``weighted-energy function''  which can be extracted from data.  In fact, for $\beta >0 $, we will only need a few nonperturbative parameters (and not a whole function) to characterize the distributions.

\begin{figure}
\begin{center}
\includegraphics[scale = 0.7]{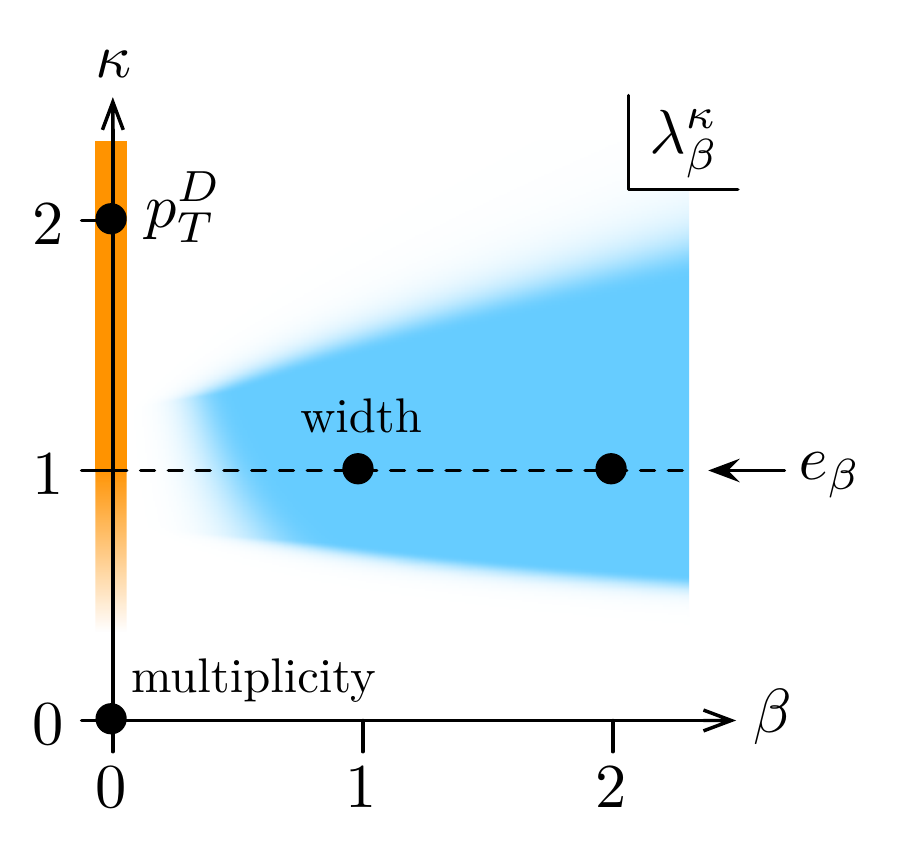}
\end{center}
\caption{The regions of the space of observables $\genang{\kappa}{\beta}$ that we calculate are shown in orange ($\beta=0$ and $\kappa \gtrsim 0.5$, \Sec{subsec:beta0}) and blue ($\beta \kappa \gtrsim 0.5$ and $\beta \kappa/ (1-\kappa)^2 \gtrsim 6$, \Sec{subsec:betaposregime}).  As explained in \Sec{subsec:betaposregime}, the funny shape of the blue region is due to a combination of perturbative and nonperturbative constraints.
}
\label{fig:lambdaspace2}
\end{figure}

Strictly speaking our calculations will only be valid for $\kappa \gtrsim 0.5$.  The reason is that as $\kappa \to 0$ the observable also becomes infrared unsafe, further complicating calculations (as discussed in the context of hadron multiplicities in e.g.~\Ref{Capella:1999ms,Bolzoni:2012ii}).  Also note that the $\beta = 0$ and $\beta > 0$ regimes are very different in how they treat collinear radiation, so we will consider them separately.  The approximate range of validity of the calculations are shown in \Fig{fig:lambdaspace2}.

%~~~~~~~~~~~~~~~~~~~~~~~~~~~~~~~~~~~~~~~~~~~~~~~~~~~~~~~~~~~~~~~~~~~~~~~~~~~~~~~
\subsection{The $\beta = 0$ Regime}
\label{subsec:beta0}
%~~~~~~~~~~~~~~~~~~~~~~~~~~~~~~~~~~~~~~~~~~~~~~~~~~~~~~~~~~~~~~~~~~~~~~~~~~~~~~~

We start with the $\beta = 0$ case with $\unsafeang{\kappa}$.  Recall from \Eq{eq:kappatoone} that $\unsafeang{1}$ effectively refers to the observable $\sum_i z_i \ln z_i$.  In order to study these observables, we need to introduce a nonperturbative object called the weighted-energy function $F^i_\kappa(x,\mu)$ that describes how the energy of a jet is distributed among its constituent hadrons.  Here, $i$ labels the flavor of the jet.  This object is similar to the charge distribution \cite{Waalewijn:2012sv} and the track function \cite{Chang:2013rca,Chang:2013iba} which describe other aspects of the fragmentation of quarks and gluons into hadrons.   

The quark weighted-energy function has the following operator definition
%%%
\begin{align}
\label{eq:Fdef}
 F^{q}_\kappa(x, \mu)&=
  \frac{1}{2 N_c} \,\sum_{H} \de\Big(x - \sum_{h \in H} (z_h)^\kappa \Big) 
  \nonumber \\ & \quad \times
  \tr\Big[ (\ga^0 + \ga^3)\,
  \big\langle 0 \big|  \big[(2\pi)^3 \de(k^- + \hat p^0 + \hat p^3) \de^2(\hat p_\perp) \psi\big] \big| H
  \big\rangle \big\langle H \big| \overline{\psi} \big|0 \big\rangle \Big]
\,.\end{align}
%%%
Here $\psi$ is the quark field, with momentum fixed by the $\delta$ functions involving the momentum operator $\hat p$, $H$ denotes a hadronic final state, and $z_h = (p_h^0+p_h^3)/k^-$ is the momentum fraction carried by the hadron $h \in H$. (The only dependence on $k^-$ is through $z_h$.) There is a similar definition for the gluon weighted-energy function, and we have suppressed eikonal Wilson lines needed for gauge invariance.  These functions are normalized such that
%%%
\be
\int_0^\infty\! \df x  \, F^i_\kappa(x, \mu) = 1.
\ee
%%%
 As a point of reference, if the hadrons were weighted by their charge, then $F^{i}_\kappa(x)$ would be the jet charge function $D_i(x,\kappa,\mu)$ \cite{Waalewijn:2012sv}.  Alternatively, for $\kappa=1$ and restricted to charged particles, this would be the track function $T_i(x,\mu)$~\cite{Chang:2013rca,Chang:2013iba}.  

At LO, the cross section differential in $\unsafeang{\kappa}$ for a parton of flavor $i$ is simply
%%%
\be
\frac{1}{\si_i}\frac{\df \si_i}{\df \unsafeang{\kappa}} = \int \df x \, F^i_\kappa(x,\mu) \, \delta(\unsafeang{\kappa} - x),
\ee
%%%
meaning that at this order, $F^i_\kappa(x,\mu)$ gives the $\unsafeang{\kappa}$ distribution directly, with $x = \unsafeang{\kappa}$. 
The dependence on the  jet $p_T$ and jet radius $R_0$ enters through the scale choice $\mu = p_T R_0$.  
At NLO, the cross section is~\cite{Waalewijn:2012sv}
%%%
\begin{align} \label{eq:unsafe_NLO}
\frac{1}{\si_i}\frac{\df \si_i}{\df \unsafeang{\kappa}} &= \frac{1}{2} \sum_{j,k} \int\! \df x_1\, \df x_2\, \df z\, \frac{\cJ_{ij}(p_T R_0,z,\mu)}{2(2\pi)^3 J_i(p_T R_0,\mu)}
\nn \\ & \quad \times
 F^j_\kappa(x_1,\mu) F^k_\kappa(x_2,\mu) \, \de\big(\unsafeang{\kappa}-z^\kappa x_1-(1-z)^\kappa x_2\big)
\,.\end{align}
%%%
The ratio $\cJ_{ij}/J_i$ describes the perturbative splitting $i \to jk$, where $j$ has momentum fraction $z$ (see also \Refs{Procura:2009vm,Jain:2011xz}). By including these NLO corrections, the perturbative uncertainty ($\mu$-dependence) is reduced.

\begin{figure}
\begin{center}
\subfloat[]{
\includegraphics[width = 0.45\textwidth]{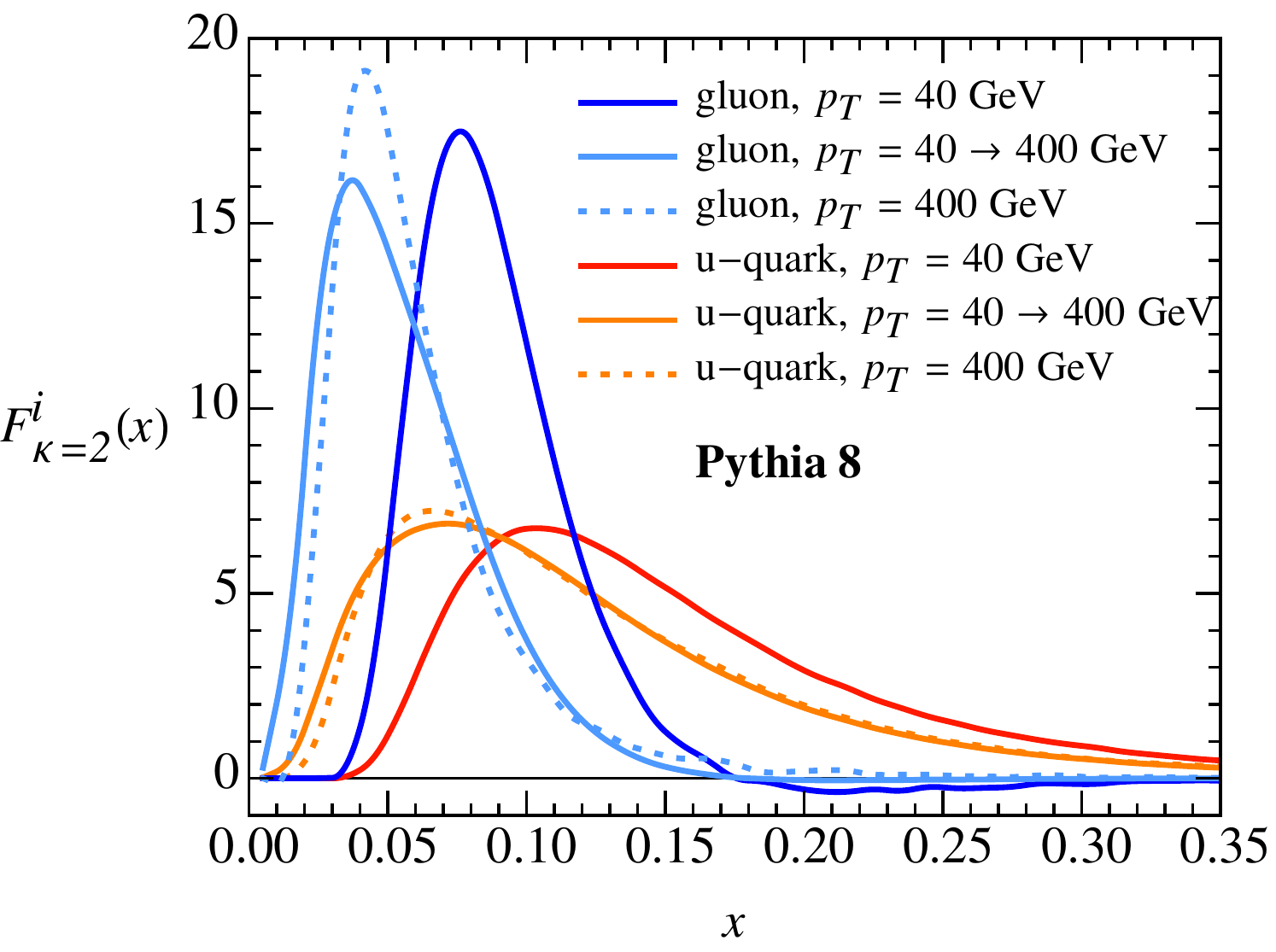}
}
$\qquad$
\subfloat[]{
\includegraphics[width = 0.45\textwidth]{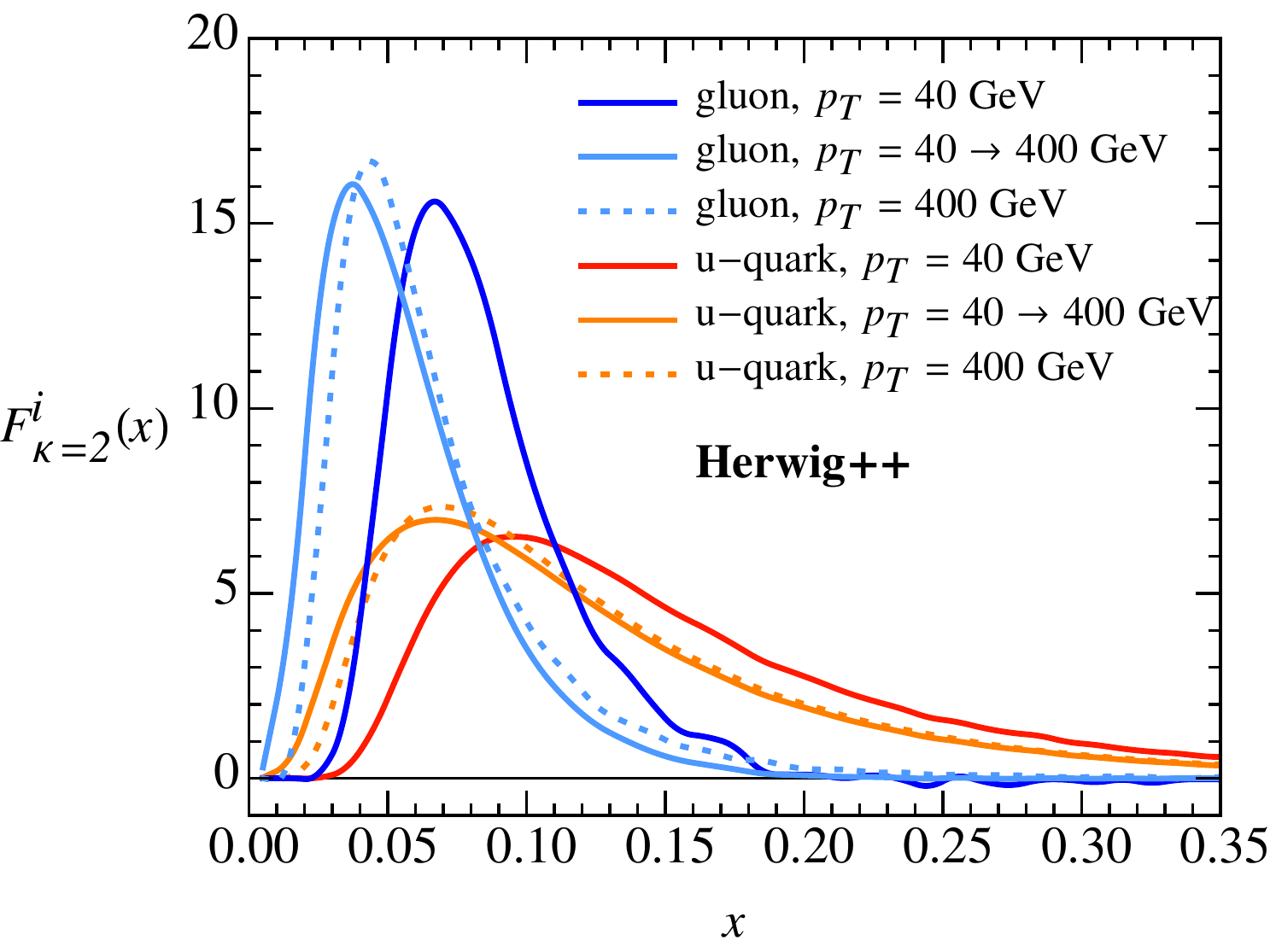}
}
\end{center}
\caption{The weighted-energy function $F^i_{\kappa=2}(x)$ for $p_T^D$ for $u$-quarks (red) and gluons (blue), extracted from \pythia{8} (left) and \herwigpp\ (right).  The darker solid curve is the parton shower results extracted at the scale 40 GeV, the lighter solid curve is the evolution from 40 GeV to 400 GeV using \Eq{eq:RGE}, and the dotted curve is the parton shower results at 400 GeV.  In all cases, we are incorporating the NLO corrections in \Eq{eq:unsafe_NLO}, which is why $F^i_{\kappa=2}(x)$ can be negative.}
\label{fig:ptd_evo}
\end{figure}

The weighted-energy functions are purely nonperturbative, so in that sense, we are not really able to predict the quark/gluon discrimination power of the $\unsafeang{\kappa}$ variables.  But $F^i_\kappa(x,\mu)$ does have a perturbative renormalization group evolution~\cite{Waalewijn:2012sv},  
%%%
\begin{align} \label{eq:RGE}
  \mu \frac{\partial}{\partial \mu}\, F_\kappa^i(x, \mu) &= \frac{1}{2} \sum_{j,k} \int\! \df z\, \df x_1\, \df x_2\, \frac{\al_s}{\pi} P_{i \to j k}(z)
  \nn \\ & \quad \times 
  F_\kappa^j(x_1,\mu) F_\kappa^k(x_2,\mu)\,\de\big(x \!-\! (1\!-\!z)^\kappa x_1 \!-\! z^\kappa x_2\big). 
\end{align}
%%%
Thus, one can measure $F^i_\kappa(x,\mu)$ at one scale (ideally in pure quark/gluon samples), and then evolve to a different scale.  This DGLAP~\cite{Gribov:1972ri, Georgi:1951sr, Gross:1974cs, Altarelli:1977zs, Dokshitzer:1977sg} evolution of $F^i_\kappa$ corresponds to the emissions described by a parton shower in Monte Carlo programs.
We have implemented these evolution equations for $p_T^D$ ($\unsafeang{2}$), reproducing the dependence on the jet $p_T$ observed in \pythia{8} and \herwigpp, as shown in \Fig{fig:ptd_evo}. 

The weighted-energy functions are sufficient for understanding a single $\unsafeang{\kappa}$, but if we want to study correlations between a pair of $\unsafeang{\rho}$ and $\unsafeang{\kappa}$, then we would need a double weighted-energy function:
%%%
\be
\label{eq:doubleweightedenergyfunction}
F_{\rho,\kappa}^i(x_1,x_2;\mu).
\ee
%%%
This is defined analogously to \Eq{eq:Fdef}, albeit with the double measurement
%%%
\be
\de\Big(x_1 - \sum_{h \in H} (z_h)^\rho \Big) \, \de\Big(x_2 - \sum_{h \in H} (z_h)^\kappa \Big).
\ee
%%%
This object also has a renormalization group evolution analogous to \Eq{eq:RGE}.  In  \Sec{subsec:unsafepairwise}, we will use the fact that 
%%%
\be
\label{eq:marginalonexinF}
\int \df x_1 \, F_{\rho, \kappa}^i(x_1,x_2;\mu) = F_\kappa^i(x_2, \mu)
\ee
%%%
when we study correlations between the $\beta > 0$ angularities.

\begin{figure}
\begin{center}
\includegraphics[width = 8cm]{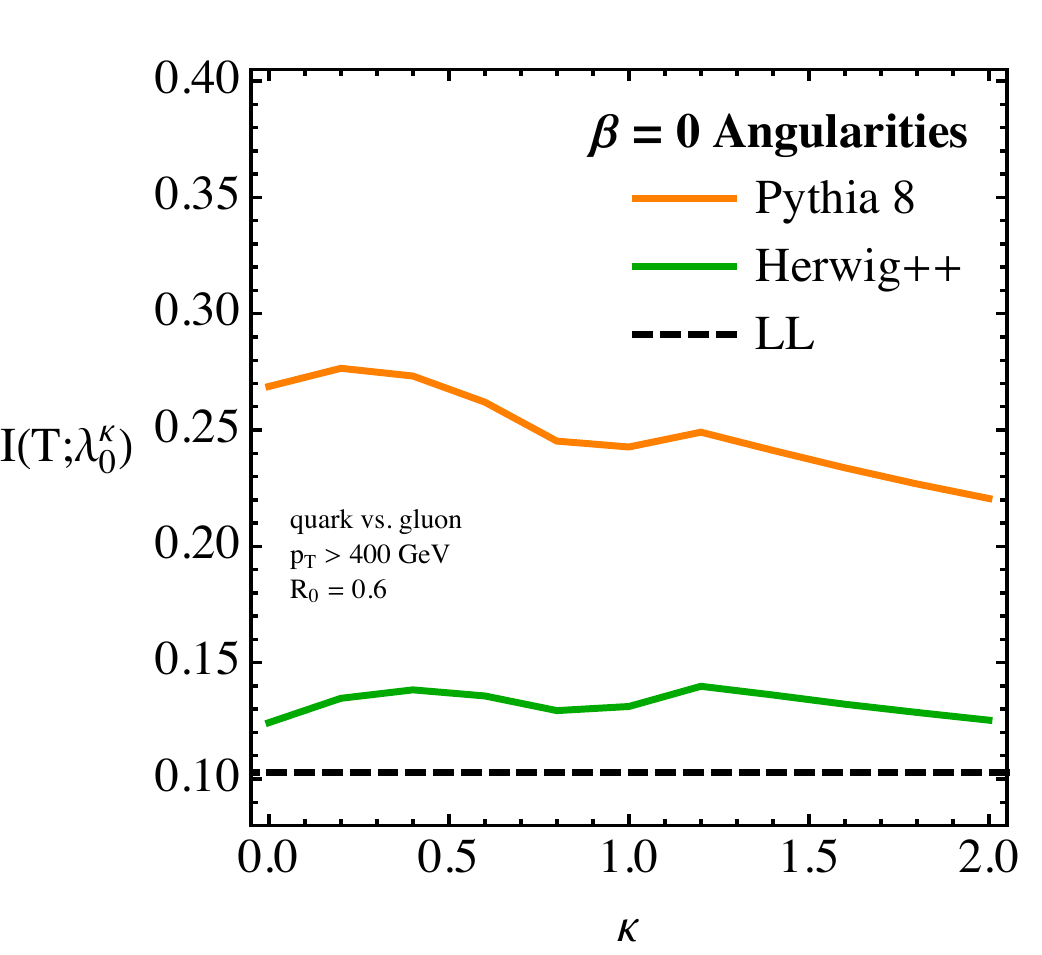}
\end{center}
\caption{The quark/gluon truth overlap for an individual generalized angularity $\unsafeang{\kappa}$ as a function of the energy-weighting power $\kappa$.  Here, we are comparing the \pythia{8} and \herwigpp\ samples to the LL baseline.}
\label{fig:b0_genang}
\end{figure}

\begin{figure}
\begin{center}
\subfloat[]{
\includegraphics[width = 7cm]{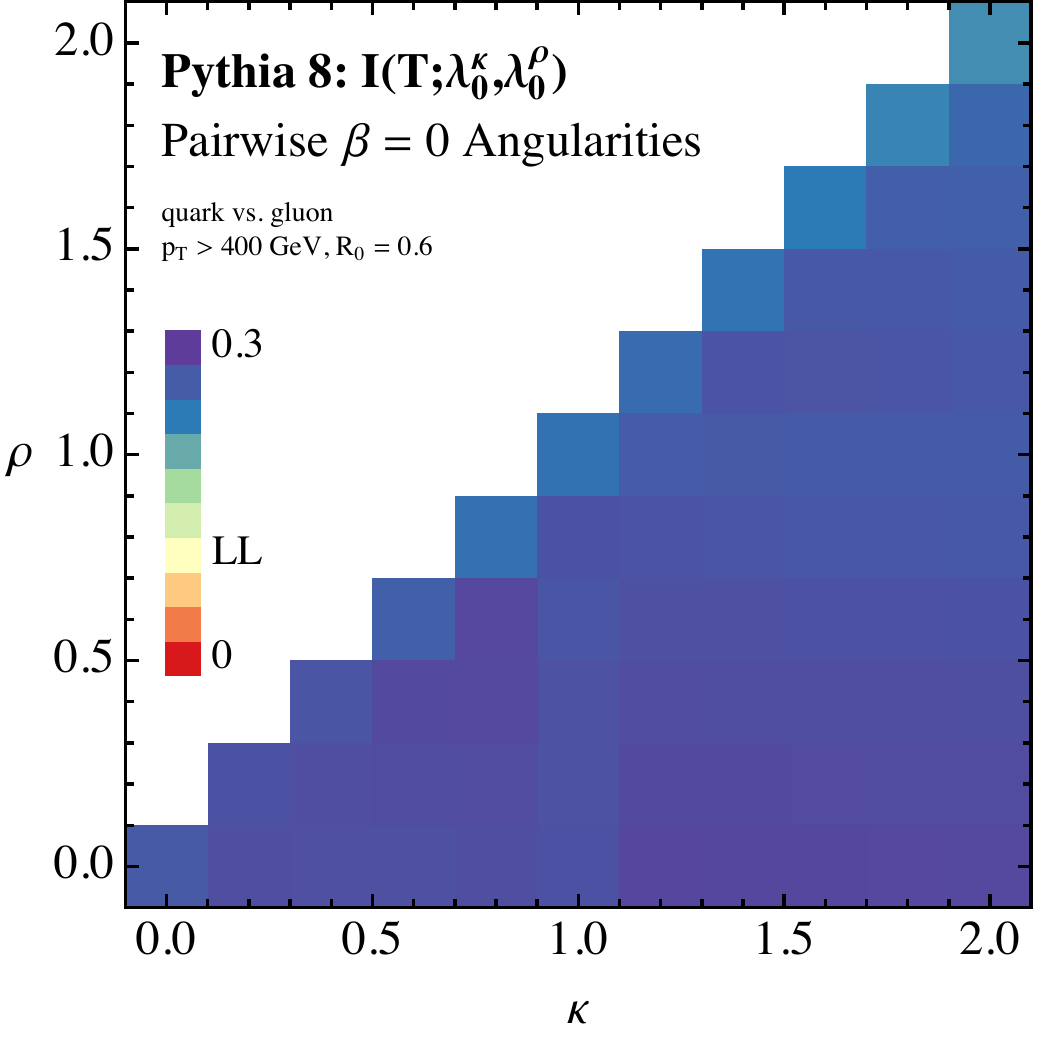}
}
$\quad$
\subfloat[]{
\includegraphics[width = 7cm]{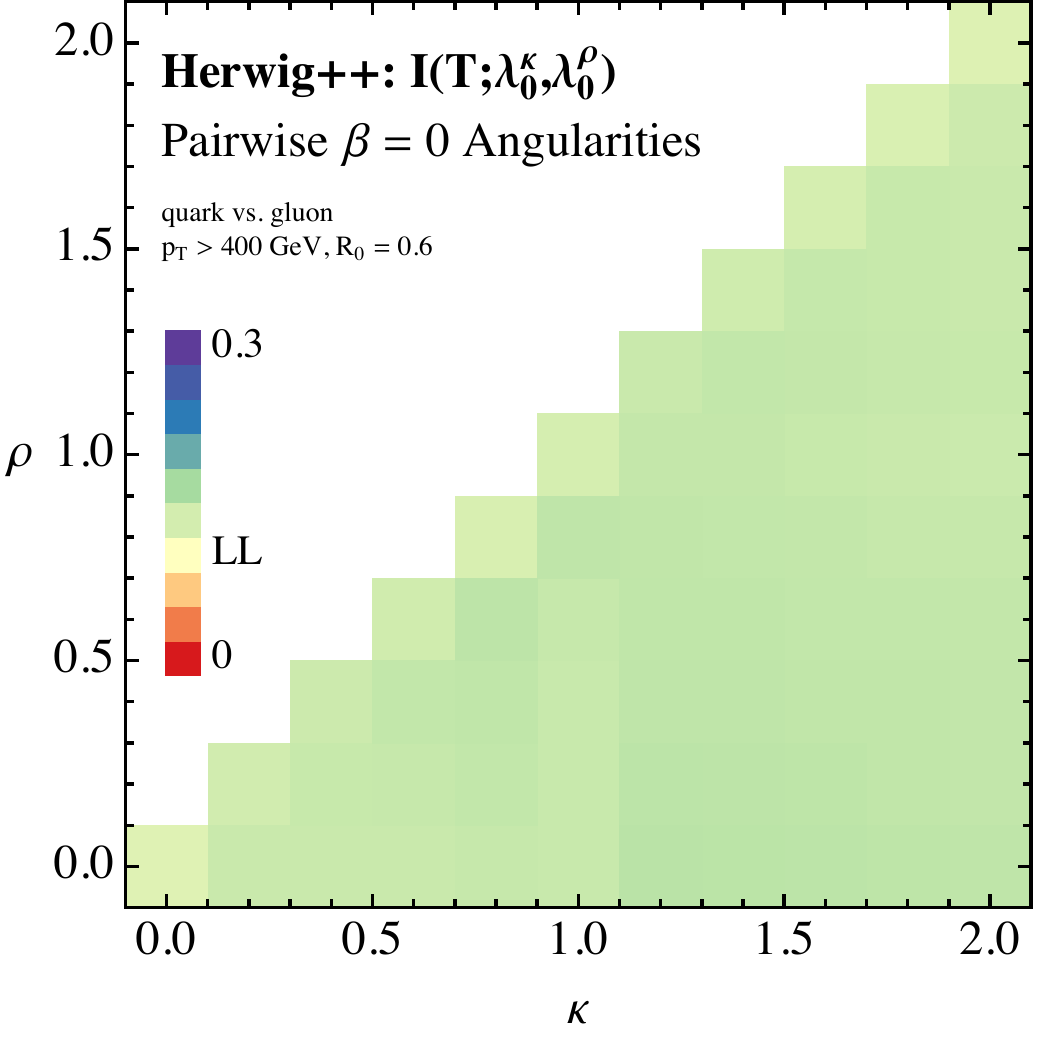}
}
\end{center}
\caption{The quark/gluon truth overlap for pairs of $\beta = 0$ angularities $(\unsafeang{\rho},\unsafeang{\kappa})$, comparing the \pythia{8} and \herwigpp\ parton showers.  
}
\label{fig:betazerocomp}
\end{figure}

Since our analytic calculations are limited by our lack of knowledge of the nonperturbative function $F_\kappa^i$, we close our discussion of $\beta = 0$ by simply showing the quark/gluon truth overlap extracted from \pythia{8} and \herwigpp.  In \Fig{fig:b0_genang}, we show the truth overlap of a single $\unsafeang{\kappa}$ and in \Fig{fig:betazerocomp} for a pair of  the generalized angularities, $\unsafeang{\kappa}$ and $\unsafeang{\rho}$.  Both figures are quite striking in illustrating the substantial difference in discrimination power predicted by \pythia{8} and \herwigpp.  Interestingly, the truth overlaps of \pythia{8} and \herwigpp\ seem to be related by a simple scaling of the value, with otherwise similar structures visible over the range of observables.  This might point to the source of the discrepancy in the physics descriptions between \pythia{8} and \herwigpp, but is beyond the scope of this paper.

%~~~~~~~~~~~~~~~~~~~~~~~~~~~~~~~~~~~~~~~~~~~~~~~~~~~~~~~~~~~~~~~~~~~~~~~~~~~~~~~
\subsection{The $\beta > 0$ Regime}
\label{subsec:betaposregime}
%~~~~~~~~~~~~~~~~~~~~~~~~~~~~~~~~~~~~~~~~~~~~~~~~~~~~~~~~~~~~~~~~~~~~~~~~~~~~~~~

For $\beta > 0$, the generalized angularities $\genang{\kappa}{\beta}$ are collinear safe with respect to emissions at $\theta = 0$, but collinear unsafe with respect to wide-angle emissions. This is precisely the same situation as for the track thrust study in \Ref{Chang:2013iba}, so we can adapt those methods here.  In particular, the jet can be described by a number of perturbative gluon emissions that can then be matched onto separate weighted-energy functions $F^i_\kappa(x,\mu)$.  At (N)LL order, the emissions (including the weighted-energy functions) exponentiate, allowing us to predict the performance of $\genang{\kappa}{\beta}$ for quark/gluon discrimination.

A jet with a single parton has $\genang{\kappa}{\beta} = 0$, since the reference axis will align with that parton.  For two partons, the winner-take-all axis will align with the harder parton.  Ignoring fixed-order corrections, we can assume that the harder parton is the initiating parton (i.e.~quark for a quark jet) such that the $\genang{\kappa}{\beta}$ distribution is determined by the emitted soft gluon.  In the LO approximation for narrow jets, we can use splitting functions 
%%%
\be
\label{eq:LOunsafe}
\frac{1}{\si_i}\frac{\df \si_i}{\df \genang{\kappa}{\beta}} =  \int_0^{1} \frac{\df \theta}{\theta} \int_0^1 \df z \, \frac{\alpha_s}{\pi} P_{i \to i g}(z) \int_0^\infty \df x  \, F^g_\kappa(x,\mu) \, \delta(\genang{\kappa}{\beta} - x z^\kappa \theta^\beta), 
\ee
%%%
where $P_{i \to i g}(z)$ is the splitting function for parton flavor $i$ to emit a soft gluon with momentum fraction $z$.

We can achieve (N)LL resummation by considering the strongly-ordered limit where the $\genang{\kappa}{\beta}$ distribution is determined  by the hardest emissions in the jet.  For the strongly-ordered limit to make sense, we assume that $F^g_\kappa(x,\mu)$ is non-singular and does not have support over a hierarchically large range in $x$, such that the emission with, say, the largest value of $z^\kappa \theta^\beta$ also typically has the largest value of $\genang{\kappa}{\beta}$.  This then allows us to use the logic of CAESAR \cite{Banfi:2004yd} to determine the $\genang{\kappa}{\beta}$ distribution up to NLL order (ignoring non-global effects \cite{Dasgupta:2001sh}).

In the CAESAR approach, the one-gluon distribution in \Eq{eq:LOunsafe} is interpreted as the radiator function:
%%%
\be
\label{eq:unsaferadiator}
R_i(\genang{\kappa}{\beta}) =  \int_0^{1} \frac{\df \theta}{\theta} \int_0^1\! \df z \, 
\frac{\alpha_s(p_T z \theta)}{\pi} P_{i \to i g}(z)   \int_0^\infty\! \df x \,  F^g_\kappa(x,\mu) \, \Theta(x z^\kappa \theta^\beta - \genang{\kappa}{\beta}),
\ee
%%%
where $\alpha_s$ is now a running coupling evaluated to two-loop order in the CMW scheme \cite{Catani:1990rr}.  This yields the cumulative distribution accurate to NLL
%%%
\be
\label{eq:CAESARmaster}
\Sigma_i(\genang{\kappa}{\beta}) = \frac{e^{-\gamma_E R_i'(\genang{\kappa}{\beta})} }{\Gamma(1 + R_i'(\genang{\kappa}{\beta}))}e^{-R_i(\genang{\kappa}{\beta})},
\ee
%%%
where $\gamma_E$ is the Euler-Mascheroni constant, $\Gamma$ is the gamma function, and  $R_i' \equiv - \df R_i/ \df \ln \genang{\kappa}{\beta}$ is the logarithmic derivative.  The cross section is obtained via 
%%%
\be
\frac{1}{\sigma_i} \frac{\df\sigma_i}{\df \genang{\kappa}{\beta}} = \frac{\df}{\df \genang{\kappa}{\beta}}\Sigma_i(\genang{\kappa}{\beta}).
\ee
%%%

\begin{figure}
\begin{center}
\subfloat[]{
\includegraphics[width=7cm]{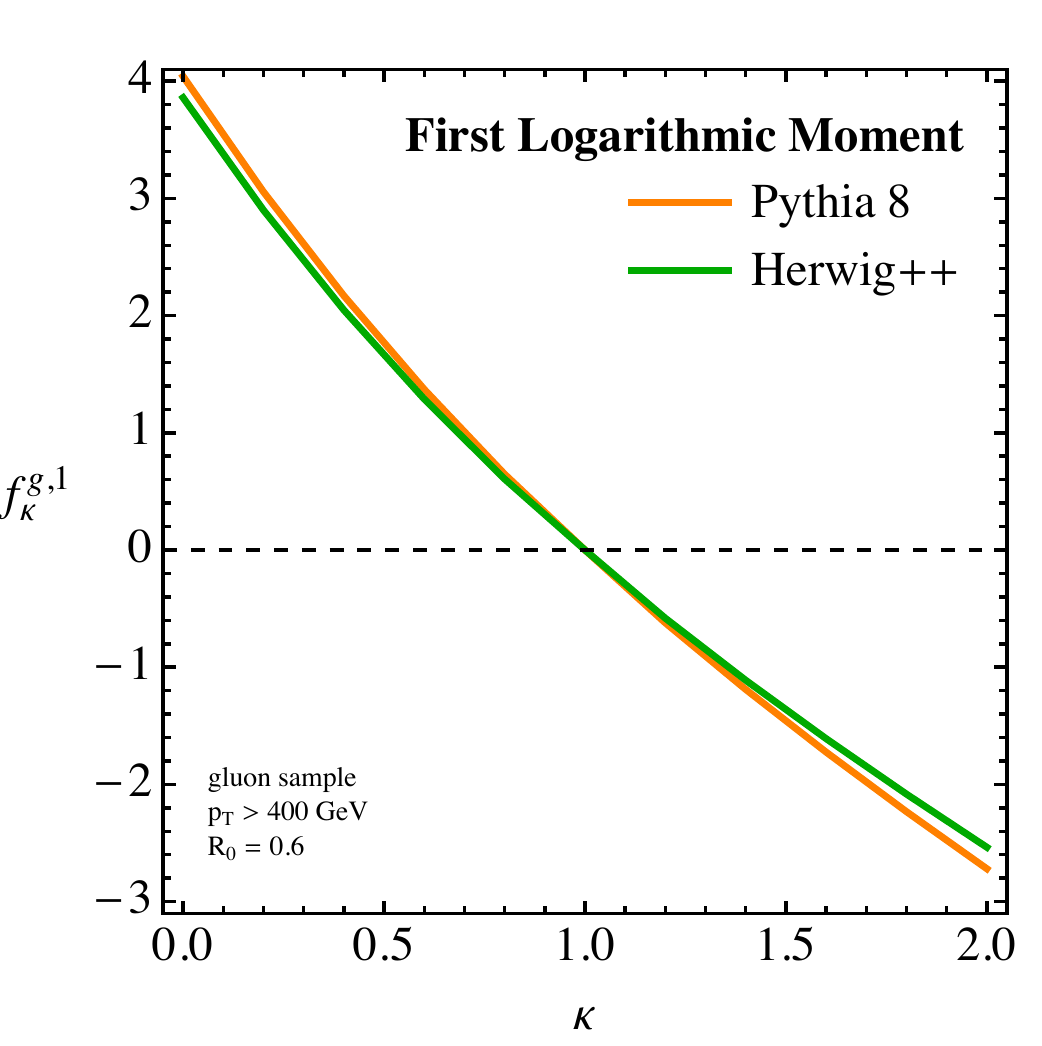}
}
$\quad$
\subfloat[]{
\includegraphics[width=7cm]{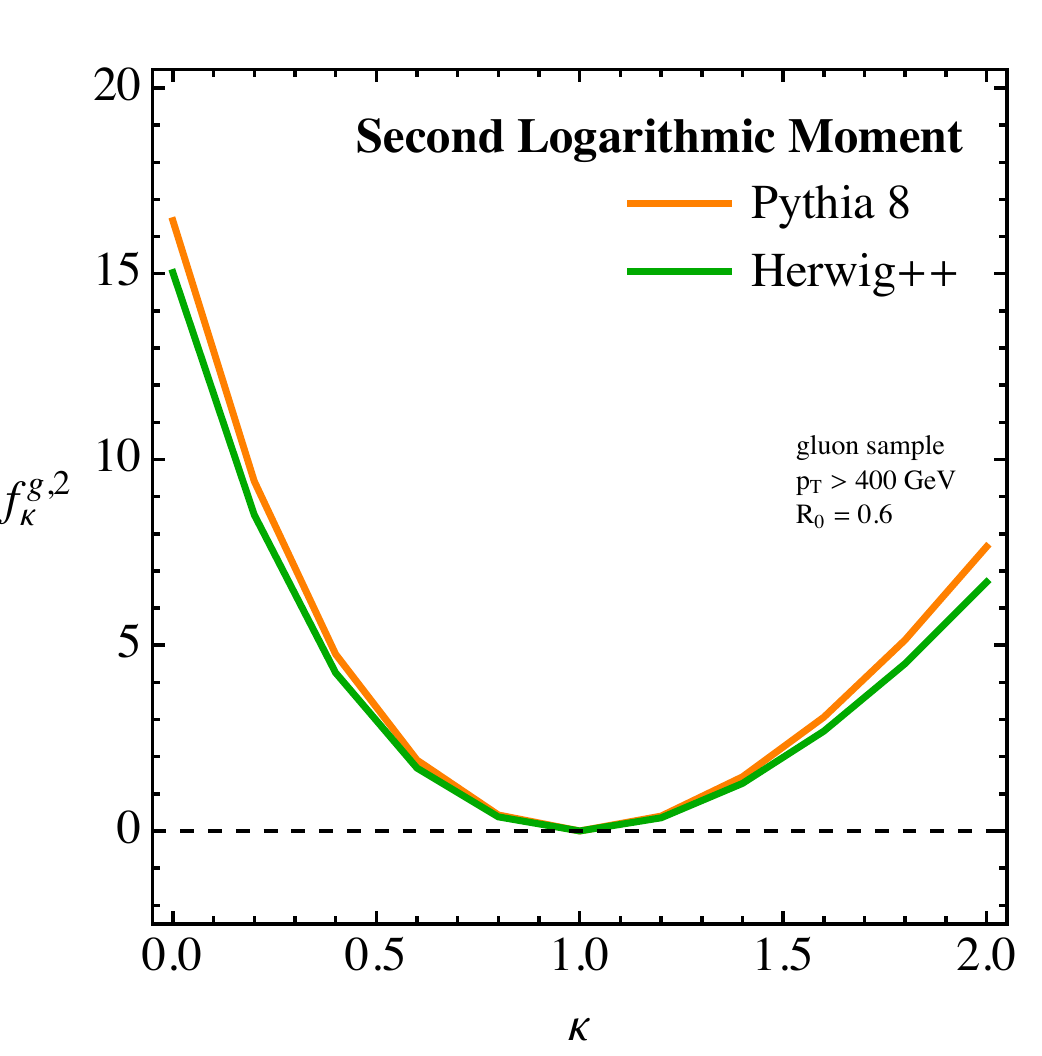}
}
\end{center}
\caption{
Extracting the logarithmic moments $f_{\kappa}^{g,1}$ (left) and $f_{\kappa}^{g,2}$ (right) defined in \Eq{eq:logmomentdef} from the \pythia{8} and \herwigpp\ gluon samples.}
\label{fig:fg1extract}
\end{figure}

We can already learn a lot from \Eq{eq:CAESARmaster} by considering just the LL limit.  In that limit, we can drop the prefactor terms involving $R'$, fix the coupling $\alpha_s$, and take $P_{i \to i g}(z) = 2 C_i / z$.  In that case we find 
%%%
\begin{equation}
R_i(\genang{\kappa}{\beta}) \stackrel{\text{LL}}{\simeq} \frac{\alpha_s}{\pi} \frac{C_i }{\beta \kappa} \int_{\genang{\kappa}{\beta}}^\infty \df x \, F^g_\kappa(x,\mu) \, \ln^2 \frac{\genang{\kappa}{\beta}}{x}.
\end{equation}
%%%
Because the bounds of integration for $x$ depends on the value of the observable $\genang{\kappa}{\beta}$, this integral cannot be simplified.  However, because we assume that $F^g_\kappa(x,\mu)$ is non-singular, for small $\genang{\kappa}{\beta}$ we can expand the integral in powers of $\genang{\kappa}{\beta}$.  To leading logarithmic accuracy we then have
%%%
\begin{align}
\label{eq:gen_LL}
R_i(\genang{\kappa}{\beta}) &\stackrel{\text{LL}}{\simeq}  \frac{\alpha_s}{\pi} \frac{C_i}{\beta \kappa} \left( \ln^2 \genang{\kappa}{\beta} -  2 f_\kappa^{g,1}  \ln \genang{\kappa}{\beta} + f_\kappa^{g,2} \right),
\end{align}
%%%
where other terms are suppressed by powers of $\genang{\kappa}{\beta}$.  The logarithmic moments are defined as
%%%
\be
\label{eq:logmomentdef}
f_{\kappa}^{g,n} \equiv \int_0^\infty \df x \, F^g_\kappa(x,\mu) \ln^n x.
\ee
%%%
 In \Fig{fig:fg1extract}, we show values of $f_{\kappa}^{g,i}$ extracted from \pythia{8} and \herwigpp\ for a range of $\kappa$ values.   These moments are quite similar between the two parton showers, suggesting that their extraction is robust.

Like in \Ref{Chang:2013iba}, we find that logarithmic moments of the nonperturbative function appears in the exponent of the cumulative distribution.  Note that the only difference between quark and gluon jets is the color factor $C_i$, since the same gluon-based $f_{\kappa}^{g,n}$ appears for both kinds of jets.  Thus, this observable satisfies Casimir scaling in the LL limit, yielding the mutual information discussed in \Sec{sec:casimir}.   Strictly speaking, these $f_{\kappa}^{g,n}$  terms are only relevant at NLL order, since they multiply at most single logarithms in the observable.\footnote{\label{footnote:logenhance}One could imagine power counting $f_{\kappa}^{g,n}$ as being logarithmically enhanced instead of as $\mathcal{O}(1)$.   In that case, however,  one would need to keep track of every $f_{\kappa}^{g,n}$ starting at NLL order, so we effectively return to the $\beta = 0$ case where the full $F^g_\kappa(x,\mu)$ function is needed.}  Therefore, we will drop the $f_{\kappa}^{g,n}$ terms at LL order:
\be
\label{eq:gen_LL_trunc}
R_i^{\text{LL}}(\genang{\kappa}{\beta}) =  \frac{\alpha_s}{\pi} \frac{C_i}{\beta \kappa} \ln^2 \genang{\kappa}{\beta}.
\ee

Doing the full NLL calculation using \Eq{eq:CAESARmaster} is straightforward with the help of two tricks.  First, using the fact that
\be
\Theta(x z^\kappa \theta^\beta - \genang{\kappa}{\beta}) = \Theta\left(z \theta^{\beta/\kappa} -\left( \frac{\genang{\kappa}{\beta}}{x} \right)^{1/\kappa}\right),
\ee
we can rewrite the radiator function in \Eq{eq:unsaferadiator} as
\be
R_i(\genang{\kappa}{\beta}) = \int_0^\infty\! \df x \,  F^g_\kappa(x,\mu) \, \hat{R}_i\left( \safeang{\beta/\kappa} = \left( \frac{\genang{\kappa}{\beta}}{x} \right)^{1/\kappa} \right),
\ee
where $\hat{R}_i$ is the radiator for the IRC safe angularity with exponent $\beta/\kappa$.  Second, we only need to keep the first logarithmic moment $f_{\kappa}^{g,1}$ at NLL order, so when we do the $x$ integral weighted by $F^g_\kappa(x,\mu)$, we can effectively replace
\be
x \to \exp(f_{\kappa}^{g,1})
\ee
in the argument of $\hat{R}_i$, up to log-suppressed terms.\footnote{Note that this rescaling would replace $f_\kappa^{g,2}$ with $(f_\kappa^{g,1})^2$ in \Eq{eq:gen_LL}.  Because the $f_\kappa^{g,2}$ term is formally beyond NLL accuracy, this is an allowed replacement.}  Thus, the IRC unsafe radiator is simply
\be
\label{eq:IRCunsafeNLLradiator}
R_i^{\text{NLL}}(\genang{\kappa}{\beta}) = \hat{R}_i^{\text{NLL}}\left( \safeang{\beta/\kappa} = \left(\frac{\genang{\kappa}{\beta}}{\exp(f_{\kappa}^{g,1})} \right)^{1/\kappa} \right),
\ee
where the IRC safe radiator $\hat{R}_i$ is given in \App{app:IRCsafeNLL}.  We find it quite remarkable that we can relate an IRC unsafe distribution to an IRC safe one in this way, and we show that these same two tricks are valid in SCET in \App{app:scet_sing_gen}.\footnote{It is perhaps even more remarkable that we can take logarithmic derivatives of the resulting $R_i$ and get the right NLL expression for the multiple emissions piece.  Ultimately, the only way we are able to justify this is via SCET, since the original CAESAR approach was only proven for IRC safe observables.  Note that there is a Jacobian factor in $R'_i$, so you cannot directly relate the cumulative distributions for $\safeang{\beta/\kappa}$ and $\genang{\kappa}{\beta}/\exp(f_{\kappa}^{g,1})$, only the radiators.  Because of the $R'_i$ term, the discrimination power is not just a function of $\beta/\kappa$, and has non-trivial $\kappa$ and $f_{\kappa}^{g,1}$ dependence at NLL.}  

Before showing analytic results, we need to comment on the range of validity of our calculation.  Because the radiator scales like $1/(\beta \kappa)$, we can only trust this perturbative expression when $\beta \kappa \gtrsim 0.5$. In addition, the validity of our approach is limited to the region where, in absolute terms, the nonperturbative parameter $f_\kappa^{g,1}$ is smaller than the typical values of $\ln \genang{\kappa}{\beta}$.\footnote{Outside of this region, the nonperturbative effects become too large to be treated using just the logarithmic moments, and we would have to include the full $F^g_\kappa(x,\mu)$ function, as also mentioned in footnote~\ref{footnote:logenhance}.  In principle, this would allow us to get beyond the ``nonperturbative barrier'' in \Eq{eq:nonperturbativebarrier}, though we have not attempted such a calculation.}  From \Fig{fig:fg1extract} we obtain the approximation $f_\kappa^{g,1} \approx 3 (1-\kappa)$, and using \Eq{eq:gen_LL_trunc}, we expect the distribution to peak at $\ln^2 \genang{\kappa}{\beta} = \beta \kappa/(2 \al_s C_i)$.  This suggests that our calculation holds for
%%%
\be
\label{eq:nonperturbativebarrier}
\frac{\beta \kappa}{(1-\kappa)^2} > c,
\ee
%%%
where $c = 18 \al_s C_i$, which is $\simeq 2.7$ for quarks and $\simeq 6.0$ for gluons.  We take the more restrictive value $c = 6.0$ when assessing the validity of the quark/gluon truth overlap, and this is the reason why the blue region in \Fig{fig:lambdaspace2} is missing the upper left and lower right corners. 

\begin{figure}
\begin{center}
\subfloat[]{
\includegraphics[width=7.5cm]{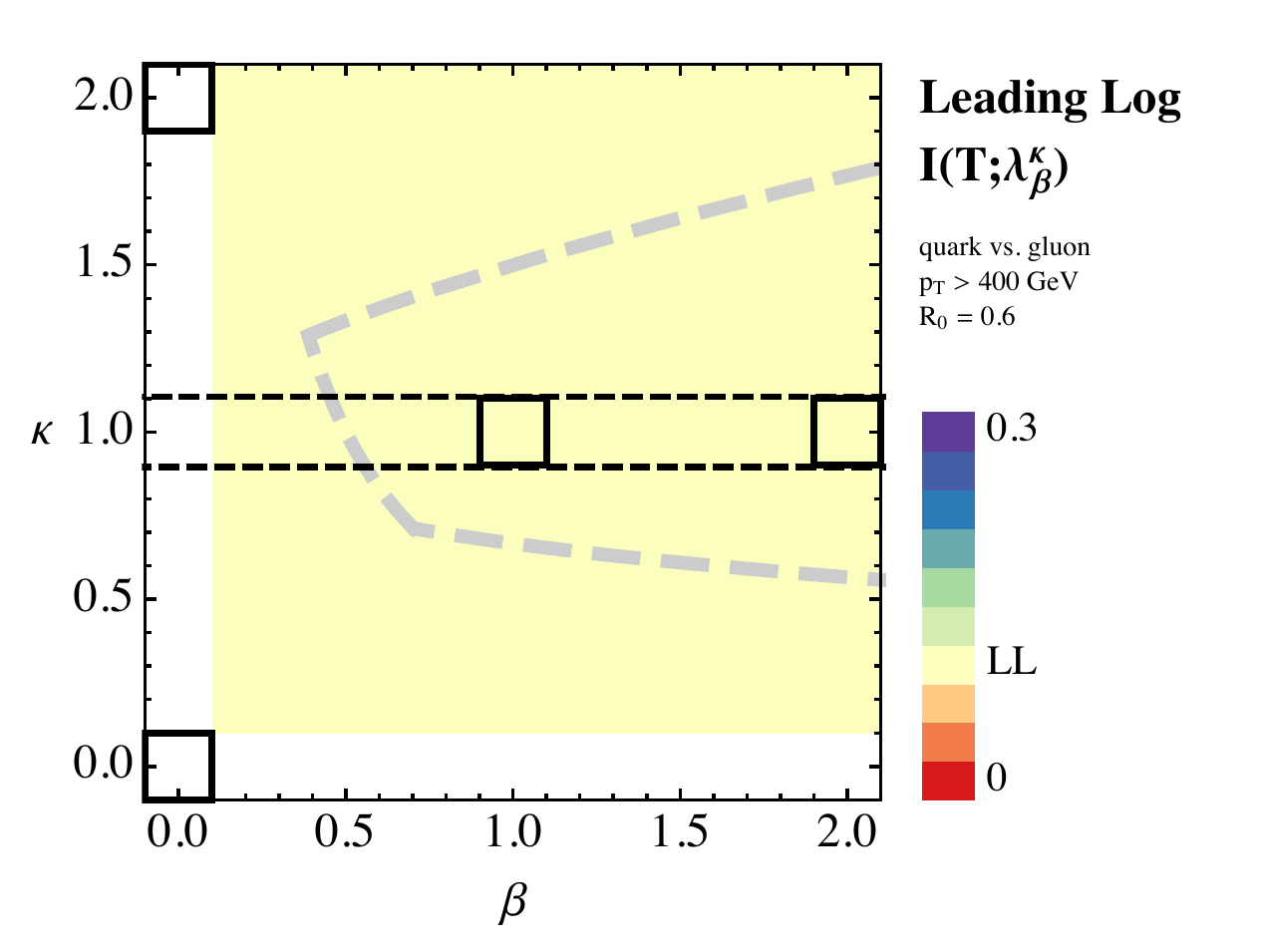}
}
\subfloat[]{
\includegraphics[width=7.5cm]{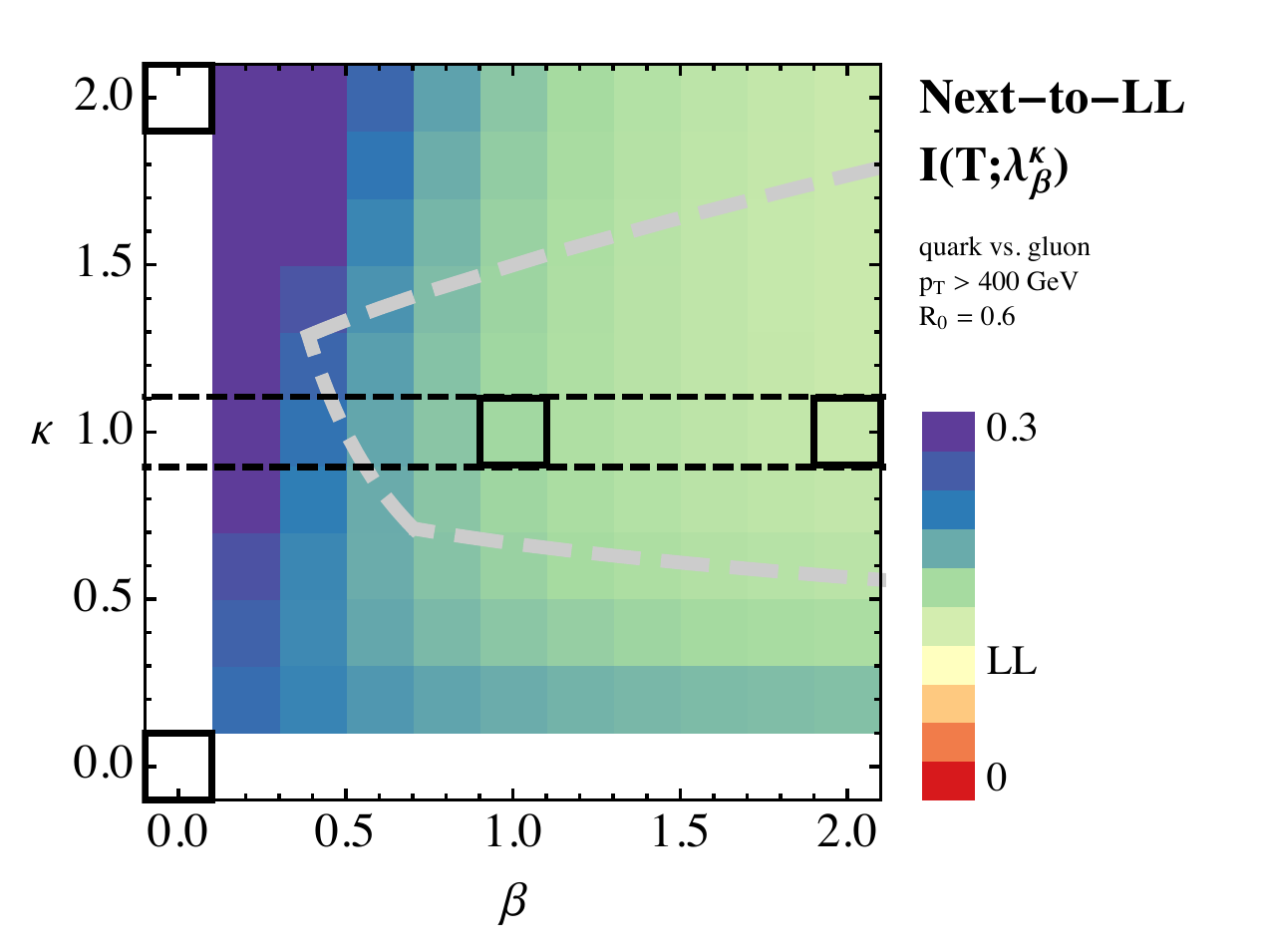}
}
\\
\subfloat[]{\label{fig:py_genang_truth}
\includegraphics[width=7.5cm]{figures/Corr_GenAng_Truth_PY.pdf}
}
\subfloat[]{\label{fig:her_genang_truth} 
\includegraphics[width=7.5cm]{figures/Corr_GenAng_Truth_HER.pdf}
}
\end{center}
\caption{
The quark/gluon truth overlap for an individual generalized angularity $\genang{\kappa}{\beta}$.  Top:  the LL and NLL analytic calculations.  Note that these calculations are singular at $\kappa = 0$ or $\beta = 0$.  Bottom: the \pythia{8} and \herwigpp\ parton showers (identical to \Figs{fig:pythiasinglelambda}{fig:herwigsinglelambda}).   The solid boxes correspond to dots indicated in \Fig{fig:lambdaspace}, the dashed box corresponds to the IRC safe angularities $e_\beta$, and the grey dashed curve in the LL/NLL plots marks the range of validity of our calculations (i.e. the edge of the blue region in \Fig{fig:lambdaspace2}).
}
\label{fig:genang_mutinf_truth_single}
\end{figure}

In \Fig{fig:genang_mutinf_truth_single}, we show the truth overlap $I(T; \genang{\kappa}{\beta})$, comparing LL, NLL, \pythia{8}, and \herwigpp.  As expected, there is no difference between different choices of $\kappa$ and $\beta$ at LL, with differences showing up first at NLL order.  The NLL calculation breaks down in the upper left region (due to large values of $f_\kappa^{g,1}$) and in the lower left region (due to $\beta \kappa$ being too small).  Unfortunately, these are exactly the same regions where there is interesting behavior in the \pythia{8} and \herwigpp\ predictions.  If we naively trust the NLL results outside of their range of validity, then starting from broadening ($\genang{1}{1}$) and approaching multiplicity ($\genang{0}{0}$), the NLL results show an increase in discrimination power, in agreement with the parton showers.  However approaching $p_T^D$ ($\genang{2}{0}$), the NLL results also show an increase in discrimination power, which is the opposite behavior from the parton showers (until one reaches the actual $\beta = 0$ line).  Of course, one should be wary of this extrapolation, since our calculations are lacking important nonperturbative corrections.\footnote{Intuitively, the parton shower results make sense.  Compared to gluons, quarks typically have smaller values of $\genang{1}{1}$ but larger values of $\genang{2}{0}$.  Thus, interpolating between $\genang{1}{1}$ and $\genang{2}{0}$ should yield a poor discriminant variable.  In the NLL approach, small $\beta$ is always favored, and calculational control is lost before the $f_\kappa^{g,1}$ parameter has a chance to reverse that trend.}

%~~~~~~~~~~~~~~~~~~~~~~~~~~~~~~~~~~~~~~~~~~~~~~~~~~~~~~~~~~~~~~~~~~~~~~~~~~~~~~~
\subsection{Two $\beta>0$ Angularities}
\label{subsec:unsafepairwise}
%~~~~~~~~~~~~~~~~~~~~~~~~~~~~~~~~~~~~~~~~~~~~~~~~~~~~~~~~~~~~~~~~~~~~~~~~~~~~~~~

Because the resummed $\genang{\kappa}{\beta}$ distributions for $\beta > 0$ only depend on logarithmic moments of the weighted-energy function, we have an opportunity to analytically study the correlations between two generalized angularities $\genang{\rho}{\alpha}$ and $\genang{\kappa}{\beta}$.  We can already gain a lot of insight from a LL study, and we can use the same tricks as for \Eq{eq:IRCunsafeNLLradiator} to obtain an NLL result.

With the help of the double weighted-energy function in \Eq{eq:doubleweightedenergyfunction}, we can define a double radiator function 
%%%
\begin{align}
\label{eq:doubleradiatorUnsafe}
R_i(\genang{\rho}{\alpha},\genang{\kappa}{\beta}) & = \int_0^{1} \frac{\df \theta}{\theta} \int_0^1\! \df z \, \frac{\alpha_s(p_T z \theta)}{\pi} P_{i \to i g}(z)   \int_0^1\! \df x_1  \int_0^1\! \df x_2 \,  F^g_{\rho,\kappa}(x_1,x_2;\mu) \nonumber \\
&\qquad \times \left[1-\Theta(\genang{\rho}{\alpha}-x_1 z^\rho \theta^\alpha) \Theta( \genang{\kappa}{\beta}-x_2 z^\kappa \theta^\beta )\right].
\end{align}
%%%
Again assuming that $F^g_{\rho,\kappa}$ does not have any large hierarchies (such that the observables are dominated by the hardest emissions), then we can follow the logic of \Ref{Larkoski:2013paa} and say that at LL accuracy 
%%%
\be
\frac{1}{\sigma_i} \frac{\df^2 \sigma_i}{\df \genang{\rho}{\alpha} \, \df \genang{\kappa}{\beta}} = \left( \frac{\partial^2}{\partial \genang{\rho}{\alpha} \, \partial \genang{\kappa}{\beta}} e^{-R_i(\genang{\rho}{\alpha},\genang{\kappa}{\beta})}\right) \Theta_0(\genang{\rho}{\alpha},\genang{\kappa}{\beta}),
\ee
%%%
where $\Theta_0(\genang{\rho}{\alpha},\genang{\kappa}{\beta})$ enforces the phase space restrictions\footnote{These assume that nonperturbative physics do not affect the phase space, which is fine for LL accuracy.  At NLL, we use \Eq{eq:NLLphasespacescaling} to adjust the phase space given the first logarithmic moments of the weighted-energy function.}
%%%
\begin{equation}
\label{eq:unsafedoublephasespace}
\left(
\frac{(\genang{\rho}{\alpha})^{\kappa}}{(\genang{\kappa}{\beta})^{\rho}}
\right)^{\text{sign}(\alpha/\rho-\beta/\kappa)} \leq 1 , \qquad 
\left(
\frac{(\genang{\kappa}{\beta})^\alpha}{(\genang{\rho}{\alpha})^\beta}
\right)^{\text{sign}(\alpha/\rho-\beta/\kappa)} \leq 1 . 
\end{equation}
%%%
This expression does not immediately generalize to NLL accuracy, since there is no (known) factorization theorem for double differential distributions over the full phase space.  Instead we will exploit the interpolation technique of \Ref{Larkoski:2014tva} to help find the NLL expression, as we did in \Sec{sec:twoAng}.

Calculating the double radiator in the LL limit, we find for $\alpha/ \rho >  \beta/\kappa$
%%%
\begin{align}
R(\genang{\rho}{\alpha},\genang{\kappa}{\beta}) & \stackrel{\text{LL}}{\simeq}  \frac{\alpha_s}{\pi} \frac{C_i}{\alpha \kappa - \beta \rho}\int \df x_1 \, \df x_2 \, F^g_{\rho,\kappa}(x_1, x_2;\mu) \, \left(\frac{\alpha}{\beta} \ln^2 \frac{\genang{\kappa}{\beta}}{x_2} + \frac{\kappa}{\rho} \ln^2 \frac{\genang{\rho}{\alpha}}{x_1} - 2 \ln \frac{\genang{\kappa}{\beta}}{x_2}  \ln \frac{\genang{\rho}{\alpha}}{x_1} \right) \nonumber \\
& =   \frac{\alpha_s}{\pi}\frac{C_i}{\alpha \kappa - \beta \rho}  \Bigl[
\frac{\alpha}{\beta}
\left( \ln^2 \genang{\kappa}{\beta} -  2 f_\kappa^{g,1}  \ln \genang{\kappa}{\beta} + f_\kappa^{g,2} \right)
+
\frac{\kappa}{\rho}
\left( \ln^2 \genang{\rho}{\alpha} -  2 f_\rho^{g,1}  \ln \genang{\rho}{\alpha} + f_\rho^{g,2} \right) \nonumber \\
& \qquad \qquad
- 2 
\left( \ln \genang{\kappa}{\beta} \ln \genang{\rho}{\alpha} -  f_\kappa^{g,1} \ln \genang{\rho}{\alpha} - f_\rho^{g,1} \ln \genang{\kappa}{\beta} + f_{\rho,\kappa}^{g,1,1} \right)
\Bigr],
\end{align}
%%%
where we have used \Eq{eq:marginalonexinF} to simplify the last line (accurate to leading power in $\genang{\rho}{\alpha}$ and $\genang{\kappa}{\beta}$) and we have defined 
\be
f_{\rho,\kappa}^{g,1,1} = \int \df x_1 \, \df x_2 \, F^g_{\rho,\kappa}(x_1, x_2;\mu) \ln x_1  \ln x_2.
\ee
As in the case of a single $\genang{\kappa}{\beta}$, strictly speaking the nonperturbative parameters $f_{\rho,\kappa}^{g,n}$ only  show up at subleading logarithmic order, and can be ignored to LL accuracy,
\be
R_i^{\rm LL}(\genang{\rho}{\alpha},\genang{\kappa}{\beta}) = \frac{\alpha_s}{\pi}\frac{C_i}{\alpha \kappa - \beta \rho}  \left(
\frac{\alpha}{\beta} \ln^2 \genang{\kappa}{\beta} +
\frac{\kappa}{\rho}
\ln^2 \genang{\rho}{\alpha} - 2 \ln \genang{\kappa}{\beta} \ln \genang{\rho}{\alpha}
\right).
\ee
Note that the exponents $\rho$ and $\kappa$ still have an effect on the discrimination power even though we are not accounting for the nonperturbative parameters at this order.  The overall prefactor implies that when $\alpha$ and $\beta$ are sufficiently different, we should only trust this distribution for
\be
\label{eq:doubletrust}
\alpha \kappa - \beta \rho \gtrsim 0.5.
\ee

To achieve NLL accuracy, we need to combine the interpolation technique of \Ref{Larkoski:2014tva} with the two tricks as we used to find the single generalized angularity distribution in \Eq{eq:IRCunsafeNLLradiator}.  Following \Ref{Larkoski:2014tva}, we take the following ansatz for the NLL distribution:
%%%
\begin{equation}
\label{eq:twoCAESARmaster}
\Sigma_i(\genang{\rho}{\alpha},\genang{\kappa}{\beta}) =\frac{e^{-\gamma_E \widetilde{R}_i(\genang{\rho}{\alpha},\genang{\kappa}{\beta})}}{\Gamma(1+\widetilde{R}_i(\genang{\rho}{\alpha},\genang{\kappa}{\beta}))}\, e^{-R_i(\genang{\rho}{\alpha},\genang{\kappa}{\beta})},
\end{equation}
where $R_i$ is the double radiator from \Eq{eq:doubleradiatorUnsafe} and $\widetilde{R}_i$ is a multiple emissions term that is \emph{not} given by any simple logarithmic derivative of $R_i$.  We start by considering the double radiator $R_i$.  Looking at the theta functions in \Eq{eq:doubleradiatorUnsafe}, we see that the integration range for the double radiator is the same as for two IRC safe angularities with 
\be
e_{\alpha/\rho}  = \left( \frac{\genang{\rho}{\alpha}}{x_1} \right)^{1/\rho}, \qquad e_{\beta/\kappa} =  \left( \frac{\genang{\kappa}{\beta}}{x_2}  \right)^{1/\kappa}.
\ee
At NLL accuracy, we only need the first logarithmic moments of the weighted-energy function, so we can make the replacement
\be
x_1 \to \exp(f_\rho^{g,1}), \qquad x_2 \to \exp(f_\kappa^{g,1}).
\ee
Thus, the double radiator is
\be
\label{eq:doubleradiatorrescaling}
R_i^{\rm NLL}(\genang{\rho}{\alpha},\genang{\kappa}{\beta}) = \hat{R}_i^{\rm NLL} \left(e_{\alpha/\rho}  = \left( \frac{\genang{\rho}{\alpha}}{\exp(f_\rho^{g,1})} \right)^{1/\rho}, \, e_{\beta/\kappa} =  \left( \frac{\genang{\kappa}{\beta}}{\exp(f_\kappa^{g,1})}  \right)^{1/\kappa}   \right),
\ee
where $\hat{R}_i^{\rm NLL}$ is the IRC safe double cumulative distribution for angularities with exponents $\alpha/\rho$ and $\beta/\kappa$, defined in \App{app:twoIRCsafeNLL}.  Turning to the multiple emissions term $\widetilde{R}_i$, we need to find a function that interpolates between the logarithmic derivative functions $R'(\genang{\rho}{\alpha})$ and $R'(\genang{\kappa}{\beta})$ on the boundaries of phase space.  Since we already found the single radiator to NLL accuracy in \Eq{eq:IRCunsafeNLLradiator}, this interpolation is straightforward, and we give the explicit expressions in \App{app:twoIRCunsafeNLL}.  Finally, to have a properly normalized distribution, we have to apply the rescaling
\be
\label{eq:NLLphasespacescaling}
\genang{\rho}{\alpha} \to \frac{\genang{\rho}{\alpha}}{\exp(f_\rho^{g,1})}, \qquad \genang{\kappa}{\beta} \to \frac{\genang{\kappa}{\beta}}{\exp(f_\kappa^{g,1})}
\ee
to the phase space constraints in \Eq{eq:unsafedoublephasespace} as well.  Again, we find it remarkable that there is such a close relationship between IRC safe and IRC unsafe calculations at NLL order, and we validate this method in SCET in \App{app:scet_pair_interpolation}.

\begin{figure}
\begin{center}
\subfloat[]{
\includegraphics[width=7cm]{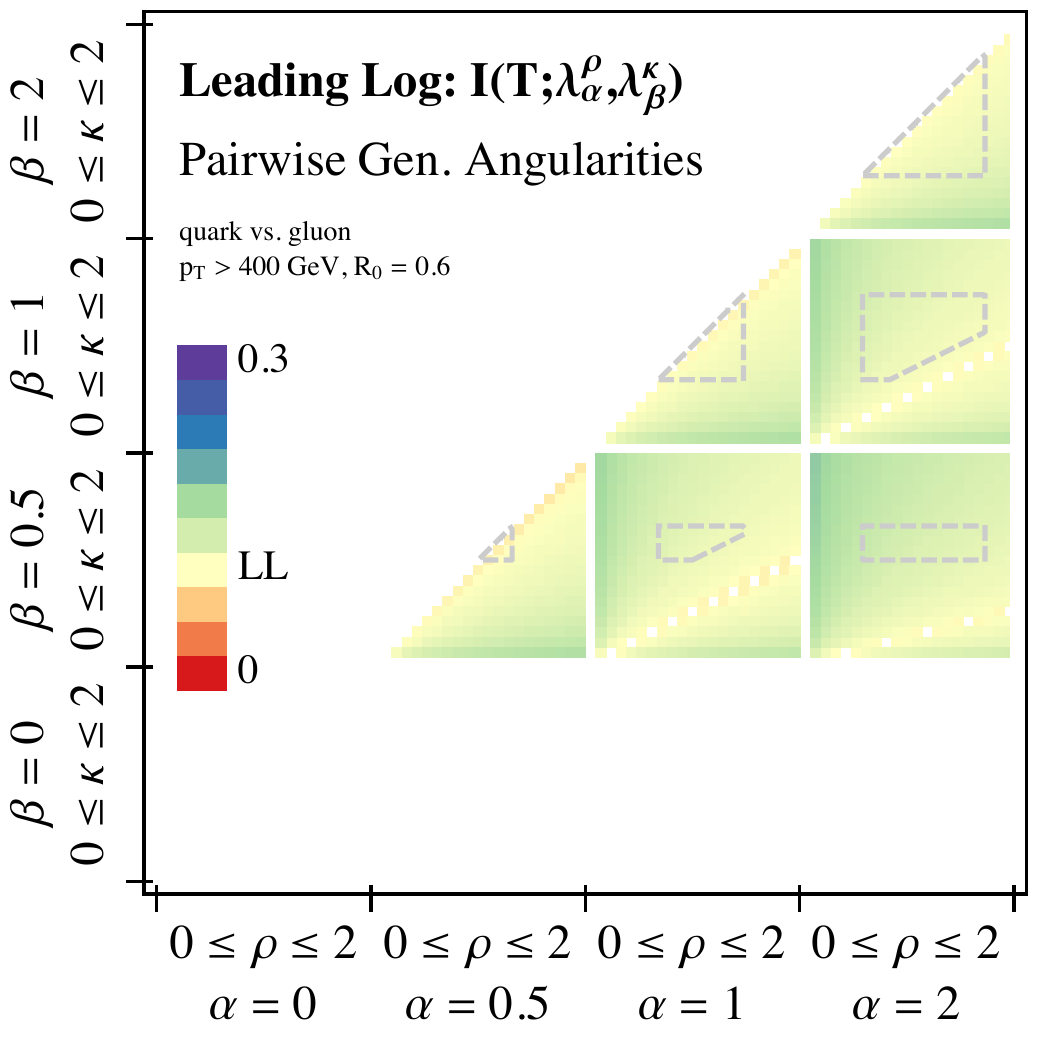}
}
$\qquad$
\subfloat[]{
\includegraphics[width=7cm]{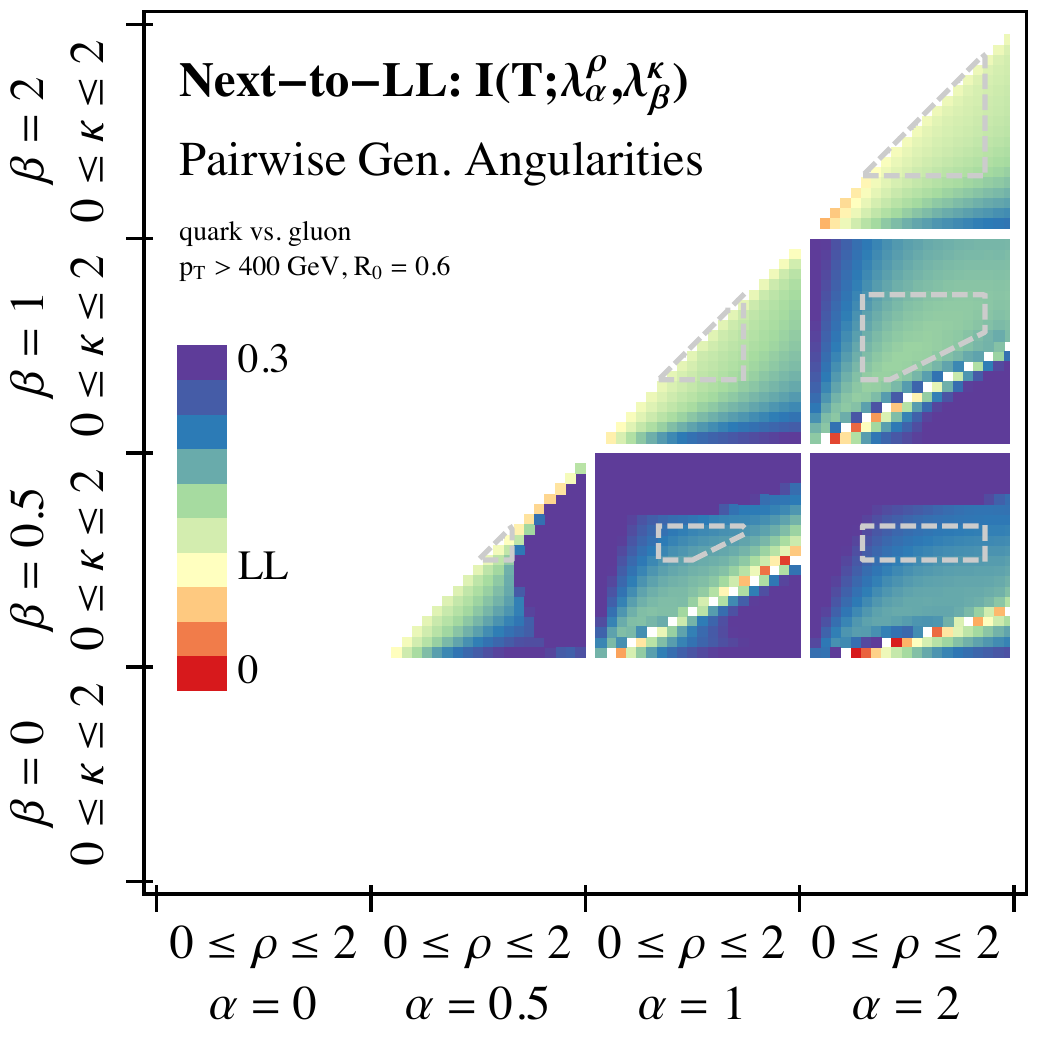}
}
\\
\subfloat[]{\label{fig:py_genang_dd} 
\includegraphics[width=7cm]{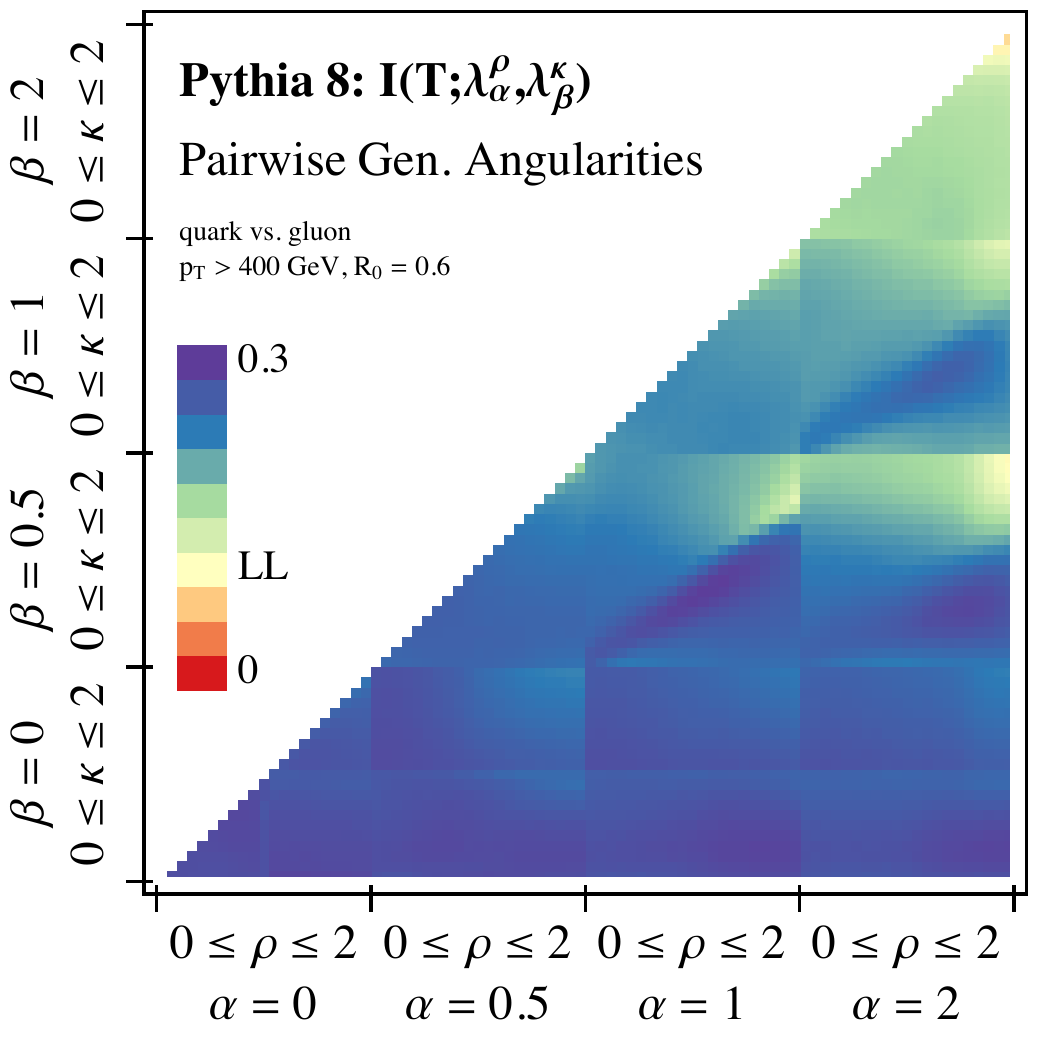}
}$\qquad$
\subfloat[]{\label{fig:her_genang_dd} 
\includegraphics[width=7cm]{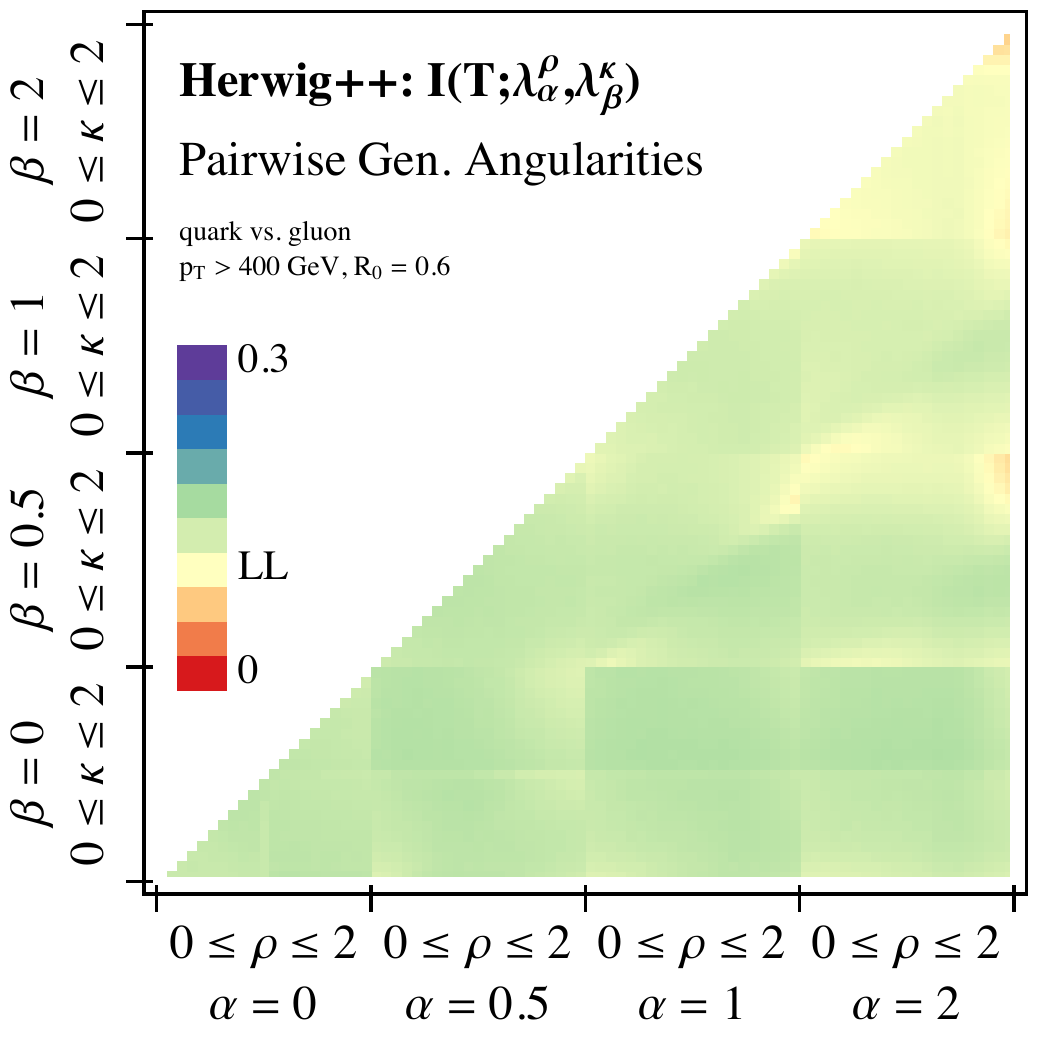}
}
\end{center}
\caption{
The quark/gluon truth overlap for pairs of generalized angularities $(\genang{\rho}{\alpha},\genang{\kappa}{\beta})$.  Top: the LL and NLL analytic calculations.  Bottom: the \pythia{8} and \herwigpp\ parton showers.  Here, we show four values of $\beta\in\{0,0.5,1,2\}$ and for each value of $\beta$, $0\leq \kappa \leq 2$ in steps of size $0.1$.  In the NLL and NLL plots, the interior of the dashed grey boxes correspond to the range of validity of our calculations.
}
\label{fig:genang_mutinf_truth_pair}
\end{figure}

In \Fig{fig:genang_mutinf_truth_pair}, we show the truth overlap for the LL and NLL calculations, compared to results obtained from \pythia{8} and \herwigpp.  The LL and NLL calculations do not extend to the region where $\al$ or $\bt$ is zero, which are left white in the plot.  We caution the reader that some of these LL and NLL results extrapolate outside the range of validity in \Eqs{eq:nonperturbativebarrier}{eq:doubletrust}.  Comparing the various predictions, we observe similarities between the regions of minimal and maximal discrimination power, though the overall discrimination power is (again) larger for \pythia{8} and NLL than for \herwigpp\ and LL. The \pythia{8} and NLL results most clearly indicate that it is advantageous to pick one of the angular exponents $\al$ or $\bt$ to be small. A notable difference between the predictions is that the LL and NLL calculations suggest that one should avoid the diagonal $\rho=\kappa$, whereas for \pythia{8}, and to a lesser extent \herwigpp, the maximum discrimination power is sometimes (surprisingly) close to it (see e.g.~$\al=1$, $\bt=0.5$).

%%%%%%%%%%%%%%%%%%%%%%%%%%%%%%%%%%%%%%%%%%%%%%%%%%%%%%%%%%%%%%%%%%%%%%%%%%%%%%%%
\section{Conclusions}
\label{sec:conclude}
%%%%%%%%%%%%%%%%%%%%%%%%%%%%%%%%%%%%%%%%%%%%%%%%%%%%%%%%%%%%%%%%%%%%%%%%%%%%%%%%

Robust quark/gluon discrimination is a key goal for the jet substructure community, so to the extent possible, it is important to use first principles calculations to assess the challenges and opportunities.  In this paper, we showed that mutual information is a powerful way to understand how variables are correlated, and whether or not that (lack of) correlation pertains to discrimination power.  We also made progress in gaining analytic control over the tagging performance of the generalized angularities $\genang{\kappa}{\beta}$.  For the IRC safe angularities and the IRC unsafe angularities with $\beta >0$, we calculated the quark/gluon truth overlap for a single angularity $I(T;\genang{\kappa}{\beta})$ and for pairs of angularities $I(T;\genang{\rho}{\alpha},\genang{\kappa}{\beta})$ to NLL order.

Ultimately, we want to extend our analysis to higher orders, but this would require a robust ``truth'' definition for a quark jet versus a gluon jet.  While the strategy of \Ref{Banfi:2006hf} is one option to define the truth flavor of a jet, we would prefer a definition for which the jet constituents are the same as for traditional flavor-less jet algorithms.  Of course, quark and gluon jets do not exist in isolation, and at some point, the color correlations between the jets will be relevant for characterizing the discrimination power.  The techniques introduced recently in \Ref{Stewart:2014nna} should help in gaining analytic control over those color correlations.  Since our NLL results are subject to large changes from scale variation, higher-order calculations will be crucial for robust uncertainty estimates.

Assuming we did have a suitable quark/gluon truth definition, then a key challenge for calculations beyond our present order is dealing with soft radiation, in particular the effect of non-global logarithms \cite{Dasgupta:2001sh}.  One option is to do quark/gluon tagging in concert with soft drop declustering \cite{Larkoski:2014wba} (a generalization of modified mass drop tagging \cite{Butterworth:2008iy,Dasgupta:2013ihk}).  The soft drop procedure removes soft radiation, and therefore removes non-global contributions to the jet.  The cumulative distributions for a single soft-dropped angularity were already calculated in \Ref{Larkoski:2014wba}, where the distributions exhibited Casimir scaling at LL order.  Using soft-dropped jet shapes for quark/gluon discrimination seems promising from both a theoretical and experimental point of view, and we leave a more detailed study to future work.

Based on our studies, we have two recommendations to the ATLAS and CMS experiments.  The first recommendation is to make (unfolded) measurements of the recoil-free angularities distributions for a range of $\beta$ values, ideally in purified quark/gluon samples.\footnote{Initial results along these lines appear in \Ref{Aad:2014gea} for the energy correlation function ratio $C_1^{(\beta)}$ \cite{Larkoski:2013eya}, though only the final quark/gluon performance is shown, not the extracted quark and gluon distributions.}  The differences seen between Monte Carlo programs in \Fig{fig:pythiainitialstudy} is worrisome, and while eventually calculations might be a guide to what these distributions should look like, in the short term $\safeang{\beta}$ measurements can be a key reference for tuning Monte Carlo programs, especially because the differences arise from effects that are formally beyond LL accuracy.  The second recommendation is to measure more double differential distributions.\footnote{There are double differential results in \Ref{TheATLAScollaboration:2013tia}, but measured only in simulation and not in data.}  While we focused on mutual information with the truth in this paper, one would still like to understand the full correlation structure.  Angularities are a good place to start, and double differential distributions of $\safeang{\alpha}$ and $\safeang{\beta}$ would be quite valuable, especially with new calculational tools available to predict these correlations \cite{Larkoski:2013paa,Larkoski:2014tva,Procura:2014cba} (see \App{app:correlations} for example plots along these lines).

Finally, in the spirit of \Refs{Krohn:2012fg,Waalewijn:2012sv,Chang:2013rca,Chang:2013iba}, we have provided another example of how collinear unsafe (but soft safe) observables can be made calculable with the help of new nonperturbative objects.  We introduced the weighted-energy functions $F_\kappa^i(x)$, which allowed us to understand many aspects of the $\kappa \not= 1$ regime.  Because the $\beta > 0$ angularities only depend on logarithmic moments of $F_\kappa^i(x)$, they are the simplest to understand.  But even the $\beta = 0$ angularities are within calculational control, since we can study the renormalization group behavior of, say, the $p_T^D$ ($\unsafeang{2}$) distribution.  Of course, hadron multiplicity ($\unsafeang{0}$) is not captured within our framework due to the presence of the soft singularity, but perhaps hadron multiplicities could be made analytically tractable by using soft drop declustering to remove soft radiation.  We expect future studies will continue to improve (and improve our understanding of) quark/gluon discrimination.

\begin{acknowledgments}
  We thank the participants of the Boost 2013 workshop for many inspiring discussions that led to this work.  We also thank Jason Gallicchio, Aram Harrow, and Matthew Headrick for conversations about mutual information and T.J.~Wilkason for coding help.
  A.L. and J.T. are supported by the U.S. Department of
  Energy (DOE) under cooperative research agreement
  DE-FG02-05ER-41360. J.T. is also supported by the DOE Early Career
  research program DE-FG02-11ER-41741 and by a Sloan Research
  Fellowship from the Alfred P. Sloan Foundation. 
  W.W. is supported by a Marie Curie International Incoming Fellowship within the 7th European Community Framework Program (PIIF-GA-2012-328913).
\end{acknowledgments}

\appendix

%%%%%%%%%%%%%%%%%%%%%%%%%%%%%%%%%%%%%%%%%%%%%%%%%%%%%%%%%%%%%%%%%%%%%%%%%%%%%%%%
\section{Properties of Mutual Information}
\label{app:properties}
%%%%%%%%%%%%%%%%%%%%%%%%%%%%%%%%%%%%%%%%%%%%%%%%%%%%%%%%%%%%%%%%%%%%%%%%%%%%%%%%

%~~~~~~~~~~~~~~~~~~~~~~~~~~~~~~~~~~~~~~~~~~~~~~~~~~~~~~~~~~~~~~~~~~~~~~~~~~~~~~~
\subsection{Relationship to the ROC Curve}
\label{app:rocMI}
%~~~~~~~~~~~~~~~~~~~~~~~~~~~~~~~~~~~~~~~~~~~~~~~~~~~~~~~~~~~~~~~~~~~~~~~~~~~~~~~

The mutual information of an observable $A$ with the truth $T$ can be derived from the ROC curve of $A$. Although we phrase this discussion in terms of quark/gluon discrimination, it obviously carries over to other cases as well. 

At the position $(q,g)$ on the ROC curve, the region $\df q$ has a fraction $\df q$ of the quark jets. The fraction of gluon jets is given by the slope of the ROC curve, $\df g$ = $g'(q)\, \df q$.  Looking at the definition of $I(T;A)$ in \Eq{eq:ITA}, we can write the integrals as
%%%
\be
\int \df a \, p_q(a) \Rightarrow \int \df q, \quad \int \df a \, p_g(a) \Rightarrow \int \df g = \int \df q \, g'(q).
\ee
%%%
For a sample with quark fraction $f$, the ratios of the probability distributions are
%%%
\be
\frac{p_q(a)}{p_{\rm tot}(a)} \Rightarrow \frac{\df q}{\df q f + \df g (1-f)}, \qquad \frac{p_g(a)}{p_{\rm tot}(a)} \Rightarrow \frac{\df g}{\df q f + \df g (1-f)}.
\ee
%%%
This leads to
%%%
\be
I(T;A) = \int\! \df q \,  \left( f\, \log_2 \frac{1}{f + (1-f)g'(q)} + (1-f) \, g'(q) \log_2 \frac{g'(q)}{f+(1-f)g'(q)} \right). \label{eq:ROCcurve}
\ee
%%%
As a simple test of this formula, note that the ROC curve for an observable that satisfies Casimir scaling is $g(q) = q^{(C_A/C_F)}$.  Plugging this into \Eq{eq:ROCcurve}, we recover  $I(T;A)$ from \Eq{eq:mutualinfocasiiarscaling}. 

Inverting this relationship to obtain the ROC curve from the mutual information is not so easy, suggesting that mutual information is an easier concept to work with.  Nevertheless it seems in principle possible, though we do not claim that the following simple-minded approach is optimal.  Consider discretizing $g'(q)$ by treating it as a constant $g'_i$ on the interval $q \in [i/n_\text{bins},(i+1)/n_\text{bins}]$ with $i=0,1, \dots, n_\text{bins}-1$.  Because an ideal ROC curve is not only monotonically increasing (i.e.~$g'(q) \ge 0$) but also convex (i.e.~$g''(q) \ge 0$), this means that $g'_{i+1} \ge g'_i$, and we have a chance to find a set of equations that (uniquely) determine $g'_i$.  Then, we can integrate $g'(q)$ in the usual way to find $g(q)$.  One set of equations is given by the $n$-th derivative of $I(T;A)$ evaluated at $f=1$ (for $n \ge 2$)
%%%
\be
\label{eq:ROC_invert}
  \frac{\df^n I(T;A)}{\df f^n}\bigg|_{f=1} = 
  -\frac{(n-2)!}{\ln 2}\, \int_0^1\! \df q\, (g'(q)-1)^n.
\ee
%%%
while for the special case of $n = 1$ we can use
%%%
\be 
\label{eq:ROC_invert_n_1}
\int_0^1\! \df q\, (g'(q)-1) = 0.
\ee
%%%
In discrete form, these become a system of polynomial equations
%%%
\begin{align}
 \frac{1}{n_\text{bins}} \sum_i (g_i'-1) &=  0, \nn \\
 \frac{1}{n_\text{bins}} \sum_i (g_i'-1)^n &=   -\frac{\ln 2}{(n-2)!}  \frac{\df^n I(T;A)}{\df f^n}\bigg|_{f=1} 
\,.\end{align}
%%%
Because these equation are non-linear, this quickly becomes numerically unstable for $n_\text{bins}>4$, but gives a proof of principle that a solution can be found.  Note that the condition that the ROC curve is convex is crucial, since otherwise each permutation of $g'_i$ would also constitute a solution.  For special functional forms, there are simpler strategies to find the ROC curve.   For example, if it is known that $g(q) = q^c$, then a practical way to estimate the exponent $c$ is via
%%%
\begin{align}
  \frac{\df I(T;A)}{\df f}\bigg|_{f=0} 
 = - \frac{1}{\ln 2} \int_0^1\! \df q\, \ln g'(q)
= \frac{c - 1 - \ln c}{\ln 2}.
 \end{align}
%%%

\subsection{Better ROC Curve Implies Greater Truth Overlap}
\label{app:betterroc}

When one observable $b$ has a larger truth overlap than another observable $w$, i.e.
%%%
\be
I(T;B) > I(T;W),
\ee
%%%
we interpreted this to mean that $b$ is a better discriminant variable and $w$ is worse.  While this is generically true, there can be cases where the ``worse'' observable can have better performance at a given operating point.  This occurs when the ROC curves of $b$ and $w$ intersect, for example when $b$ has better background rejection at low signal efficiency but $w$ performs better at high signal efficiency.  Thus, we cannot conclude from the truth overlap alone whether $b$ or $w$ is better, since even from a ROC curve perspective, ``better'' is ill-defined.

What we can prove is that if the ROC curve for $b$ is everywhere better than for $w$, then the corresponding truth overlap is strictly larger.  In information theory language, we would say that $b$ is Pareto optimal with respect to $w$.  Consider the ROC curves $g_b(q)$ and $g_w(q)$.  A better observable's ROC curve will take a smaller value (less background) at each value of signal efficiency.  Thus we can define the better observable as
\be
g_b(q)\equiv g_w(q) - \Delta(q),
\ee
where $\Delta(q)\geq0$ for all $q\in[0,1]$.  We will now use a variational method to show that $I(T;B) > I(T;W)$.

From these two ROC curves, we can calculate the difference between their truth overlaps.  Using the relationship between mutual information and the ROC curve from \Eq{eq:ROCcurve} we have:
\begin{align}
I(T;B)-I(T;W) &= \int\! \df q \,  \left( f\, \log_2 \frac{1}{f + (1-f)g_b'(q)} + (1-f) \, g_b'(q) \log_2 \frac{g_b'(q)}{f+(1-f)g_b'(q)} \right)\nonumber \\
&-\int\! \df q \,  \left( f\, \log_2 \frac{1}{f + (1-f)g_w'(q)} + (1-f) \, g_w'(q) \log_2 \frac{g_w'(q)}{f+(1-f)g_w'(q)} \right)\nonumber \\
&=-\, \int\! dq \, \Delta'(q) \log_2 \frac{(1-f)g_w'(q)}{f+(1-f)g_w'(q)} + {\cal O}(\Delta^2) \, .
\end{align}
In the last step, we expanded the integrand assuming that $\Delta(q)$ is small.  Because any ROC curve $g$ must satisfy $g(0)=0$ and $g(1)=1$, we can perform integration by parts on $\Delta'(q)$ without introducing any boundary terms.
\begin{align}
I(T;B)-I(T;W) &= \frac{1}{\ln 2} \int\! dq \, \Delta(q) \frac{g_w''(q)}{g_w'(q)} \frac{f(1-f)}{f+(1-f)g_w'(q)}+{\cal O}(\Delta^2) \,.
\end{align}
This term is manifestly positive because for any (ideal) ROC curve $g'(q)\geq0$ and $g''(q)\geq0$ for all $q\in[0,1]$, and $\Delta(q)$ is positive by assumption.  Therefore,
\begin{align}
I(T;B)-I(T;W) > 0 \,
\end{align}
up to corrections of order $\Delta^2$.  However, we can make these $\Delta^2$ corrections arbitrarily small by applying this variational logic to a sequence of ROC curves that smoothly interpolate between $g_b(q)$ and $g_w(q)$.\footnote{Concretely, one can always build an interpolating strategy based on randomly selecting $b$ or $w$ as the discriminant variable with predetermined probability.}   Thus, we have proven that an observable with an everywhere-better ROC curve also has a larger truth overlap.

%~~~~~~~~~~~~~~~~~~~~~~~~~~~~~~~~~~~~~~~~~~~~~~~~~~~~~~~~~~~~~~~~~~~~~~~~~~~~~~~
\subsection{Subtleties of Finite Statistics and Binning}
\label{app:entropystats}
%~~~~~~~~~~~~~~~~~~~~~~~~~~~~~~~~~~~~~~~~~~~~~~~~~~~~~~~~~~~~~~~~~~~~~~~~~~~~~~~

When dealing with a sample with a finite number of events, the Shannon entropy $H$ is sensitive to the way in which the sample is binned.  We will compute the leading effect this has on the entropy.  Let $\nbins$ be the number of bins and let $\lambda_i$ be the average number of events in bin $i$.  The total number of events in the sample is
%%%
\begin{equation}
\sum_{i=1}^{\nbins} \lambda_i = \Nev .
\end{equation}
%%%
For sufficiently large $\Nev$, the fluctuations within the bins should be independent and the number of events in bin $i$ should be Poisson distributed about $\lambda_i$.  Thus, the expected entropy of the binned sample is
%%%
\begin{equation}
H = \sum_{i=1}^{\nbins} \sum_{n_i=0}^\infty \frac{\lambda_i^{n_i} e^{-\lambda_i}}{n_i!}\left( -\frac{n_i}{\Nev} \log_2 \frac{n_i}{\Nev} \right),
\end{equation}
%%%
where $n_i$ ranges over the possible observed events in bin $i$.

Assuming that $\lambda_i\gg1$ for all bins, we can expand the entropy about the average value $\langle n_i \rangle =\lambda_i$ in each bin.  The lowest order term is the expected value of the entropy in the limit of  infinite statistics
%%%
\begin{equation}
H_\infty= -\sum_{i=1}^{\nbins} \frac{\lambda_i}{\Nev} \log_2 \frac{\lambda_i}{\Nev}.
\end{equation}
%%%
The first order term vanishes because $\lambda_i$ is the average value. The second order term is the first non-trivial effect from finite sample size.  The second derivative of the entropy factor is
%%%
\begin{equation}
\left. \frac{\df^2}{\df n_i^2}\left( -\frac{n_i}{\Nev} \log_2 \frac{n_i}{\Nev} \right)\right|_{n_i=\lambda_i} =-\frac{1}{\lambda_i \Nev \ln 2} . 
\end{equation}
%%%
Then, the entropy is
%%%
\begin{align}
H&=H_\infty +  \frac{1}{\ln 2}\sum_{i=1}^{\nbins} \sum_{n_i=0}^\infty \frac{\lambda_i^{n_i} e^{-\lambda_i}}{n_i!}\frac{(n_i-\lambda_i)^2}{2}\left( -\frac{1}{\lambda_i \Nev}\right) +\ldots\nonumber \\
&=H_\infty- \frac{1}{2\ln 2}\frac{{\nbins}}{\Nev} +\ldots \ ,
\end{align}
%%%
where we have used the fact that a Poisson distribution with average $\lambda_i$ has variance $\lambda_i$.  Higher terms in the expansion will depend on the distribution of $\lambda_i$, but these contributions are suppressed by inverse powers of $\lambda_i$.

In calculating the mutual information $I(A;B)$ on a single sample, the leading effect from finite statistics is
%%%
\begin{align}
  I(A;B) &= H(A) + H(B) - H(A,B)
  \nn \\
  &= H_\infty(A) + H_\infty(B) -H_\infty(A,B) - \frac{1}{2\ln 2} \left(  \frac{\nbins^A}{\Nev^A} + \frac{\nbins^B}{\Nev^B} - \frac{\nbins^{AB}}{\Nev^{AB}}  \right) + \ldots,
  \end{align}
%%%
where $\nbins^{X}$ is the number of bins used to calculate $H(X)$ and $\Nev^X$ is the number of events in the corresponding sample.  To avoid the leading bias term from binning when the sample sizes are the same (i.e. $\Nev^A = \Nev^B = \Nev^{AB}$), it is important to take $\nbins^{AB}$ (approximately) equal to $\nbins^{AB} = \nbins^A + \nbins^B$, rather than the naive choice $\nbins^{AB} = \nbins^A \nbins^B$.  This is the strategy we used to make \Figs{fig:ang_corr_q}{fig:ang_corr_g}.  Alternatively, if one wants to use the same binning in the $A$ and $B$ observables such that $\nbins^{AB} = \nbins^A \nbins^B$, then one has to adjust the number of events accordingly.  This is the strategy we used to make \Figs{fig:genang_mutinf_corr_q}{fig:genang_mutinf_corr_g}, where we took $\nbins^A = \nbins^B \equiv \nbins$, $\Nev^A = \Nev^B \equiv \Nev$, but $ \Nev^{AB} = \Nev \nbins / 2$.

For the calculation of the truth overlap $H(T;A)$ of an observable $A$, we have to deal with separate quark and gluon event samples.  The truth overlap is defined in terms of the Shannon entropies as
%%%
\begin{equation}
I(T;A) = H^{q+g}(A) - f H^q(A) - (1-f) H^g(A),
\end{equation}
%%%
where $H^{q+g}(A)$ is the entropy of the combined quark and gluon sample, $H^q(A)$ ($H^g(A)$) is the entropy of the quark (gluon) sample, and $f$ is the fraction of quarks in the combined sample.  For finite bins and sample size, the expected entropy of the combined sample is
%%%
\begin{equation}
H^{q+g}(A)=-\sum_{i=1}^{\nbins^{q+g}} \sum_{n_i^q=0}^\infty  \sum_{n_i^g=0}^\infty \frac{\left(\lambda_i^q\right)^{n_i^q}e^{-n_i^q}}{n_i^q!} \frac{\left(\lambda_i^g\right)^{n_i^g}e^{-n_i^g}}{n_i^g!}\left(
\frac{n_i^q+n_i^g}{N_q+N_g}
\right)\log_2\left(
\frac{n_i^q+n_i^g}{N_q+N_g}
\right).
\end{equation}
%%%
Here, $\lambda_i^q$ ($\lambda_i^g$) is the average number of quark (gluon) jets in bin $i$, $n_i^{q}$  ($n_i^{g}$) is the corresponding number of observed jets, $\nbins^{q+g}$ is the total number of bins, and $N_q$ ($N_g$) is the total number of quark (gluon) jets in the combined sample.  Note that
%%%
\begin{equation}
f=\frac{N_q}{N_q+N_g} .
\end{equation}
%%%
Using the same technique as before, we can find the leading effect from finite statistics on the entropy by expanding $n_i^q$ and $n_i^g$ about $\lambda_i^q$ and $\lambda_i^g$ to quadratic order.  We find
%%%
\begin{equation}
H^{q+g}(A)=H^{q+g}_\infty(A)-\frac{1}{2\ln 2}\frac{\nbins^{q+g}}{N_q+N_g}+\ldots.
\end{equation}
%%%
The resulting truth overlap $I(T;A)$ is
%%%
\begin{align}\label{eq:tbfbias}
I(T;A) &= H^{q+g}_\infty(A)- f H^q_\infty(A) - (1-f) H^g_\infty(A)\nonumber \\
&\qquad-\frac{1}{2\ln 2}\frac{\nbins^{q+g}}{N_q+N_g}+\frac{f}{2\ln 2}\frac{\nbins^{q}}{N_q'}+\frac{1-f}{2\ln 2}\frac{\nbins^{g}}{N_g'}+\ldots , 
\end{align}
%%%
where $\nbins^{q}$ ($\nbins^{g}$) is the number of bins in the pure quark (gluon) sample and $N_q'$ ($N_g'$) is the number of jets in the pure quark (gluon) sample.

This finite statistics bias can be reduced by using samples and binning such that the second line of \Eq{eq:tbfbias} is zero.  There are two useful cases to consider.  When the combined sample is created by simply merging the pure samples (i.e.~$N_q'=N_q$ and $N_g'=N_g$), the bias terms reduce to
%%%
\begin{equation}
\delta I(T;A) = \frac{1}{2\ln 2}\frac{\nbins^{q}+\nbins^{g}-\nbins^{q+g}}{N_q+N_g},
\end{equation}
%%%
and so the bias can be removed by binning such that $\nbins^{q}+\nbins^{g}=\nbins^{q+g}$.  Alternatively, if we choose to use the same binning for each sample (i.e.~$\nbins^{q}=\nbins^{g}=\nbins^{q+g}\equiv \nbins$), the bias terms are
%%%
\begin{equation}
\delta I(T;A) = \frac{1}{2\ln 2}\frac{\nbins}{N_q+N_g}\left(
\frac{N_q}{N_q'}+\frac{N_g}{N_g'}-1
\right).
\end{equation}
%%%
This is zero, if, for example, we take the combined sample to have half the number of events of the pure samples (i.e.~$N_q/N_q'=N_g/N_g'=1/2$).  This later strategy is the one we used for all of the truth overlap plots in this paper.

%%%%%%%%%%%%%%%%%%%%%%%%%%%%%%%%%%%%%%%%%%%%%%%%%%%%%%%%%%%%%%%%%%%%%%%%%%%%%%%%
\section{Calculational Details}
\label{app:details}
%%%%%%%%%%%%%%%%%%%%%%%%%%%%%%%%%%%%%%%%%%%%%%%%%%%%%%%%%%%%%%%%%%%%%%%%%%%%%%%%

%~~~~~~~~~~~~~~~~~~~~~~~~~~~~~~~~~~~~~~~~~~~~~~~~~~~~~~~~~~~~~~~~~~~~~~~~~~~~~~~
\subsection{One IRC Safe Angularity at NLL}
\label{app:IRCsafeNLL}
%~~~~~~~~~~~~~~~~~~~~~~~~~~~~~~~~~~~~~~~~~~~~~~~~~~~~~~~~~~~~~~~~~~~~~~~~~~~~~~~

The cross section for a single recoil-free angularity $e_\beta$ was derived in \Refs{Larkoski:2014uqa,Banfi:2004yd} (see also \Ref{Ellis:2010rwa}), and we summarize the results here.  To NLL order, the cumulative distribution for $e_\beta$ can be expressed as
%%%
\begin{equation}
\label{eq:singleangNLL}
\Sigma_i(e_\beta) =\frac{e^{-\gamma_E R_i'(e_\beta)}}{\Gamma(1+R_i'(e_\beta))}\, e^{-R_i(e_\beta)-\gamma_i T_i(e_\beta)}.
\end{equation}
%%%
Here, we are using a slightly different notation from \Ref{Banfi:2004yd} and the body of the text (see~\Eq{eq:CAESARmaster}), with
%%%
\be
\label{eq:change_r_def}
R_i^{\rm text}(e_\beta) = R_i(e_\beta) +\gamma_i T_i(e_\beta).
\ee
%%%
The reason for separating out the $T$ function is that it contains terms that formally start at NLL order, such that we do not need to consider the logarithmic derivative $T'(e_\beta)$ in the prefactor.

In this notation, the radiator $R(e_\beta)$ consists of the cusp pieces of the jet and soft function anomalous dimensions, while the function $T(e_\beta)$ contains the non-cusp terms (for details on these anomalous dimensions see \App{app:scet_sing_gen}).  The logarithmic derivative $R'(e_\beta)$ is given by  
%%%
\begin{equation}
R'(e_\beta) \equiv -\frac{\df}{\df \ln e_\beta}\, R(e_\beta) .
\end{equation}
%%%
To NLL accuracy, the cusp anomalous dimensions are evaluated at two-loop order and the radiator is
%%%
\begin{align}
R(e_\beta) &= \frac{C_i}{2\pi \alpha_s \beta_0^2}\frac{1}{\beta-1}\left[
\left( 1+\lambda  \right) \ln(1+\lambda)-(\beta+\lambda)\ln\left(1+\frac{\lambda}{\beta}\right)
\right] 
\nonumber \\ &\quad
+\frac{C_i}{4\pi^2 \beta_0^2}\frac{1}{\beta-1}\bigg[
\bigg( \frac{\Gamma^1_{\text{cusp}}}{\Gamma^0_{\text{cusp}}} -2\pi\, \frac{\beta_1}{\beta_0} \bigg)\left(
\beta \ln\left( 1+\frac{\lambda}{\beta} \right)-\ln(1+\lambda)
\right)
\nonumber \\ &\qquad 
+\, \pi\,\frac{\beta_1}{\beta_0}\left(   
\ln^2(1+\lambda) - \beta \ln^2\left( 1+\frac{\lambda}{\beta}  \right)
\right)
\bigg] .
\end{align}
Here, the strong coupling constant is evaluated at the hard scale
\be
\alpha_s \equiv \alpha_s (p_T R),
\ee
using two-loop running with $n_f = 5$ from $\alpha_s(m_Z) = 0.12$.  The observable is contained in $\lambda = 2\alpha_s \beta_0 \ln  e_\beta$, the color factor of the jet is $C_i$, the one- and two-loop beta functions are
%%%
\begin{equation}
\beta_0 = \frac{11}{12\pi}\, C_A - \frac{1}{6\pi}\, n_f  \ , \qquad \beta_1 = \frac{17}{24\pi^2}\,C_A^2-\frac{5}{24\pi^2}\, C_A n_f -\frac{1}{8\pi^2}\,C_F n_f ,
\end{equation}
%%%
and the ratio of the two-loop to the one-loop cusp anomalous dimensions is
%%%
\begin{equation}
\frac{\Gamma^1_{\text{cusp}}}{\Gamma^0_{\text{cusp}}} = \left( \frac{67}{18}-\frac{\pi^2}{6}  \right) C_A - \frac{5}{9}\,n_f .
\end{equation}
%%%
For the non-cusp terms, the function $T(e_\beta)$ is
%%%
\begin{equation}
T(e_\beta) = \frac{1}{\pi \beta_0}\ln\left(  1+\frac{\lambda}{\beta}  \right),
\end{equation}
%%%
and $\gamma_i$ is the non-cusp anomalous dimension to one-loop.  For quarks and gluons, they are
%%%
\begin{equation}
\gamma_q = \frac{3}{4}C_F \ , \qquad \gamma_g = \frac{11}{12} C_A - \frac{1}{6} n_f .
\end{equation}
%%%
For NLL accuracy, the logarithmic derivative $R'(e_\beta)$ only needs to be evaluated at one-loop: 
%%%
\begin{equation}
R'(e_\beta)_\text{NLL} = \frac{C_i}{\pi \beta_0}\frac{1}{\beta-1}\left(\ln\left( 1+\frac{\lambda}{\beta}  \right)-\ln\left( 1+\lambda  \right) \right).
\end{equation}
%%%
The cross section is obtained from the cumulative distribution in the standard way, 
%%%
\be
\frac{1}{\sigma} \frac{\df \sigma}{\df e_\beta} = \frac{\df }{\df  e_\beta} \, \Sigma(e_\beta) .
\ee
%%%
We will verify this single differential calculation in the language of SCET in \App{app:scet_sing_gen}.

%~~~~~~~~~~~~~~~~~~~~~~~~~~~~~~~~~~~~~~~~~~~~~~~~~~~~~~~~~~~~~~~~~~~~~~~~~~~~~~~
\subsection{Two IRC Safe Angularities at NLL}
\label{app:twoIRCsafeNLL}
%~~~~~~~~~~~~~~~~~~~~~~~~~~~~~~~~~~~~~~~~~~~~~~~~~~~~~~~~~~~~~~~~~~~~~~~~~~~~~~~

For two angularities $e_\alpha$ and $e_\beta$, the form of the double cumulative cross section to NLL order was conjectured in \Ref{Larkoski:2014tva}.  Assuming $\alpha > \beta$, to NLL order it can be written as
%%%
\begin{equation}
\Sigma(e_\alpha,e_\beta) =\frac{e^{-\gamma_E \widetilde{R}(e_\alpha,e_\beta)}}{\Gamma(1+\widetilde{R}(e_\alpha,e_\beta))}\, e^{-R(e_\alpha,e_\beta)-\gamma_i T(e_\alpha,e_\beta)} \ ,
\end{equation}
%%%
where the functions $R(e_\alpha,e_\beta)$, $T(e_\alpha,e_\beta)$, and $\widetilde{R}(e_\alpha,e_\beta)$ are given below.  As in \Eq{eq:change_r_def}, we are changing the notation from the text (see \Eq{eq:twoCAESARmaster}) to separate out the $T$ piece from the radiator.  Note that this has a similar form to \Eq{eq:singleangNLL}, albeit with functions that depend on two arguments.  Unlike the single angularity case, $\widetilde{R}$ is not related to any logarithmic derivative of $R$.

The radiator $R(e_\alpha,e_\beta)$ is
%%%
\begin{align} \label{eq:R_ea_eb}
R(e_\alpha,e_\beta) =&\ \frac{C_i}{2\pi \alpha_s \beta_0^2}\bigg[
 \frac{1}{\alpha-1}\ U\!\left(  2\alpha_s \beta_0 \ln e_\alpha\right)-\frac{\beta}{\beta-1}\ U\!\left(  2\alpha_s \beta_0\, \frac{\ln e_\beta}{\beta} \right)
 \nonumber \\
&\quad+
\frac{\alpha-\beta}{(\alpha-1)(\beta-1)}\ U\bigg(  2\alpha_s \beta_0\, \frac{\ln e_\alpha^{1-\beta} e_\beta^{\alpha-1}}{\alpha-\beta}\bigg)
  \bigg] \nonumber \\
  &+\frac{C_i}{4\pi^2 \beta_0^2}\bigg[
\bigg( \frac{\Gamma^1_{\text{cusp}}}{\Gamma^0_{\text{cusp}}} -2\pi\, \frac{\beta_1}{\beta_0} \bigg)\bigg(
\frac{\beta}{\beta-1} \ln\left( 1+2\alpha_s \beta_0\,\frac{\ln e_\beta}{\beta} \right)
\nonumber \\
& \quad -\frac{1}{\alpha-1}\ln(1+2\alpha_s \beta_0 \ln e_\alpha)
-\frac{\alpha-\beta}{(\alpha-1)(\beta-1)}\ln\bigg(
1+2\alpha_s\beta_0\, \frac{\ln e_\alpha^{1-\beta}e_\beta^{\alpha-1}}{\alpha-\beta}
\bigg)
\bigg)
\nonumber \\
&\quad+\, \pi\frac{\beta_1}{\beta_0}\left(   
\frac{1}{\alpha-1}\ln^2(1+2 \alpha_s \beta_0 \ln e_\alpha) 
- \frac{\beta}{\beta-1} \ln^2\left( 1+2\alpha_s\beta_0\,\frac{\ln e_\beta}{\beta}  \right)\right.
\nonumber \\
& \quad+\, \frac{\alpha-\beta}{(\alpha-1)(\beta-1)}\ln^2\bigg(
1+2\alpha_s\beta_0\, \frac{\ln e_\alpha^{1-\beta}e_\beta^{\alpha-1}}{\alpha-\beta}
\bigg)
\bigg)
\bigg] 
,\end{align}
%%%
where $U(z) = (1+z)\ln(1+z)$.
The non-cusp piece $T(e_\alpha,e_\beta)$ is
%%%
\begin{equation}
T(e_\alpha,e_\beta) = \frac{1}{\pi \beta_0} \ln\left( 1+2\alpha_s \beta_0\, \frac{\ln e_\beta}{\beta} \right) 
-2\frac{\alpha_s}{\pi}\, \frac{\alpha-\beta}{\alpha}\, \frac{e_\alpha^{-\frac{\beta}{\alpha-\beta}}e_\beta^{\frac{\alpha}{\alpha-\beta}}}{\beta+2\alpha_s \beta_0 \ln e_\beta} \ .
\end{equation}
The multiple emissions piece $\widetilde{R}(e_\alpha,e_\beta)$ is
\begin{align}\label{eq:rtildeircsafe}
\widetilde{R}(e_\alpha,e_\beta) &= \frac{C_i}{\pi \beta_0} \Bigg[
\frac{1}{\beta-1}\ln\left( 1+2\alpha_s \beta_0\, \frac{\ln e_\beta}{\beta}  \right) - \frac{1}{\alpha-1} \ln\left(  1+2\alpha_s\beta_0 \ln e_\alpha \right)
 \\ &\quad
-\frac{\alpha-\beta}{(\alpha-1)(\beta-1)}\ln\bigg(
1+2\alpha_s\beta_0\, \frac{\ln e_\alpha^{1-\beta}e_\beta^{\alpha-1}}{\alpha-\beta} 
\bigg)
+ 2\alpha_s\beta_0\, \frac{\alpha-\beta}{\alpha}\, \frac{e_\alpha^{-\frac{\beta}{\alpha-\beta}}e_\beta^{\frac{\alpha}{\alpha-\beta}}}{\beta+2\alpha_s \beta_0 \ln e_\beta}
\Bigg]  \ .
\nn\end{align}
%%%
The power suppressed terms have been chosen such that the sum of the exponents of $e_\alpha$ and $e_\beta$ is 1.  The cross section is obtained from the cumulative distribution by differentiation and imposing the phase space constraint in \Eq{eq:safephasespace}, 
%%%
\be
\frac{1}{\sigma_i} \frac{\df^2 \sigma_i}{\df \safeang{\alpha} \, \df \safeang{\beta}} = \left( \frac{\partial^2}{\partial \safeang{\alpha} \, \partial \safeang{\beta}} \Sigma(e_\alpha,e_\beta) \right) \Theta(\safeang{\beta} - \safeang{\alpha}) \,  \Theta \big((\safeang{\alpha})^\beta - (\safeang{\beta})^\alpha \big).
\ee
%%%
We will verify this double differential calculation in the language of SCET in \App{app:scet_pair_interpolation}.

%~~~~~~~~~~~~~~~~~~~~~~~~~~~~~~~~~~~~~~~~~~~~~~~~~~~~~~~~~~~~~~~~~~~~~~~~~~~~~~~
\subsection{Two IRC Unsafe Angularities at NLL}
\label{app:twoIRCunsafeNLL}
%~~~~~~~~~~~~~~~~~~~~~~~~~~~~~~~~~~~~~~~~~~~~~~~~~~~~~~~~~~~~~~~~~~~~~~~~~~~~~~~

For the IRC unsafe angularities, the double differential distribution takes the form
%%%
\begin{equation}
\Sigma_i(\genang{\rho}{\alpha},\genang{\kappa}{\beta}) =\frac{e^{-\gamma_E \widetilde{R}_i(\genang{\rho}{\alpha},\genang{\kappa}{\beta})}}{\Gamma(1+\widetilde{R}_i(\genang{\rho}{\alpha},\genang{\kappa}{\beta}))}\, e^{-R_i(\genang{\rho}{\alpha},\genang{\kappa}{\beta}) - \gamma_i T_i(\genang{\rho}{\alpha},\genang{\kappa}{\beta})},
\end{equation}
%%%
where we are again using the notation change in \Eq{eq:change_r_def}.  Using the rescaling trick in \Eq{eq:doubleradiatorrescaling}, we can determine the double radiator $R_i + \gamma_i T_i$.  For the multiple emissions term $\widetilde{R}_i$, we interpolate between expressions derived at the phase space boundaries \cite{Larkoski:2014tva}. This interpolation yields 
\begin{align}
\widetilde{R}_i(\genang{\rho}{\alpha},\genang{\kappa}{\beta}) &= \frac{C_i}{\pi \beta_0} \Bigg[
\frac{1}{\beta-\kappa}\ln\left( 1+2\alpha_s \beta_0\, \frac{\ln \genang{\kappa}{\beta}}{\beta}  \right) - \frac{1}{\alpha-\rho} \ln\left(  1+2\alpha_s\beta_0\, \frac{\ln \genang{\rho}{\alpha}}{\rho} \right)
 \\ &\quad
-\frac{\alpha-\beta+\kappa-\rho}{(\alpha-\rho)(\beta-\kappa)}\ln\bigg(
1+2\alpha_s\beta_0\, \frac{\ln (\genang{\rho}{\alpha})^{\kappa-\beta}(\genang{\kappa}{\beta})^{\alpha-\rho}}{\alpha \kappa-\beta\rho} 
\bigg)\nn\\
&\quad
+ 2\alpha_s\beta_0\, \frac{(\alpha-\beta)^2}{\alpha(\alpha \kappa -\beta\rho)}\, \frac{(\genang{\rho}{\alpha})^{-\frac{\beta}{\alpha-\beta}}(\genang{\kappa}{\beta})^{\frac{\alpha}{\alpha-\beta}}}{\beta+2\alpha_s \beta_0 \ln \genang{\kappa}{\beta}}
-2 \alpha_s\beta_0 \frac{(\rho-\kappa)^2}{\kappa(\alpha \kappa \!-\!\beta\rho)}\, \frac{(\genang{\rho}{\alpha})^{\frac{\kappa}{\kappa-\rho}}(\genang{\kappa}{\beta})^{-\frac{\rho}{\kappa-\rho}}}{\rho+2\alpha_s \beta_0 \ln \genang{\rho}{\alpha}}
\Bigg]
.
\nn\end{align}
The final two terms, proportional to powers of $\genang{\rho}{\alpha}$ and $\genang{\kappa}{\beta}$, are formally power-suppressed over the entire phase space, but are necessary to satisfy the boundary conditions.  When $\rho=\kappa=1$, this reduces to the IRC safe case in \Eq{eq:rtildeircsafe}.

%~~~~~~~~~~~~~~~~~~~~~~~~~~~~~~~~~~~~~~~~~~~~~~~~~~~~~~~~~~~~~~~~~~~~~~~~~~~~~~~
\subsection{Finding the ROC Curve at LL}
\label{app:ROCatLL}
%~~~~~~~~~~~~~~~~~~~~~~~~~~~~~~~~~~~~~~~~~~~~~~~~~~~~~~~~~~~~~~~~~~~~~~~~~~~~~~~

In this paper, we focused on mutual information, but one can (in principle) use the same cross sections to determine the ROC curve for quark/gluon discrimination.

At LL accuracy, the quark/gluon probability distributions for the IRC safe angularities can be obtained from differentiating the cumulative distribution from \Eq{eq:LLdoubSud} as 
%%%
\begin{align}
p_i(e_\alpha,e_\beta) &= \frac{1}{\sigma_i} \frac{\df^2\sigma_i}{\df \safeang{\alpha}\, \df \safeang{\beta}} = \frac{\partial^2}{\partial \safeang{\alpha} \, \partial \safeang{\beta}}\,\Sigma_i(\safeang{\alpha},\safeang{\beta}) 
\nonumber \\
&=\left( \frac{2\alpha_s}{\pi} \frac{C_i}{\alpha-\beta}\frac{1}{\safeang{\alpha} \safeang{\beta}} 
+ \frac{4\alpha_s^2}{\pi^2} \frac{C_i^2}{\beta(\alpha-\beta)^2}\frac{1}{\safeang{\alpha} \safeang{\beta}} \log\frac{\safeang{\beta}}{\safeang{\alpha}} \, \log\frac{(\safeang{\alpha})^\beta}{(\safeang{\beta})^\alpha}
\right)\Sigma_i(\safeang{\alpha},\safeang{\beta}).
\end{align}
%%%
To determine the ROC curve, one needs to find contours of constant discrimination power, which is the same as finding contours of constant quark over gluon probabilities.   A displacement $(\df e_\al, \df e_\bt)$ along such a contour satisfies
%%%
\begin{align}
  \frac{p_q (e_\al + \df e_\al,e_\bt + \df e_\bt)}{p_g(e_\al + \df e_\al,e_\bt + \df e_\bt)} = 
  \frac{p_q(e_\al,e_\bt)}{p_g(e_\al,e_\bt)} 
\,,\end{align}
%%%
which can be rewritten in terms of a gradient $\vec \nabla = (\partial/\partial e_\alpha, \partial/\partial e_\beta)$ as
%%%
\begin{align}
 0 &= (\df e_\al, \df e_\bt) \cdot \vec \nabla \ln \frac{p_q(e_\al,e_\bt)}{p_g(e_\al,e_\bt)}
\\
 &= (\df e_\al, \df e_\bt) \cdot \vec \nabla \bigg[ \ln \frac{1 + \frac{2\al_s}{\pi} \frac{C_q}{\bt(\al-\bt)} \ln \frac{e_\bt}{e\al} \ln \frac{(e_\al)^\bt}{(e_\bt)^\al}}{1 + \frac{2\al_s}{\pi} \frac{C_g}{\bt(\al-\bt)} \ln \frac{e_\bt}{e\al} \ln \frac{(e_\al)^\bt}{(e_\bt)^\al}} - \frac{\al_s}{\pi} (C_q - C_g) \Big(\frac{1}{\bt} \ln^2 e_\bt + \frac{1}{\al-\bt} \ln^2 \frac{e_\al}{e_\bt}\Big) \bigg]
\,,\nn\end{align}
%%%
resulting in the following differential equation for contours of constant discrimination power
%%%
\begin{align}
  \frac{\df \safeang{\beta}}{\df \safeang{\alpha}} &= 
  \frac{\beta e_\beta}{e_\alpha} \bigg\{  
3 (\alpha-\beta)^2  \beta^2 \ln e_\alpha 
 -\beta (\alpha-\beta)^2 (\alpha+2 \beta) \ln e_\beta 
 + \frac{2\al_s (C_q+C_g)}{\pi}\, \beta \Big[
    \beta (\beta-\alpha) \ln^3 e_\alpha
  \nn \\ & \quad 
 +   (\alpha-\beta) (\alpha+2 \beta) \ln ^2 e_\alpha \ln e_\beta
 +   (-2 \alpha^2+\alpha \beta+\beta^2)  \ln e_\alpha \ln ^2 e_\beta
 - \alpha (\beta-\alpha) \ln ^3 e_\beta  \Big] 
  \nn \\ & \quad 
 + \frac{4\al_s^2 C_q C_g}{\pi^2}\, \Big[
  \beta^2 \ln ^5 e_\alpha
- \beta (2 \alpha+3 \beta) \ln ^4 e_\alpha  \ln e_\beta 
 +  (\alpha^2+6 \alpha \beta+3 \beta^2) \ln ^3 e_\alpha \ln ^2 e_\beta
 \nn \\ & \quad
 - (3 \alpha^2+6 \alpha \beta+\beta^2) \ln ^2 e_\alpha \ln ^3 e_\beta 
  + \alpha  (3 \alpha+2 \beta) \ln e_\alpha \ln ^4 e_\beta
 - \alpha^2  \ln ^5 e_\beta  \Big] 
\bigg\}\bigg/
 \nn \\ & \quad 
 \bigg\{ \beta^2 (\alpha-\beta)^2 (\alpha+2 \beta) \ln e_\alpha
  -3 \alpha (\alpha-\beta)^2 \beta^2 \ln e_\beta
 + \frac{2\al_s (C_q+C_g)}{\pi}\, \bt(\al-\bt)  \nn \\ & \quad  
 \times \Big[
  -\beta^2  \ln ^3 e_\alpha  
+ \beta (2\alpha+\beta) \ln ^2 e_\alpha  \ln e_\bt
 - \alpha (\alpha+2 \beta) \ln e_\alpha \ln ^2 e_\beta
 + \alpha^2 \ln ^3 e_\beta
 \Big]
 \nn \\ & \quad  
  + \frac{4\al_s^2 C_q C_g}{\pi^2}\,\Big[
 \beta^3 \ln ^5 e_\alpha
 -\beta^2 (3 \alpha+2 \beta) \ln ^4 e_\alpha \ln e_\beta 
 + \beta (3 \alpha^2+6 \alpha \beta+\beta^2) \ln ^3 e_\alpha \ln ^2 e_\beta
 \nn \\ & \quad
 - \al (\alpha^2+6 \alpha \beta+3 \beta^2)  \ln ^2 e_\alpha \ln ^3 e_\beta
 + \alpha^2 (2 \alpha+3 \beta) \ln e_\alpha \ln ^4 e_\beta
 - \alpha^3 \ln ^5(e_\beta) \Big]
 \bigg\} 
\,.\end{align}
%%%
This equation is not easy to solve, which is one of the reasons we focused on mutual information in this paper.

%%%%%%%%%%%%%%%%%%%%%%%%%%%%%%%%%%%%%%%%%%%%%%%%%%%%%%%%%%%%%%%%%%%%%%%%%%%%%%%%
\section{Equivalent NLL Results from SCET}
\label{app:scet}
%%%%%%%%%%%%%%%%%%%%%%%%%%%%%%%%%%%%%%%%%%%%%%%%%%%%%%%%%%%%%%%%%%%%%%%%%%%%%%%%

For the IRC safe angularities, \Ref{Larkoski:2014uqa} demonstrated that the SCET approach and CAESAR approach to resummation give the same single differential cross sections to NLL accuracy.  In this appendix, we repeat the same exercise for the generalized angularities.  We also show how to perform the double differential interpolation of \Ref{Larkoski:2014tva} in the language of SCET.

%~~~~~~~~~~~~~~~~~~~~~~~~~~~~~~~~~~~~~~~~~~~~~~~~~~~~~~~~~~~~~~~~~~~~~~~~~~~~~~~
\subsection{One Generalized Angularity in SCET}
\label{app:scet_sing_gen}
%~~~~~~~~~~~~~~~~~~~~~~~~~~~~~~~~~~~~~~~~~~~~~~~~~~~~~~~~~~~~~~~~~~~~~~~~~~~~~~~

The SCET calculation for the generalized angularities $\genang{\kappa}{\beta}$ mirrors that of track thrust \cite{Chang:2013iba}.  We factorize the cross section into a hard function, jet functions, and a soft function which describe physics at the corresponding scales  
%%%
\begin{align} \label{eq:scales}
 \mu_H = p_T
 \,, \quad
 \mu_J = (\genang{\kappa}{\beta})^{1/\beta}\, p_T R_0
 \,, \quad
 \mu_S = (\genang{\kappa}{\beta})^{1/\kappa}\, p_T R_0
\,.\end{align}
%%%
At NLL order, the cross section is completely generated by renormalization group evolution. For simplicity we evolve the hard and jet function to the soft scale, which results in a cumulative distribution of~\cite{Becher:2008cf,Hornig:2009vb,Chang:2013iba,Almeida:2014uva}
%%%
\begin{align} \label{eq:nll}
\Sigma_i(\genang{\kappa}{\beta}) = 
 \frac{e^{K_H^i + K_J^i -\gamma_E\, \eta_J^i}}{\Gamma(1+\eta_J^i)}\, 
\Bigl( \frac{(\genang{\kappa}{\beta})^{1/\beta}\, p_T R_0}{\mu_J} \Bigr)^{\beta\, \eta_J^i}
 \Bigl(\frac{p_T^2}{\mu_H^2}\Bigr)^{\eta_H^i}
\,.\end{align}
%%%
The evolution kernels that enter here are
%%%
\begin{align}
K_H^i(\mu_H,\mu_S) &= -2C_i\, K_\Gamma(\mu_H,\mu_S) + K_{\gamma_H^i}(\mu_H,\mu_S)
\,, \quad &
\eta_H^i(\mu_J,\mu_S) &= C_i\, \eta_{\Gamma}(\mu_J,\mu_S)
\,, \\
K_J^i(\mu_J,\mu_S) &= \frac{2C_i \beta}{\beta-\kappa}\, K_\Gamma(\mu_J,\mu_S) + K_{\gamma_J^i}(\mu_J,\mu_S)
\,, \quad &
\eta_J^i(\mu_J,\mu_S) &= -\frac{2C_i}{\beta-\kappa}\, \eta_{\Gamma}(\mu_J,\mu_S)
\,,\nn\end{align}
%%%
which are given in terms of 
%%%
\begin{align}
K_\Gamma(\mu_0, \mu) &= -\frac{\Gamma_0}{4\beta_0^2}\,
\bigg[ \frac{4\pi}{\alpha_s(\mu_0)}\, \Bigl(1 - \frac{1}{r} - \ln r\Bigr)
   + \biggl(\frac{\Gamma_1 }{\Gamma_0 } - \frac{\beta_1}{\beta_0}\biggr) (1-r+\ln r)
   + \frac{\beta_1}{2\beta_0} \ln^2 r \bigg]
\,, \nn\\
\eta_\Gamma(\mu_0, \mu) &=
 - \frac{\Gamma_0}{2\beta_0}\, \ln r
\,, \nn\\
K_{\gamma}(\mu_0, \mu) &=
 - \frac{\gamma}{2\beta_0}\, \ln r
\,,\end{align}
%%%
where $r=\al_s(\mu)/\al_s(\mu_0)$. The coefficients of the beta function, the cusp, and the non-cusp anomalous dimensions are
%%%
\begin{align} \label{eq:ga_coeff}
\beta_0 &= \frac{11}{3}\,C_A -\frac{4}{3}\,T_F\,n_f
\,, &
\beta_1 &= \frac{34}{3}\,C_A^2  - \Bigl(\frac{20}{3}\,C_A\, + 4 C_F\Bigr)\, T_F\,n_f
\,,\\
\Gamma_0 &= 4
\,, &
\Gamma_1 &= 4 \Bigl[\Bigl( \frac{67}{9} -\frac{\pi^2}{3} \Bigr)\,C_A  -
   \frac{20}{9}\,T_F\, n_f \Bigr]
\,, \nn\\
 \ga_H^q & = -6 C_F
  \,, \qquad
  \ga_H^g  = -2 \beta_0
  \,, \qquad
  &
 \ga_J^q & = 6 C_F + \frac{8 C_F}{\beta-\kappa}\, f_\kappa^{g,1}
  \,, \qquad
  \ga_J^g  = 2 \beta_0 + \frac{8 C_A}{\beta-\kappa}\, f_\kappa^{g,1}
\,.\nn\end{align}
%%%
Note that the conventions for these coefficients differ from those in \Sec{app:IRCsafeNLL}, and that the nonperturbative effect described by $f_\kappa^{g,1}$ is included in the non-cusp anomalous dimension.

To test the agreement with the IRC safe result in \Sec{app:IRCsafeNLL}, we use the central scale choice in \Eq{eq:scales}, such that the SCET result for $\safeang{\beta}$ can be written as
\be
\Sigma_i(\safeang{\beta}) = 
 \frac{e^{-\gamma_E\, \eta_J^i}}{\Gamma(1+\eta_J^i)} e^{K_H^i + K_J^i}.
\ee
%%%
Up to terms that are beyond NLL order and ignoring logarithms of the jet radius $R_0$, we find
%%%
\begin{align}
 -2C_i\, K_\Gamma(\mu_H,\mu_S) + \frac{2C_i \beta}{\beta-1}\, K_\Gamma(\mu_J,\mu_S) & \stackrel{\text{NLL}}{\simeq} -R(e_\beta), \nn \\ 
 K_{\ga_H^i}(\mu_H,\mu_S) + K_{\ga_J^i}(\mu_J,\mu_S) =  K_{\ga_H^i}(\mu_H,\mu_J) &\stackrel{\text{NLL}}{\simeq} \frac{1}{8}\, \ga_H^i T(e_\beta) = - \gamma_i T(e_\beta), \nn \\
K_H^i(\mu_H,\mu_S) + K_J^i(\mu_J,\mu_S) &\stackrel{\text{NLL}}{\simeq} -R(e_\beta) - \gamma_i T(e_\beta), \nn \\
\eta_J^i(\mu_J,\mu_S) &\stackrel{\text{NLL}}{\simeq} R'(e_\beta),
\end{align}
%%%
so the CAESAR and SCET results indeed agree at this order.  One advantage of the SCET approach is that it separates the physics at different scales, allowing one to estimate the perturbative uncertainty of \Eq{eq:nll} by (independently) varying $\mu_H$, $\mu_J$, and $\mu_S$.

For the IRC unsafe case in \Sec{subsec:betaposregime}, we need to verify the rescaling hypothesis in \Eq{eq:IRCunsafeNLLradiator}, which is equivalent to
%%%
\begin{align}
 \label{eq:rescaleSCET}
R_i(\genang{\kappa}{\beta}) &\stackrel{\text{NLL}}{\simeq} \hat{R}_i\left(\safeang{\beta/\kappa} = \left(\frac{\genang{\kappa}{\beta}}{\exp(f_{\kappa}^{g,1})} \right)^{1/\kappa} \right), \nn \\
T_i(\genang{\kappa}{\beta}) &\stackrel{\text{NLL}}{\simeq} \hat{T}_i\left(\safeang{\beta/\kappa} = \left(\genang{\kappa}{\beta} \right)^{1/\kappa} \right), \nn \\
R'_i(\genang{\kappa}{\beta}) &\stackrel{\text{NLL}}{\simeq} \frac{1}{\kappa} \hat{R}_i'\left(\safeang{\beta/\kappa} = \left(\genang{\kappa}{\beta}\right)^{1/\kappa} \right),
\end{align}
%%%
where $X$ is the function in the IRC unsafe case and $\hat{X}$ is the function in the IRC safe case (with angular exponent $\beta/\kappa$).  Note that we can drop the $f_{\kappa}^{g,1}$ terms from $T_i$ and $R'_i$ to NLL accuracy, and that there is a $1/\kappa$ Jacobian factor from the logarithmic derivative in $R'$.  Under this rescaling, the central scales are related as
%%%
\be
 \ln \frac{\mu_H}{\mu_S} = \ln \frac{\hat{\mu}_H}{\hat{\mu}_S} - \frac{1}{\kappa} f_\kappa^{g,1}, \qquad \ln \frac{\mu_J}{\mu_S} = \ln \frac{\hat{\mu}_J}{\hat{\mu}_S} - \frac{\beta - \kappa}{\kappa \beta} f_\kappa^{g,1}.
\ee
%%%
To NLL order, this has the effect of introducing additional terms in the evolution kernels
%%%
\begin{align}
 -2C_i\, K_\Gamma(\mu_H,\mu_S) + \frac{2C_i \beta}{\beta-\kappa}\, K_\Gamma(\mu_J,\mu_S) & \stackrel{\text{NLL}}{\simeq} 
  -2C_i\, K_\Gamma(\hat \mu_H,\hat \mu_S) + \frac{2C_i (\beta/\kappa)}{(\beta/\kappa)-1}\, K_\Gamma(\hat \mu_J, \hat \mu_S)
  \nn \\ & \qquad
 - K_{\hat \ga_J^i}(\hat \mu_J, \hat \mu_S) \, \frac{8 C_i}{\beta-\kappa}\, \frac{f_\kappa^{g,1}}{\hat \ga_J^i}
 , \nn \\ 
 K_{\ga_H^i}(\mu_H,\mu_S) + K_{\ga_J^i}(\mu_J,\mu_S) &\stackrel{\text{NLL}}{\simeq} 
  K_{\ga_H^i}(\hat \mu_H,\hat \mu_S) + K_{\hat \ga_J^i}(\hat \mu_J, \hat \mu_S) \Big(1 + \frac{8 C_i}{\beta-\kappa}\, \frac{f_\kappa^{g,1}}{\hat \ga_J^i}\Big)
 , \nn \\
\eta_J^i(\mu_J,\mu_S) &  \stackrel{\text{NLL}}{\simeq} \frac{1}{\kappa} \hat{\eta}_J^i(\hat{\mu}_J,\hat{\mu}_S).
\label{eq:Kshifts}
\end{align}
%%%
Because of the $1/\kappa$ factor in the last equation, the $R'$ condition in \Eq{eq:rescaleSCET} is immediately satisfied.  The cusp and non-cusp contributions in the first and second equations of \Eq{eq:Kshifts} do not individually satisfy the rescaling, but their sum does.  In the first equation this arises from rescaling the hard and jet cusp contributions, and in the second equation it comes from the non-cusp anomalous dimension
%%%
\be
\gamma_J^i = \hat{\gamma}_J^i + \frac{8 C_i}{\beta - \kappa} f_\kappa^{g,1}.
\ee
%%%
In the CEASAR approach, this non-cusp contribution is part of the radiator $R$, such that $R$ and $T$ each satisfy the rescaling in \Eq{eq:rescaleSCET}.  Because of this shuffling of contributions, we do not expect the rescaling relation to persist beyond NLL accuracy.

%~~~~~~~~~~~~~~~~~~~~~~~~~~~~~~~~~~~~~~~~~~~~~~~~~~~~~~~~~~~~~~~~~~~~~~~~~~~~~~~
\subsection{Double Differential Interpolation in SCET}
\label{app:scet_pair_interpolation}
%~~~~~~~~~~~~~~~~~~~~~~~~~~~~~~~~~~~~~~~~~~~~~~~~~~~~~~~~~~~~~~~~~~~~~~~~~~~~~~~

We now determine the double differential cross sections with SCET, starting with two IRC safe angularities and then generalizing to the IRC unsafe case.  Following \Ref{Larkoski:2014tva}, the known results on the boundaries $e_\al = e_\bt$, $(e_\al)^\bt = (e_\bt)^\al$ are used to build an interpolation on the full phase space. As discussed in \App{app:scet_sing_gen}, the cross section up to NLL order is determined by the scales $\mu_H$, $\mu_J$, and $\mu_S$, so we can implement the interpolation in terms of these scales.

The interpolation for the cumulative distribution $\Sigma(e_\alpha,e_\beta)$ in the angularities $e_\al$ and $e_\bt$ must satisfy the following boundary conditions:  
%%%
\begin{align} \label{eq:boundaries}
  \Sigma(e_\alpha,e_\beta)|_{(e_\al)^\beta=(e_\beta)^\al} &= \Sigma(e_\alpha)  
  \,, &
  \Sigma(e_\alpha,e_\beta)|_{e_\al =e_\beta} &= \Sigma(e_\beta)  
  \,,\nn \\ 
  \frac{\partial}{\partial e_\al} \Sigma(e_\alpha,e_\beta)|_{(e_\al)^\beta=(e_\beta)^\al} &= \frac{\df}{\df e_\alpha}  \Sigma(e_\alpha)  
  \,, &
  \frac{\partial}{\partial e_\bt} \Sigma(e_\alpha,e_\beta)|_{e_\al=e_\beta} &= \frac{\df}{\df e_\beta} \Sigma(e_\beta)
  \,,\nn \\ 
  \frac{\partial}{\partial e_\bt} \Sigma(e_\alpha,e_\beta)|_{(e_\al)^\beta=(e_\beta)^\al} &= 0
  \,, &
  \frac{\partial}{\partial e_\al} \Sigma(e_\alpha,e_\beta)|_{e_\al=e_\beta} &=  0  
\,.\end{align}
%%%
We start at the boundary $(e_\al)^\beta=(e_\beta)^\al$ where we may multiply the scales by arbitrary powers of $(e_\beta)^\al/(e_\al)^\bt$. If the jet and soft scales on the one boundary maps onto the jet and soft scales on the other boundary, we find
%%%
\begin{align} \label{eq:interpolate_scales}
 \mu_{J \to J} = (e_\beta)^{1/\beta}\, p_T R_0
 \,, \qquad
 \mu_{S \to S} = e_\alpha\, p_T R_0
\,.\end{align}
%%%
If the jet and soft scales swap from one boundary to the other boundary, we get 
%%%
\begin{align}
  \mu_{J \to S} = \Big((e_\al)^{1-\bt} (e_\bt)^{\al-1}\Big)^{1/(\al-\bt)} p_T R_0
 \,, \qquad
  \mu_{S \to J} = \Big((e_\al)^{\al-1} (e_\bt)^{\al(1-\beta)/\beta}\Big)^{1/(\al-\bt)} p_T R_0
\,.\end{align}
%%%
This is of course strange from the point of view of factorization, but one should remember that the factorization theorem on the boundary does not hold in the interior.  Using these interpolating scales, we can write down a candidate form for the double radiator,
%%%
\begin{align} \label{eq:R_ansatz}
R(e_\alpha,e_\beta)
 &= p K_\Gamma(\mu_H,\mu_{S\to S}) + q K_\Gamma(\mu_H,\mu_{J\to S})
 + r K_\Gamma(\mu_H,\mu_{S\to S}) + s K_\Gamma(\mu_H,\mu_{S\to J}) 
 \nn \\ & \quad
 + t K_\Gamma(\mu_{J\to J},\mu_{S \to S}) + u K_\Gamma(\mu_{J\to S},\mu_{S \to J})
  + v K_\Gamma(\mu_{J\to J},\mu_{J \to S}) 
 \nn \\ & \quad  
  + w K_\Gamma(\mu_{J\to J},\mu_{S \to J})
  + x K_\Gamma(\mu_{S\to S},\mu_{J \to S})
  + y K_\Gamma(\mu_{S\to S},\mu_{S \to J}) .\end{align}
%%%
Imposing the  boundary conditions in \Eq{eq:boundaries} to NLL order leads to a one parameter family of solutions. The simplest one is: 
%%%
\begin{align} \label{eq:R_int}
 R(e_\alpha,e_\beta) = 2C_i \Big(K_\Gamma(\mu_H,\mu_{J\to S}) + \frac{\beta}{1-\beta}\, K_\Gamma(\mu_{J\to J},\mu_{J \to S}) + \frac{1}{\alpha-1}\,K_\Gamma(\mu_{S\to S},\mu_{J \to S}) \Big)
\,.\end{align}
%%%
This agrees with \Eq{eq:R_ea_eb} up to terms that are beyond NLL order. 

For the non-cusp piece, we have 
%%%
\begin{align} \label{eq:noncusp_int}
 -\ga^i\, T(e_\al,e_\beta) &= K_{\ga_H^i}(\mu_H,\mu_{J\to J}) 
- \al_s\, \frac{\alpha-\beta}{\alpha}\, \frac{e_\alpha^{-\beta/(\alpha-\beta)}e_\beta^{\alpha/(\alpha-\beta)}}{4\pi \beta+2\alpha_s \beta_0 \ln e_\beta} 
\,.\end{align}
%%%
(Remember that a different convention for $\beta_0$ is used here than in \App{app:details}.)
The first term satisfies the boundary conditions on the cumulative distribution and the derivative boundary conditions at $e_\al=e_\bt$. The second term is power suppressed, except at the boundary $e_\al = (e_\bt)^{\al/\bt}$, and is introduced to satisfy the derivative boundary conditions there. It is formally beyond the order we are working so other choices are possible. 

The single angularity result $R'(e_\al) = \eta_J^i(\mu_J,\mu_S) = -2C_i/(\bt-1) \eta_\Gamma(\mu_J,\mu_S)$ suggests an ansatz similar to \Eq{eq:R_ansatz} for $\tilde R$. This leads to
%%%
\begin{align} \label{eq:R_mult}
  \tilde R(e_\al,e_\bt) &= -2 C_i \bigg[ \frac{1}{\al-1} \eta_\Gamma(\mu_{J \to S},\mu_{S \to S}) +
  \frac{1}{\bt-1} \eta_\Gamma(\mu_{J\to J},\mu_{J\to S}) 
  \nn \\ & \qquad
- 4 \al_s\, \frac{\alpha-\beta}{\alpha}\, \frac{e_\alpha^{-\beta/(\alpha-\beta)}e_\beta^{\alpha/(\alpha-\beta)}}{4\pi \beta+2\alpha_s \beta_0 \ln e_\beta}   
  \bigg]
\,.\end{align}
%%%
The first two terms satisfies the boundary condition on the cumulative distribution and the derivative boundary condition at $e_\al=e_\bt$. As in \Eq{eq:noncusp_int}, the additional power suppressed terms take care of the derivative boundary conditions at $e_\al = (e_\bt)^{\al/\bt}$.

The interpolation for IRC unsafe angularities is a direct generalization of \Eq{eq:R_int}   
%%%
\begin{align}
 R(\lambda_\alpha^\rho,\lambda_\beta^\kappa) = 2C_i \Big(K_\Gamma(\mu_H,\mu_{J\to S}) + \frac{\beta}{\kappa-\beta}\, K_\Gamma(\mu_{J\to J},\mu_{J \to S}) + \frac{\rho}{\alpha-\rho}\,K_\Gamma(\mu_{S\to S},\mu_{J \to S}) \Big)
\,,\end{align}
%%%
where we assume $\al/\rho > \beta/\kappa$ and the scales are modified to  
%%%
\begin{align} 
 \mu_{J \to J} &= (\lambda_\beta^\kappa)^{1/\beta}\, p_T R_0
 \,, \nn \\
  \mu_{S \to S} &= (\lambda_\alpha^\rho)^{1/\rho}\, p_T R_0
 \,, \nn \\
  \mu_{J \to S} &= \Big((\lambda_\alpha^\rho)^{\kappa-\bt} (\lambda_\beta^\kappa)^{\al-\rho}\Big)^{1/(\al \kappa-\bt \rho)}  p_T R_0
 \,, \nn \\
  \mu_{S \to J} &= \Big((\lambda_\alpha^\rho)^{\kappa(\al-\rho)/\rho} (\lambda_\beta^\kappa)^{\al(\kappa-\beta)/\beta}\Big)^{1/(\al \kappa-\bt \rho)} p_T R_0
\,.\end{align}
%%%
The interpolation of the non-cusp piece in \Eq{eq:noncusp_int} mostly carries over.   The contribution from $\ga_H^i = -\hat \ga_J^i$ only involves the hard and jet scales, which are the same as before. However, the nonperturbative coefficients enter through $K_{\ga_J}(\mu_J,\mu_S)$.  Although one can build an interpolation similar to \Eq{eq:noncusp_int}, we also need an interpolation between $f_\rho^{g,1}/(\al-\rho)$ at $(\lambda_\alpha^\rho)^\beta = (\lambda_\beta^\kappa)^\alpha$ and $f_\kappa^{g,1}/(\bt-\kappa)$ at $(\lambda_\alpha^\rho)^\kappa = (\lambda_\beta^\kappa)^\rho$.  It is therefore much more convenient to use the rescaling trick in \Eq{eq:doubleradiatorrescaling}.  Finally, the multiple emissions contribution in \Eq{eq:R_mult} generalizes to
%%%
\begin{align}
  \tilde R(\lambda_\al^\rho,\lambda_\bt^\kappa) &= -2 C_i \bigg[ \frac{1}{\al-\rho} \eta_\Gamma(\mu_{J \to S},\mu_{S \to S}) +
  \frac{1}{\bt-\kappa} \eta_\Gamma(\mu_{J\to J},\mu_{J\to S}) 
  \\ & \qquad 
-4 \alpha_s \frac{(\alpha-\beta)^2}{\alpha(\alpha \kappa \!-\!\beta\rho)}\, \frac{(\genang{\rho}{\alpha})^{-\beta/(\alpha-\beta)}(\genang{\kappa}{\beta})^{\alpha/(\alpha-\beta)}}{4\pi \beta+2\alpha_s \beta_0 \ln \genang{\kappa}{\beta}}  \nn \\
& \qquad
-4 \alpha_s \frac{(\rho-\kappa)^2}{\kappa(\alpha \kappa \!-\!\beta\rho)}\, \frac{(\genang{\rho}{\alpha})^{\kappa/(\kappa-\rho)}(\genang{\kappa}{\beta})^{-\rho/(\kappa-\rho)}}{4\pi \rho+2\alpha_s \beta_0 \ln \genang{\rho}{\alpha}} \bigg].
\nn \end{align}
%%%
The terms on the first line satisfy the boundary condition for the cumulative distribution. The terms on the second and third line are power suppressed except at the boundaries $(\lambda_\al^\rho)^\bt=(\lambda_\bt^\kappa)^\al$ and $(\lambda_\al^\rho)^\kappa=(\lambda_\bt^\kappa)^\rho$, respectively, where their inclusion enforces the derivative boundary conditions. For $\rho=\kappa$ the third line is absent, and for $\rho=\kappa=1$ this reduces to the IRC safe case.

%%%%%%%%%%%%%%%%%%%%%%%%%%%%%%%%%%%%%%%%%%%%%%%%%%%%%%%%%%%%%%%%%%%%%%%%%%%%%%%%
\section{Additional Plots}
\label{app:correlations}
%%%%%%%%%%%%%%%%%%%%%%%%%%%%%%%%%%%%%%%%%%%%%%%%%%%%%%%%%%%%%%%%%%%%%%%%%%%%%%%%

In this appendix, we present additional plots involving mutual information to complement the truth overlap plots in the main text.  We also show some raw angularity distributions.

\begin{figure}
\begin{center}
\subfloat[]{\label{fig:LL_ang_corr_quark}
\includegraphics[width=7cm]{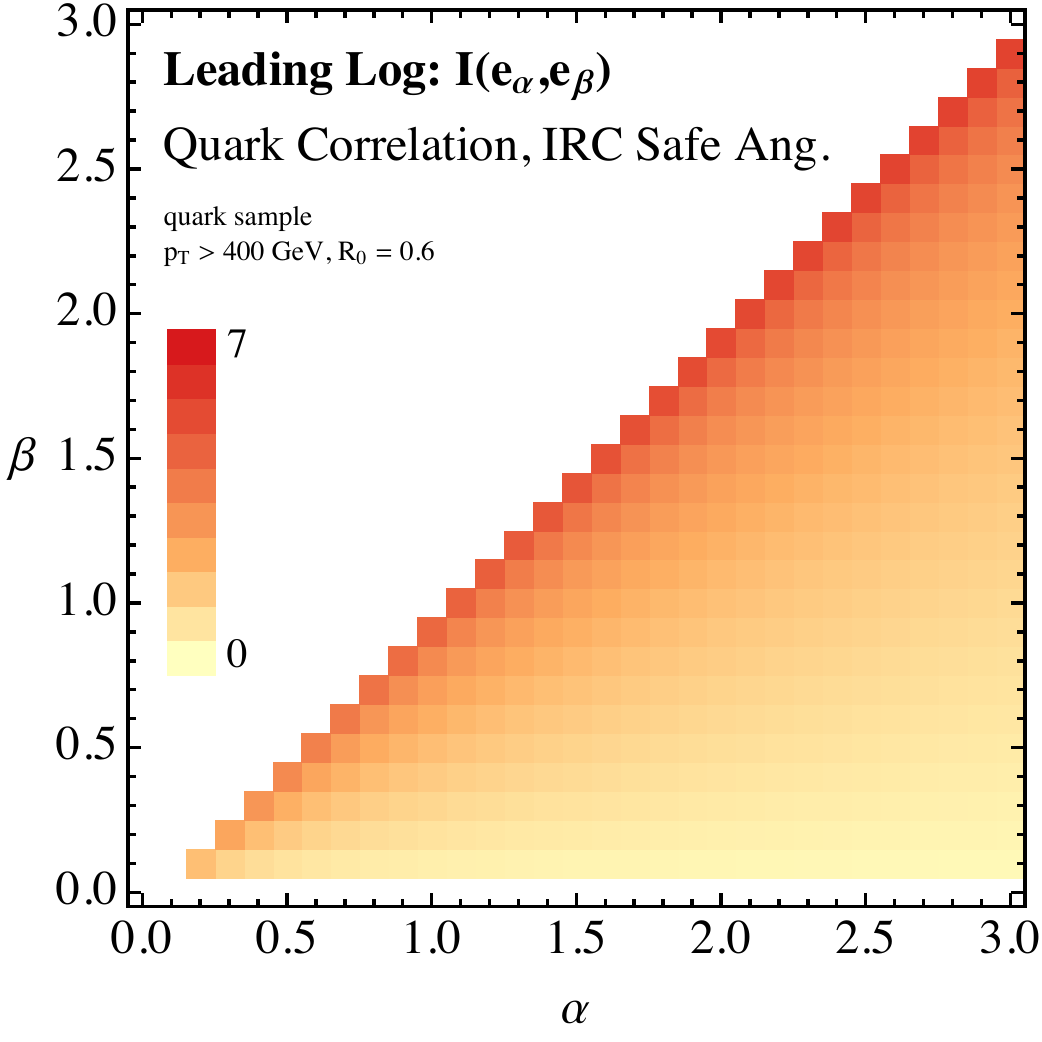}
}
$\qquad$
\subfloat[]{\label{fig:NLL_ang_corr_quark} 
\includegraphics[width=7cm]{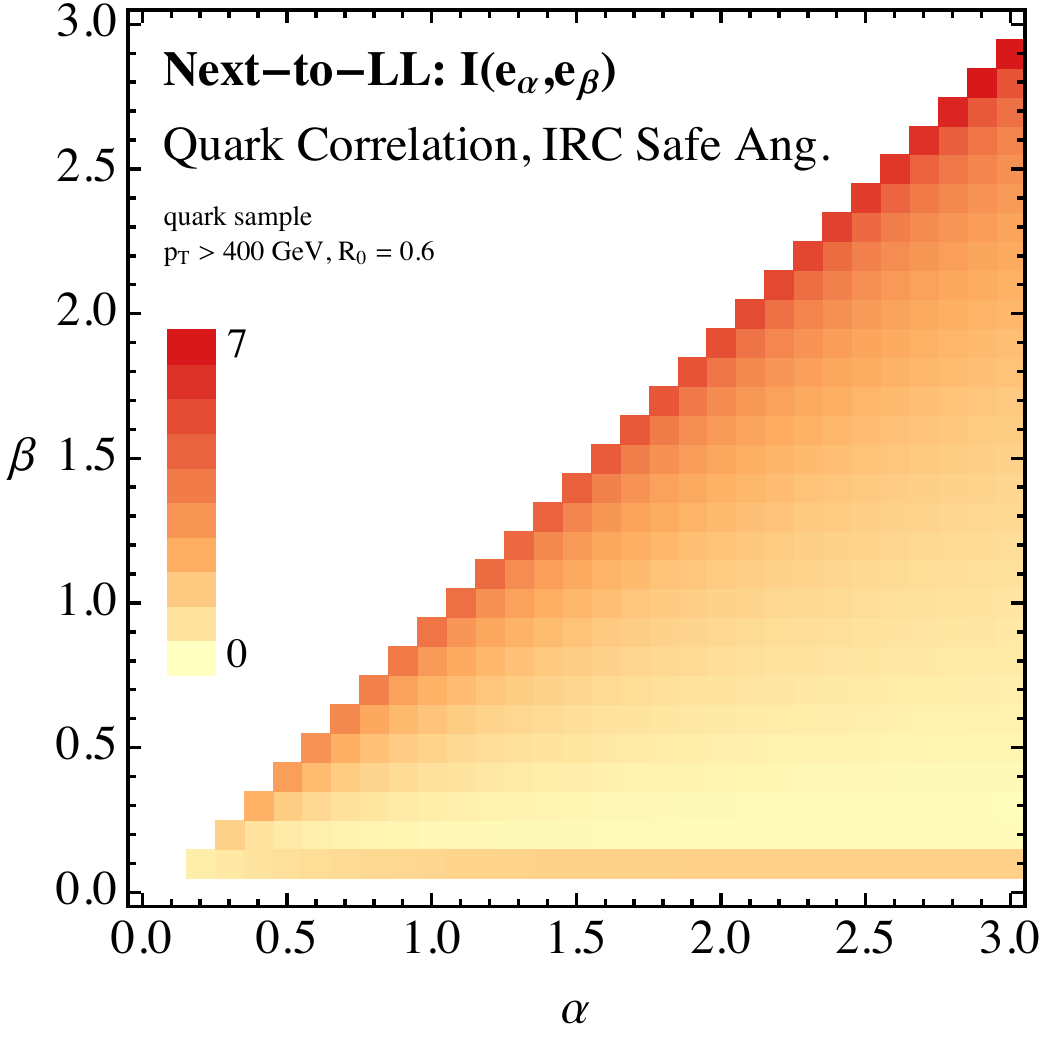}
}\\
\subfloat[]{\label{fig:pythia_ang_corr_quark} 
\includegraphics[width=7cm]{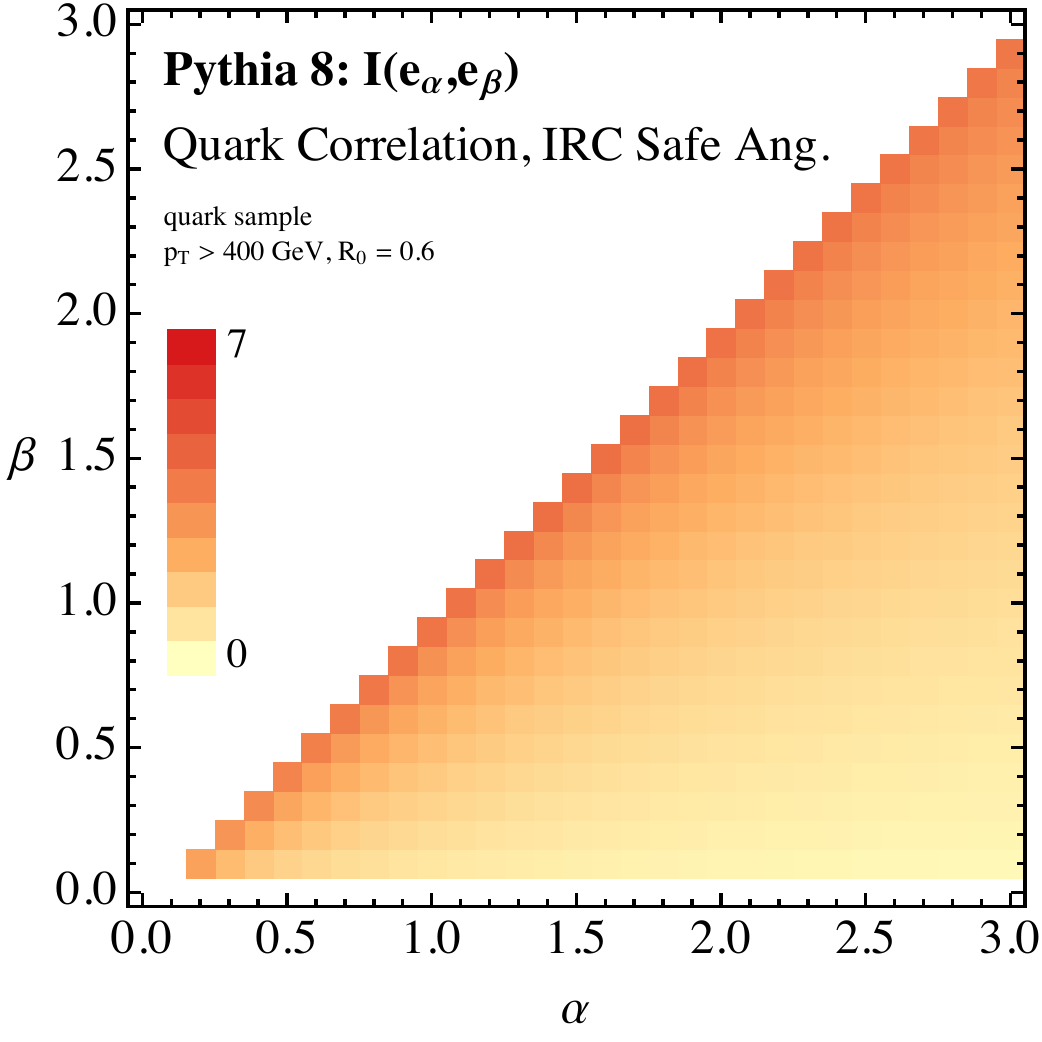}
}$\qquad$
\subfloat[]{\label{fig:herwig_ang_corr_quark} 
\includegraphics[width=7cm]{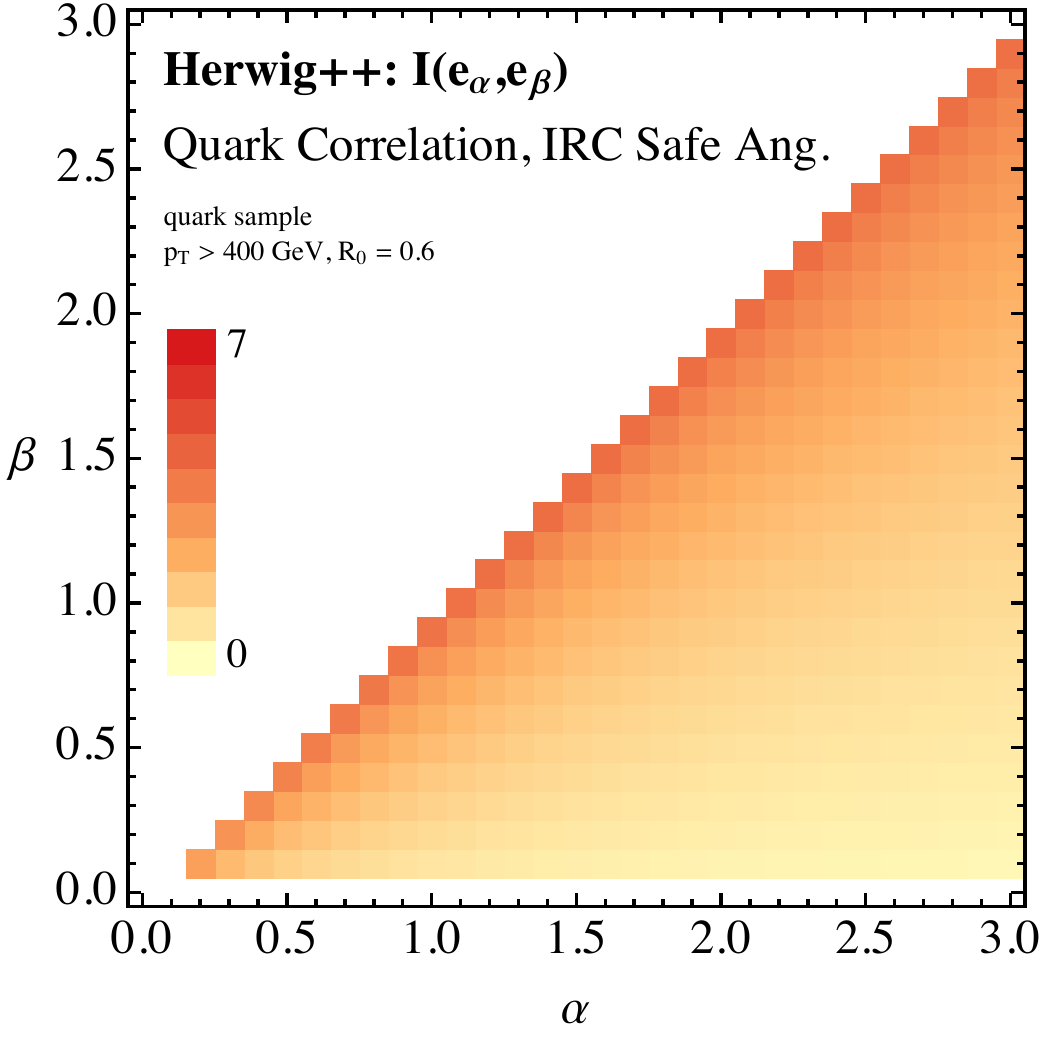}
}
\end{center}
\caption{
Correlation between two IRC safe angularities $(\safeang{\alpha},\safeang{\beta})$ on a pure quark jet sample.  Top:  the LL and NLL analytic calculations.  Bottom: the \pythia{8} and \herwigpp\ parton showers.
}
\label{fig:ang_corr_q}
\end{figure}

\begin{figure}
\begin{center}
\subfloat[]{\label{fig:LL_ang_corr_gluon}
\includegraphics[width=7cm]{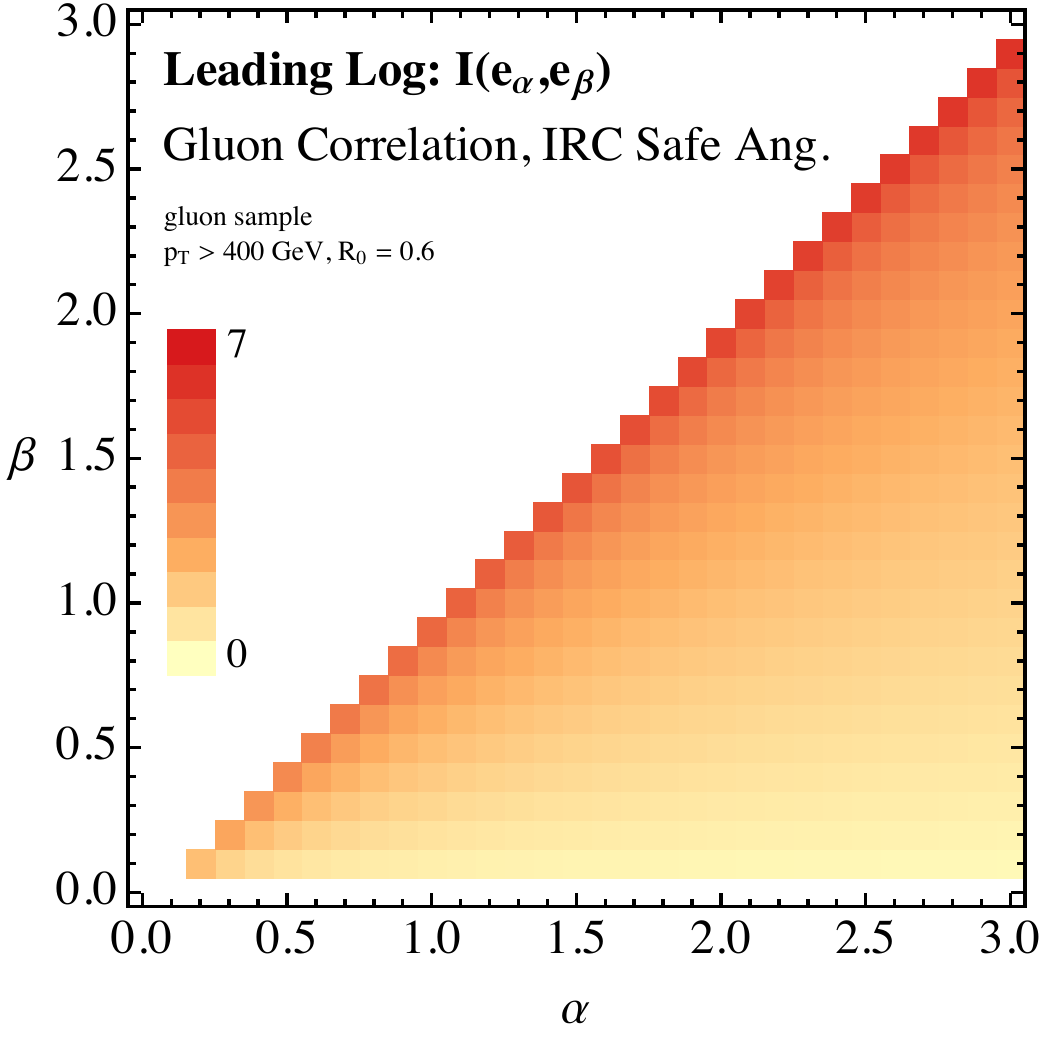}
}
$\qquad$
\subfloat[]{\label{fig:NLL_ang_corr_gluon} 
\includegraphics[width=7cm]{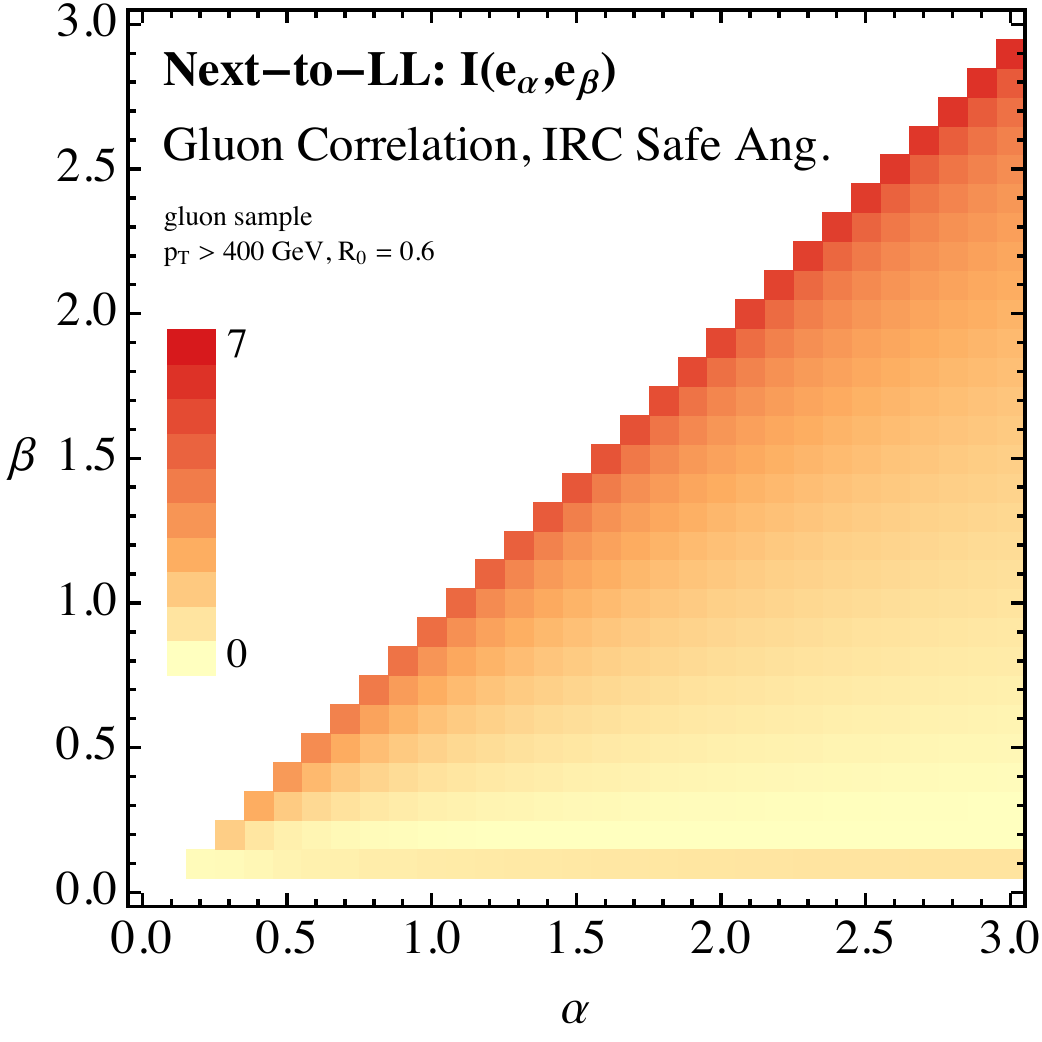}
}\\
\subfloat[]{\label{fig:pythia_ang_corr_gluon} 
\includegraphics[width=7cm]{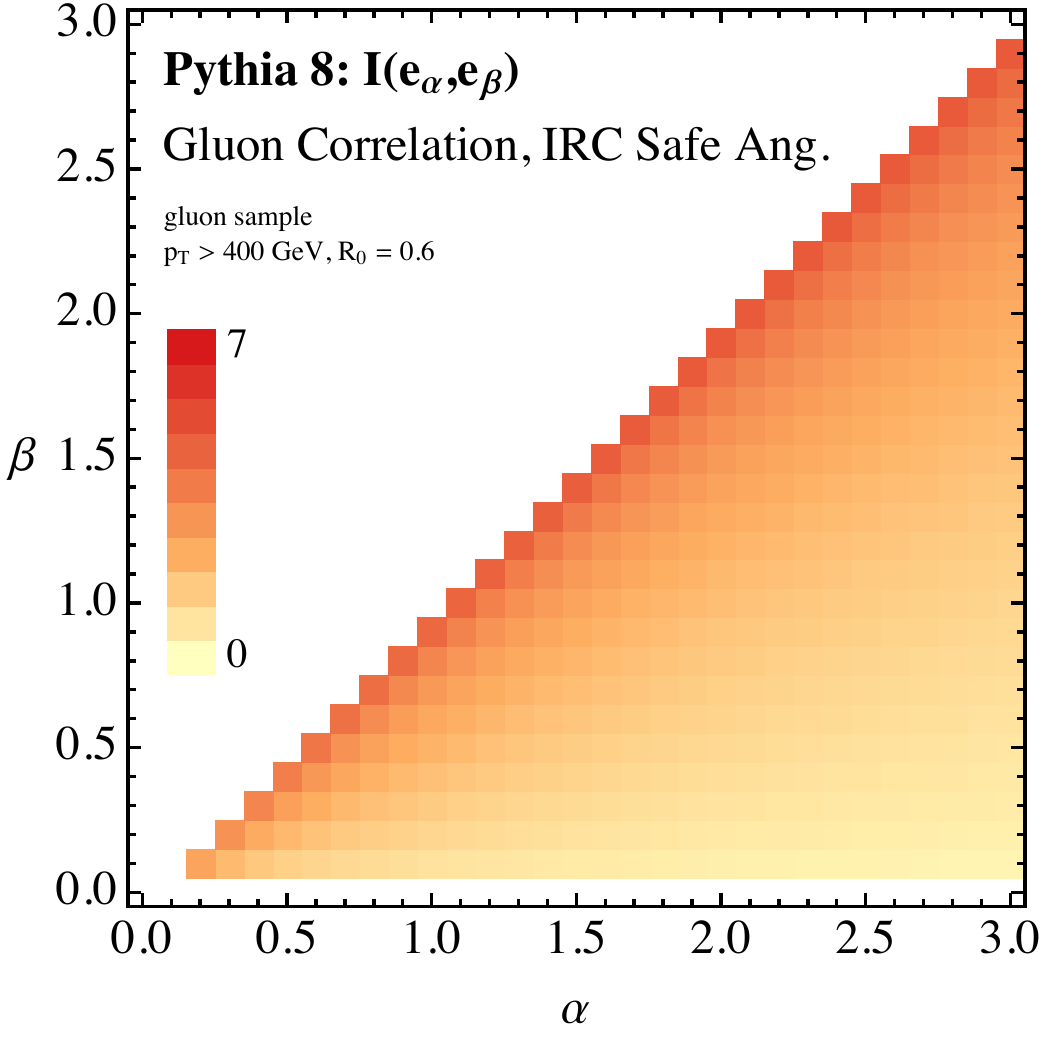}
}$\qquad$
\subfloat[]{\label{fig:herwig_ang_corr_gluon} 
\includegraphics[width=7cm]{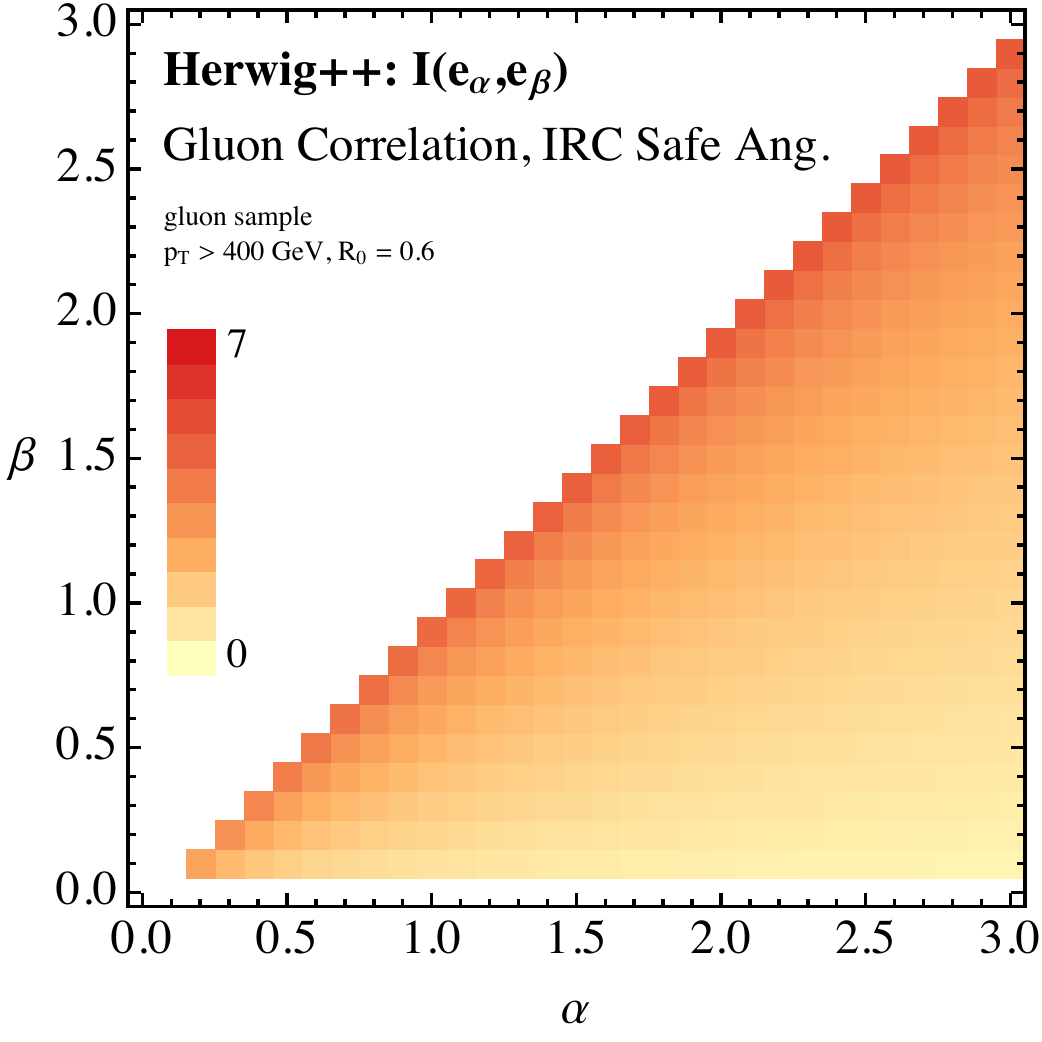}
}
\end{center}
\caption{
Same as \Fig{fig:ang_corr_q}, but on a pure gluon sample.
}
\label{fig:ang_corr_g}
\end{figure}

We first study the correlation between IRC safe angularities with different angular exponents, as measured by $I(\safeang{\alpha};\safeang{\beta})$.  This is shown for a pure sample of quark jets in \Fig{fig:ang_corr_q} and for a pure sample of gluon jets in  \Fig{fig:ang_corr_g}.  Two angularities are highly correlated when their angular exponents are close to one another, and become increasingly uncorrelated as the angular exponents move farther apart.   This behavior can be understood from the definition of the angularities.  For large values of angular exponent, the angularity is dominated by soft, wide-angle emissions because collinear emissions are suppressed by small angles raised to a high power.  By contrast, at small values of the angular exponent, the angularity is dominated by hard collinear emissions.  Thus, when two angularities have very different angular exponents, their values are dominated by different physics and so are largely uncorrelated.  Unlike the case of the truth overlap in \Fig{fig:mutinf_plot}, there is broad agreement between LL, NLL, \pythia{8}, and \herwigpp\ as far as the raw correlations are concerned.

\begin{figure}
\begin{center}
\subfloat[]{\label{fig:LL_ang_mutinf_imp}
\includegraphics[width=7cm]{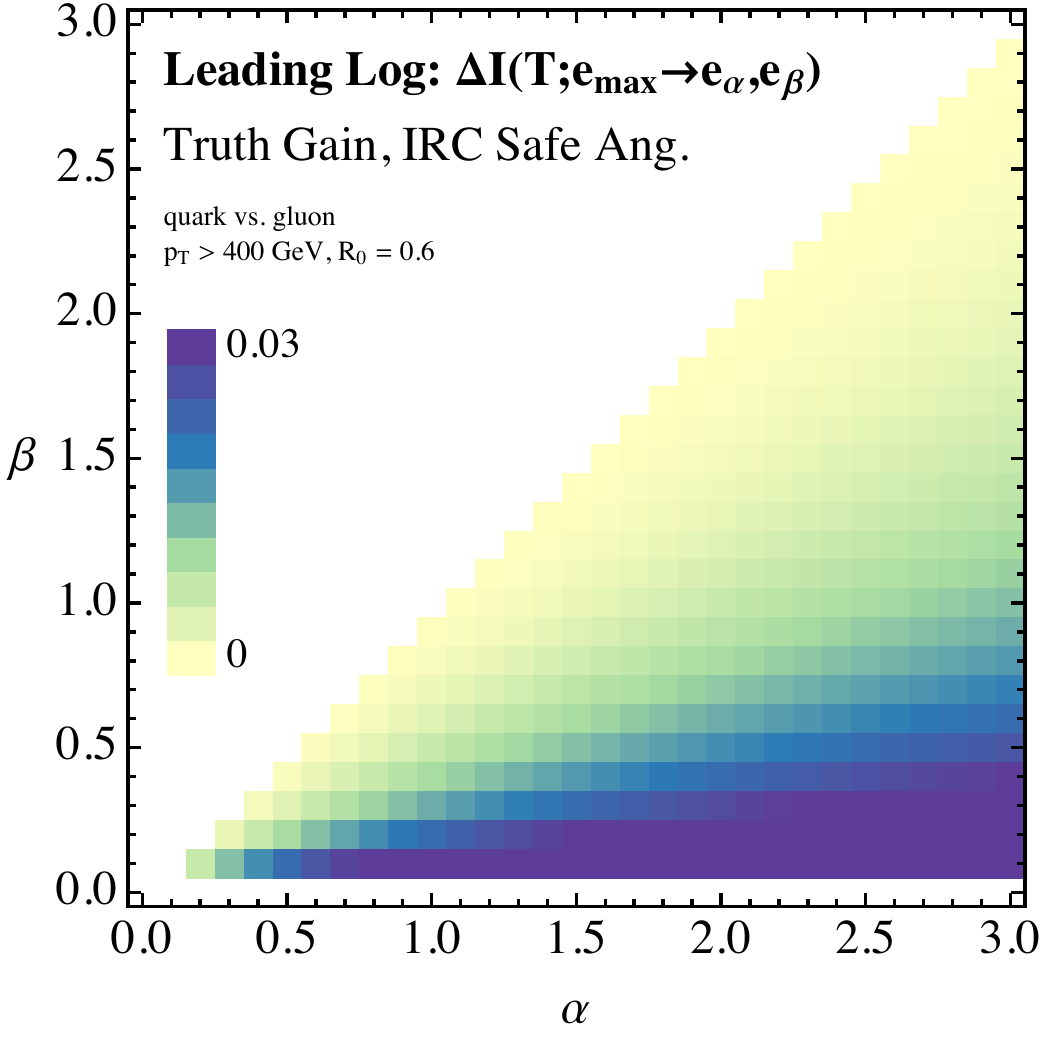}
}
$\qquad$
\subfloat[]{\label{fig:NLL_ang_mutinf_imp} 
\includegraphics[width=7cm]{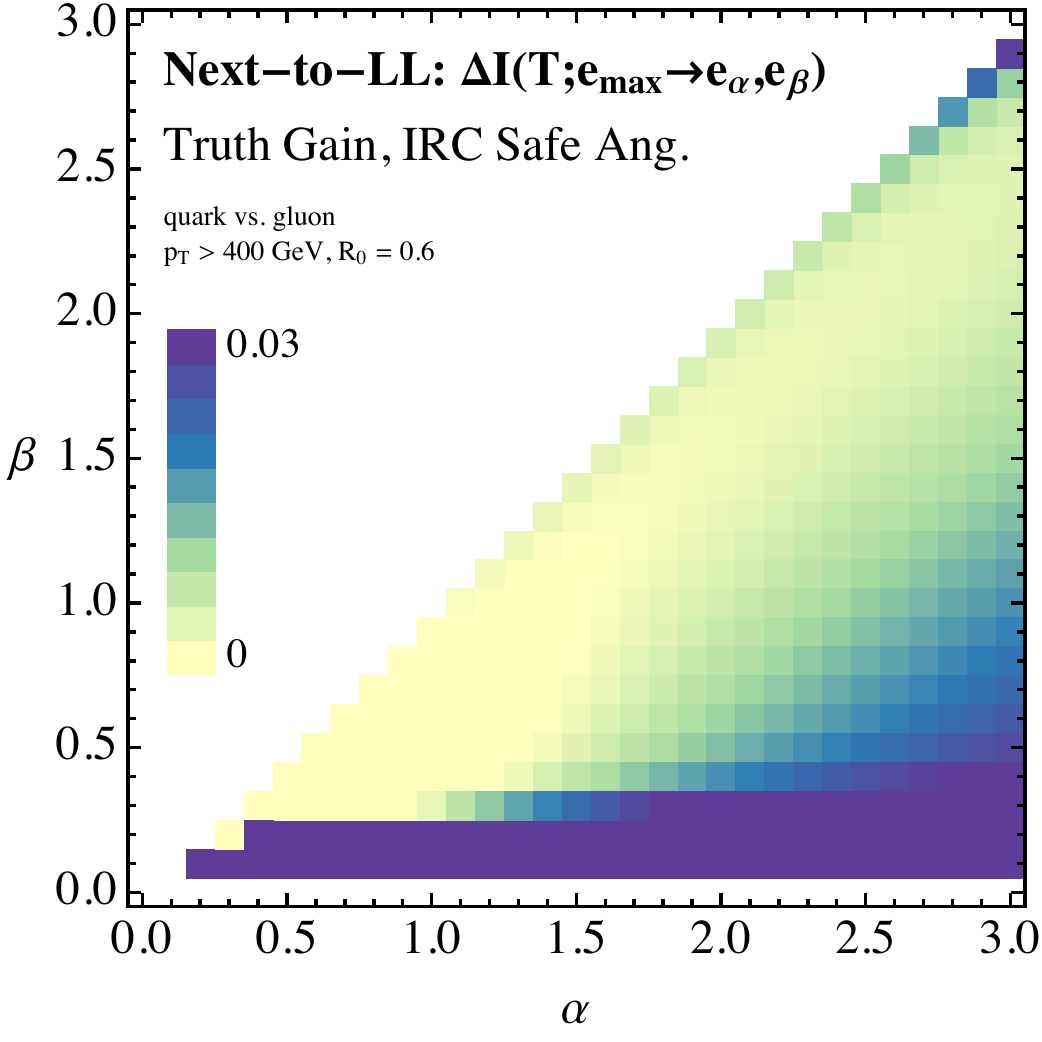}
}\\
\subfloat[]{\label{fig:pythia_ang_mutinf_imp} 
\includegraphics[width=7cm]{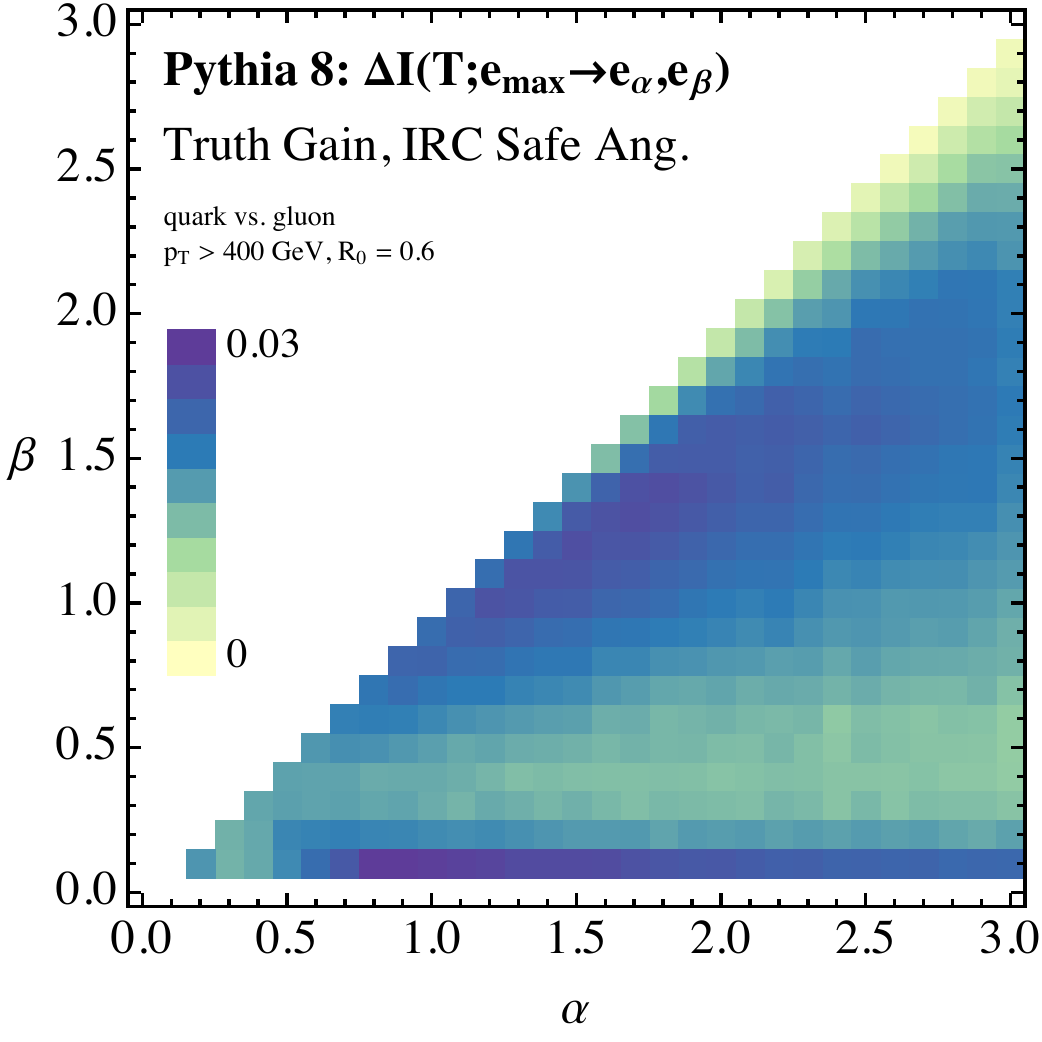}
}$\qquad$
\subfloat[]{\label{fig:herwig_ang_mutinf_imp} 
\includegraphics[width=7cm]{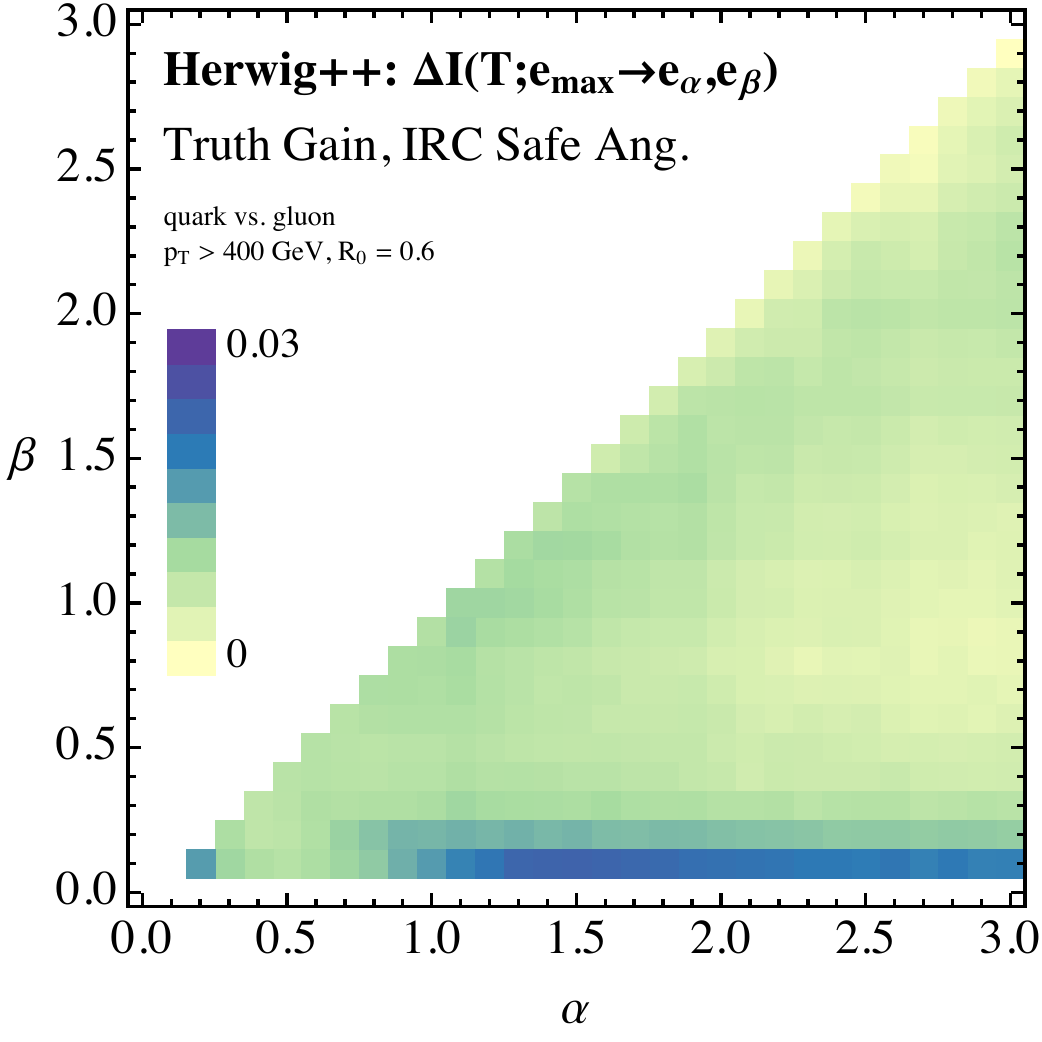}
}
\end{center}
\caption{Improvement in the truth overlap by using two IRC safe angularities $(\safeang{\alpha},\safeang{\beta})$ as compared to only one, $I(T;e_\alpha,e_\beta)-\max[I(T;e_\alpha),I(T;e_\beta)]$.  Top:  the LL and NLL analytic calculations.  Bottom: the \pythia{8} and \herwigpp\ parton showers.}
\label{fig:ang_mutinf_imp}
\end{figure}

Next, we want to understand better the degree to which two angularities have more truth overlap than one angularity.  In \Fig{fig:ang_mutinf_imp}, we plot $\Delta I (T,e_{\rm max} \to \safeang{\alpha},\safeang{\beta})$ from \Eq{eq:DeltaITemax}, namely the pairwise truth overlap $I(T;\safeang{\alpha},\safeang{\beta})$ minus the truth overlap of the stronger angularity $\max\{I(T;\safeang{\alpha}),I(T;\safeang{\beta})\}$.  The information gain is on the order of 10\% ($\mathcal{O}(0.01)$ compared to a baseline truth overlap of $\mathcal{O}(0.1)$).  As in \Fig{fig:mutinf_plot}, there are quite substantial differences between the various methods. It is interesting that in \pythia{8} one can already achieve considerable gains in performance just off the diagonal, i.e.\ for observables that are not very different.  This may be because when $\alpha$ and $\beta$ are close (compare to \Eq{eq:kappatoone}),
\be
\safeang{\alpha} - \safeang{\beta} \simeq (\alpha - \beta) \sum_i z_i \theta_i^\alpha \log \theta_i + \ldots,
\ee
and $\theta_i^\alpha \log \theta_i$ is similar to the optimal kernel found in \Refs{Gallicchio:2011xq,Gallicchio:2012ez}.

\begin{figure}
\begin{center}
\subfloat[]{
\includegraphics[width=7cm]{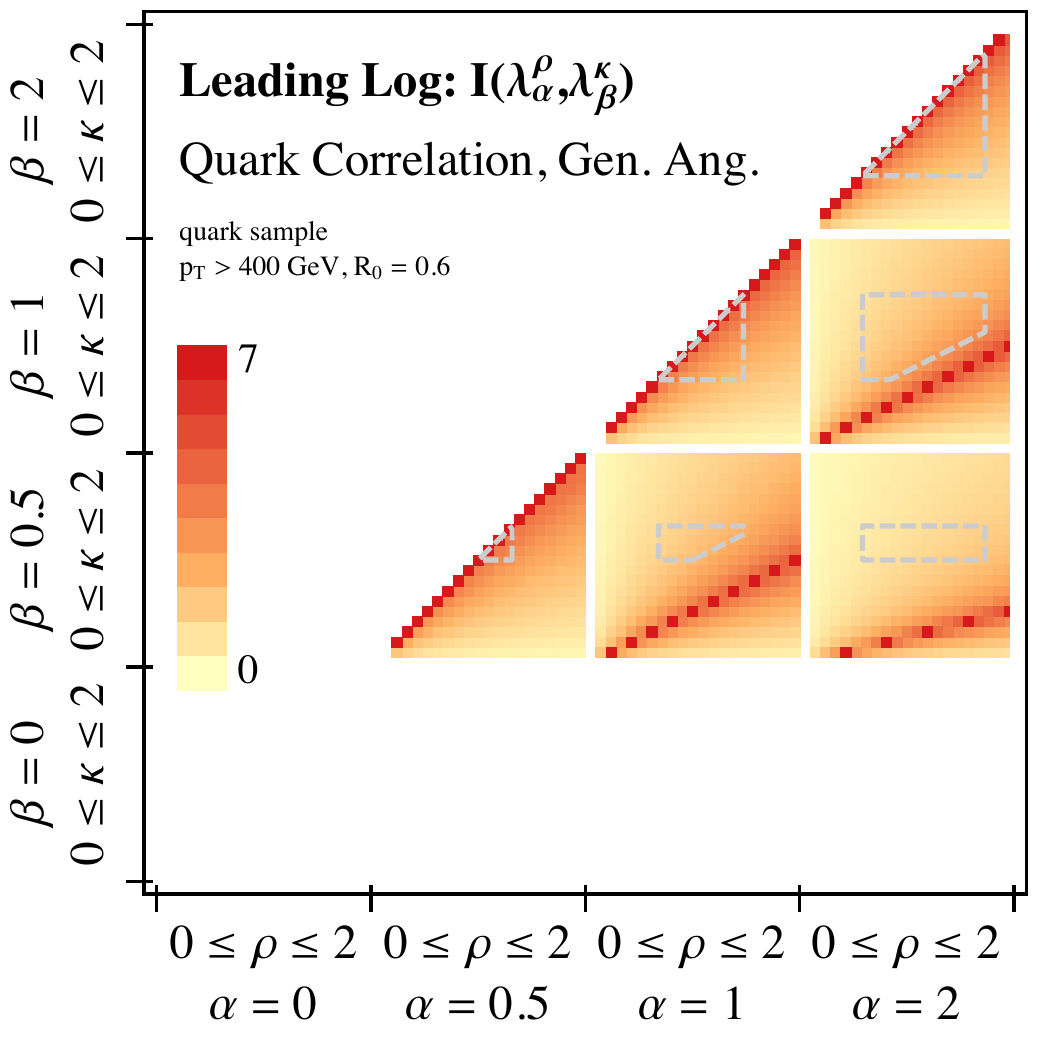}
}
$\qquad$
\subfloat[]{
\includegraphics[width=7cm]{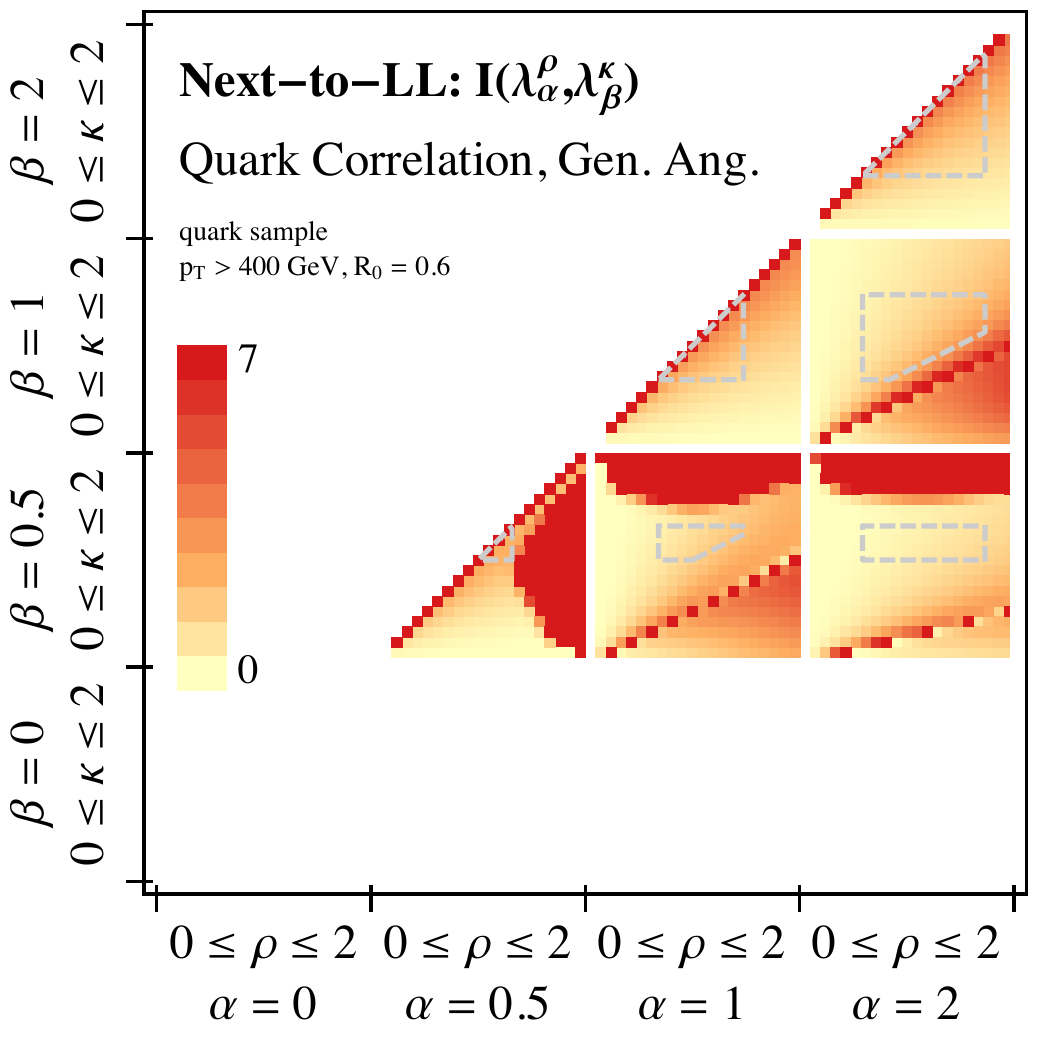}
}
\\
\subfloat[]{\label{fig:py_genang_corr_q}
\includegraphics[width=7cm]{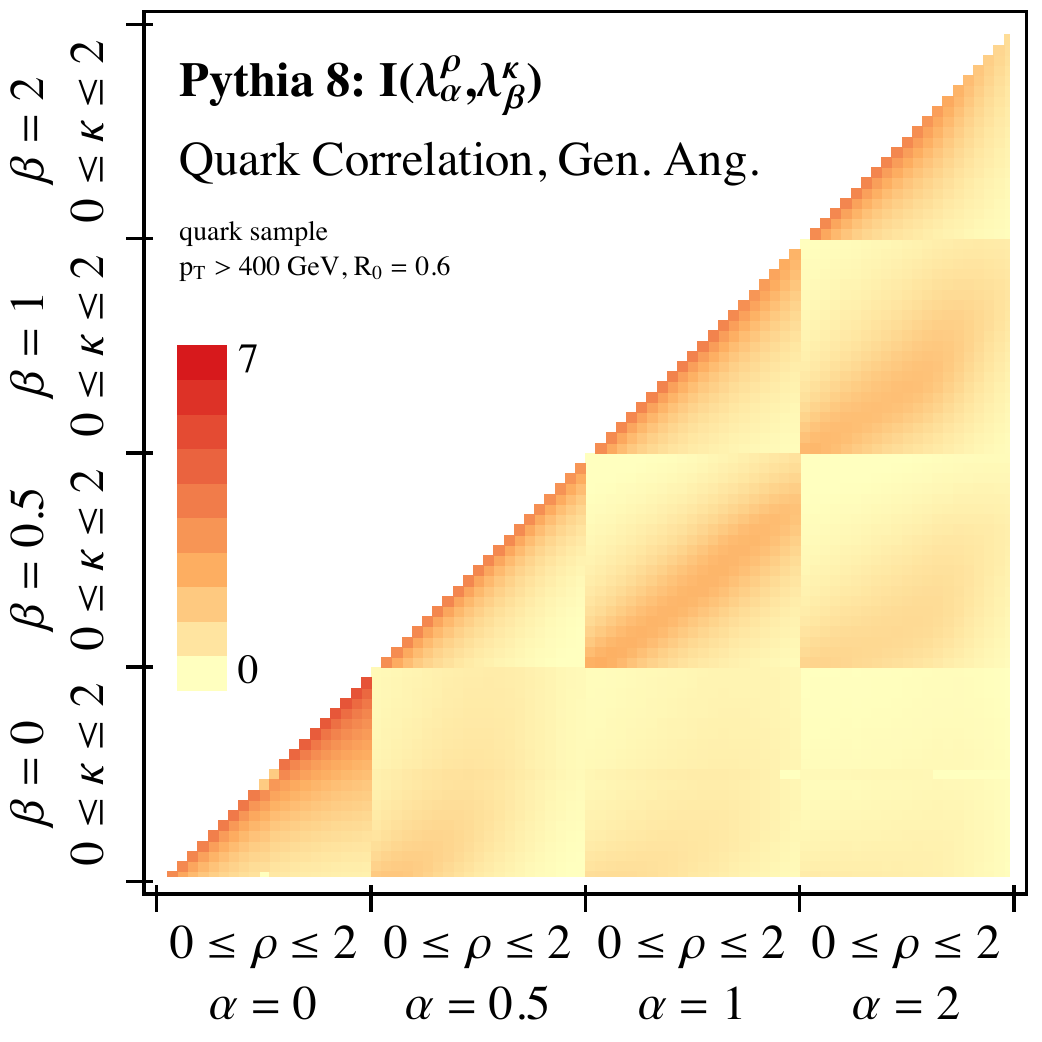}
}
$\qquad$
\subfloat[]{\label{fig:her_genang_corr_q} 
\includegraphics[width=7cm]{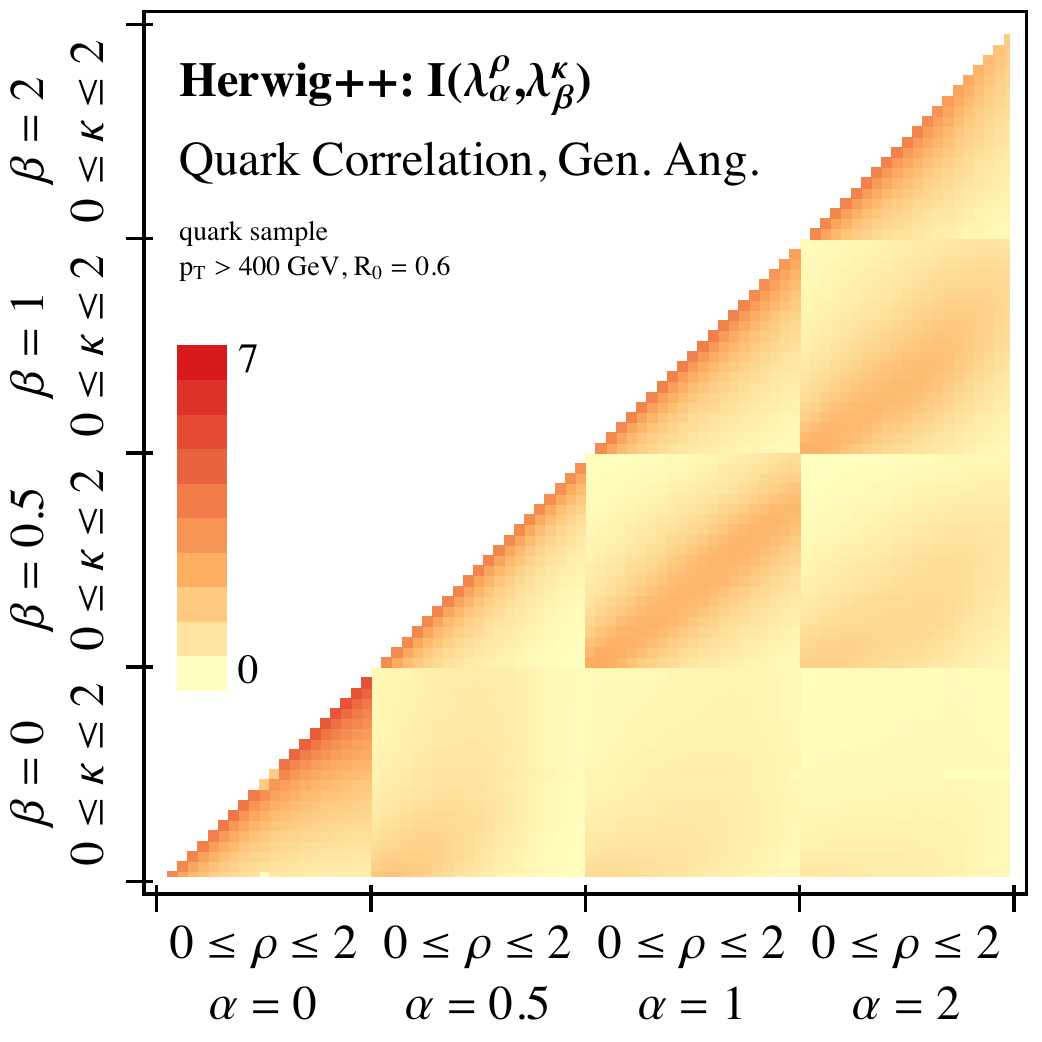}
}
\end{center}
\caption{
Correlation between two generalized angularities $(\genang{\rho}{\alpha},\genang{\kappa}{\beta})$ on a pure quark sample.  Top:  the LL and NLL analytic calculations.  Bottom: the \pythia{8} and \herwigpp\ parton showers.  As in \Fig{fig:genang_mutinf_truth_pair}, $\beta\in\{0,0.5,1,2\}$ and we sweep $0\leq \kappa \leq 2$.
}
\label{fig:genang_mutinf_corr_q}
\end{figure}

\begin{figure}
\begin{center}
\subfloat[]{
\includegraphics[width=7cm]{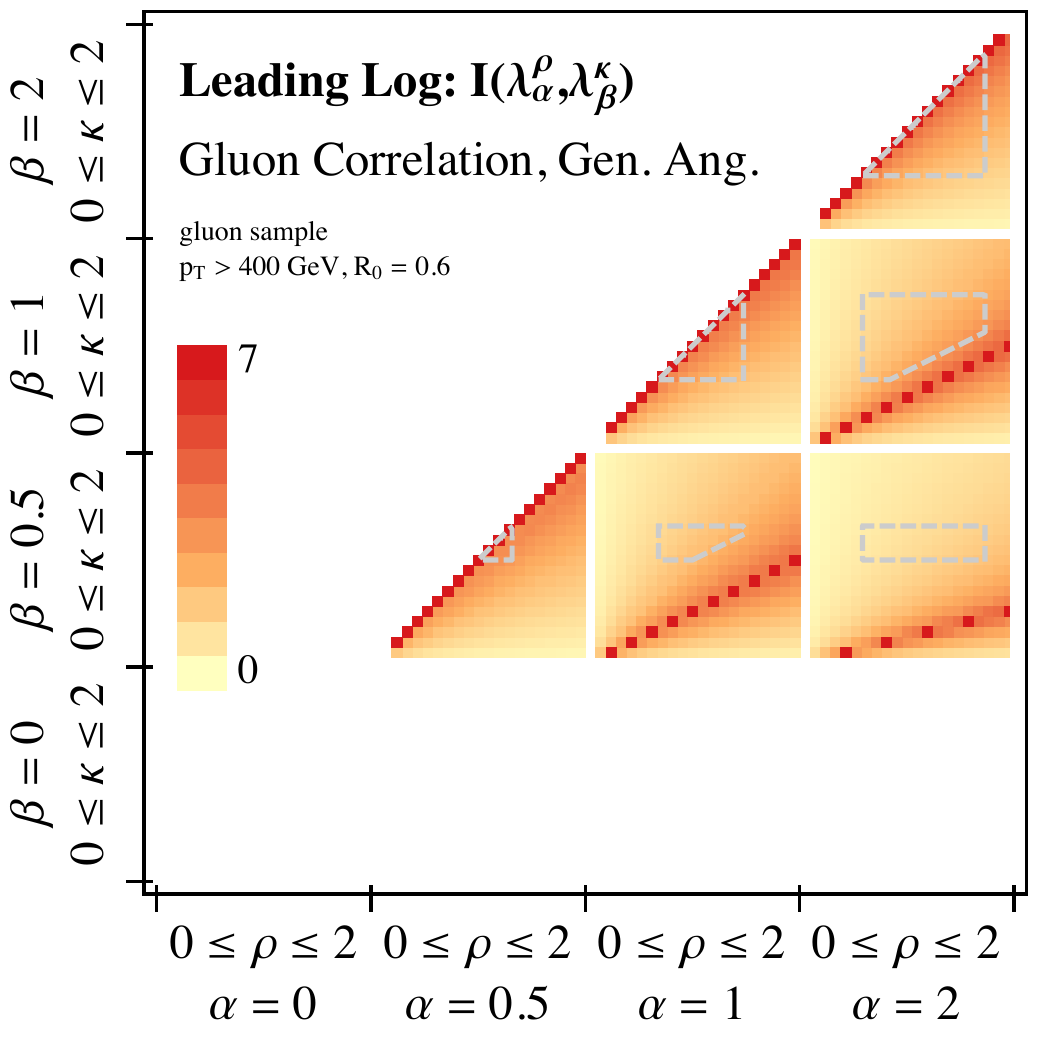}
}
$\qquad$
\subfloat[]{
\includegraphics[width=7cm]{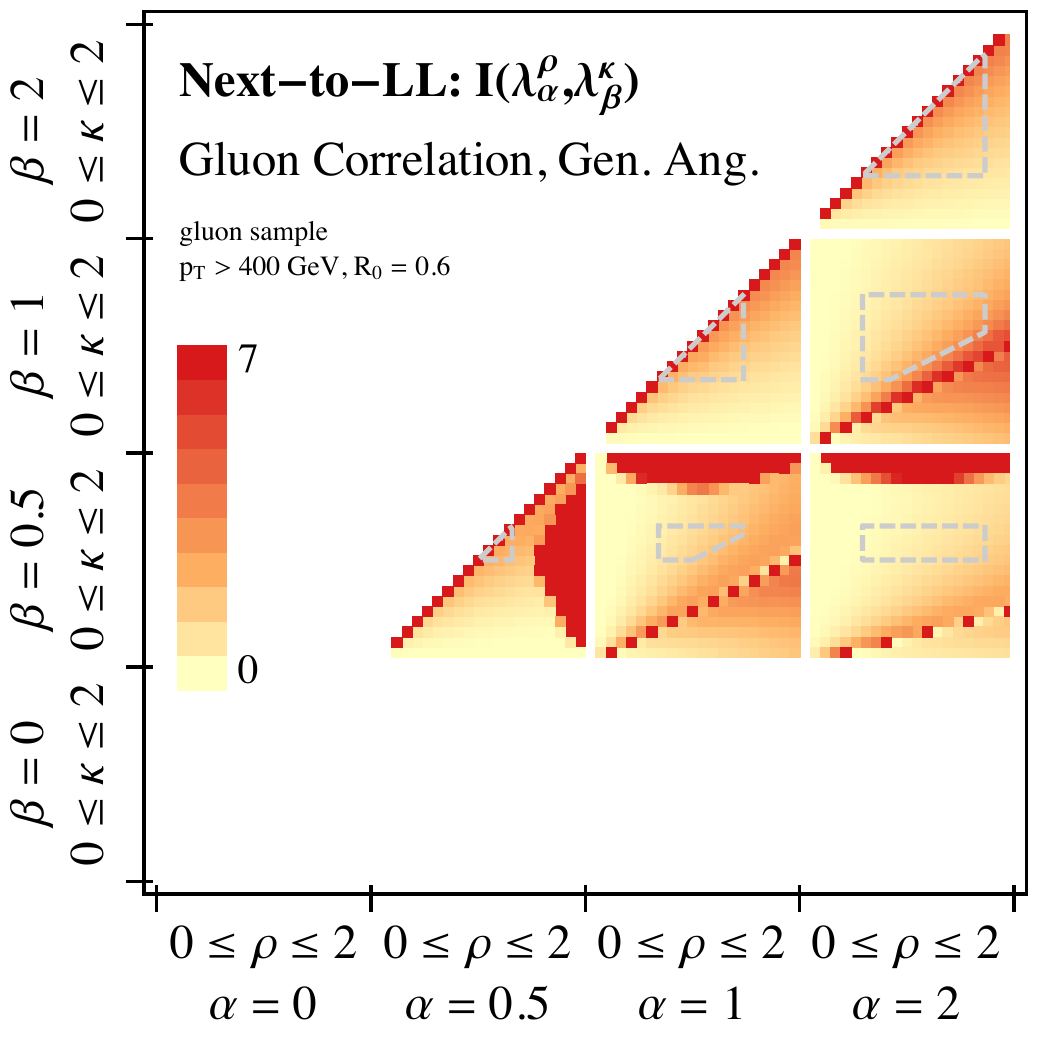}
}
\\
\subfloat[]{\label{fig:py_genang_corr_g} 
\includegraphics[width=7cm]{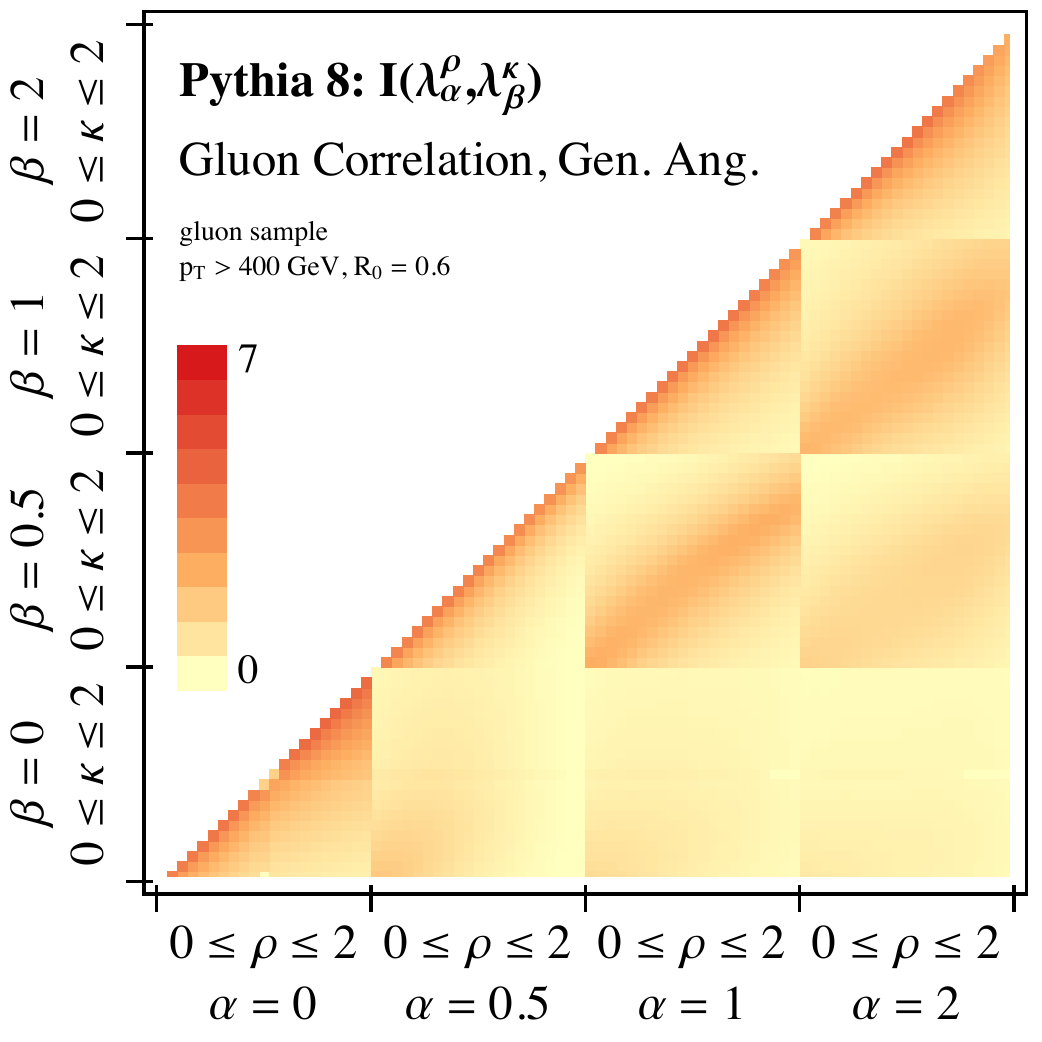}
}
$\qquad$
\subfloat[]{\label{fig:her_genang_corr_g} 
\includegraphics[width=7cm]{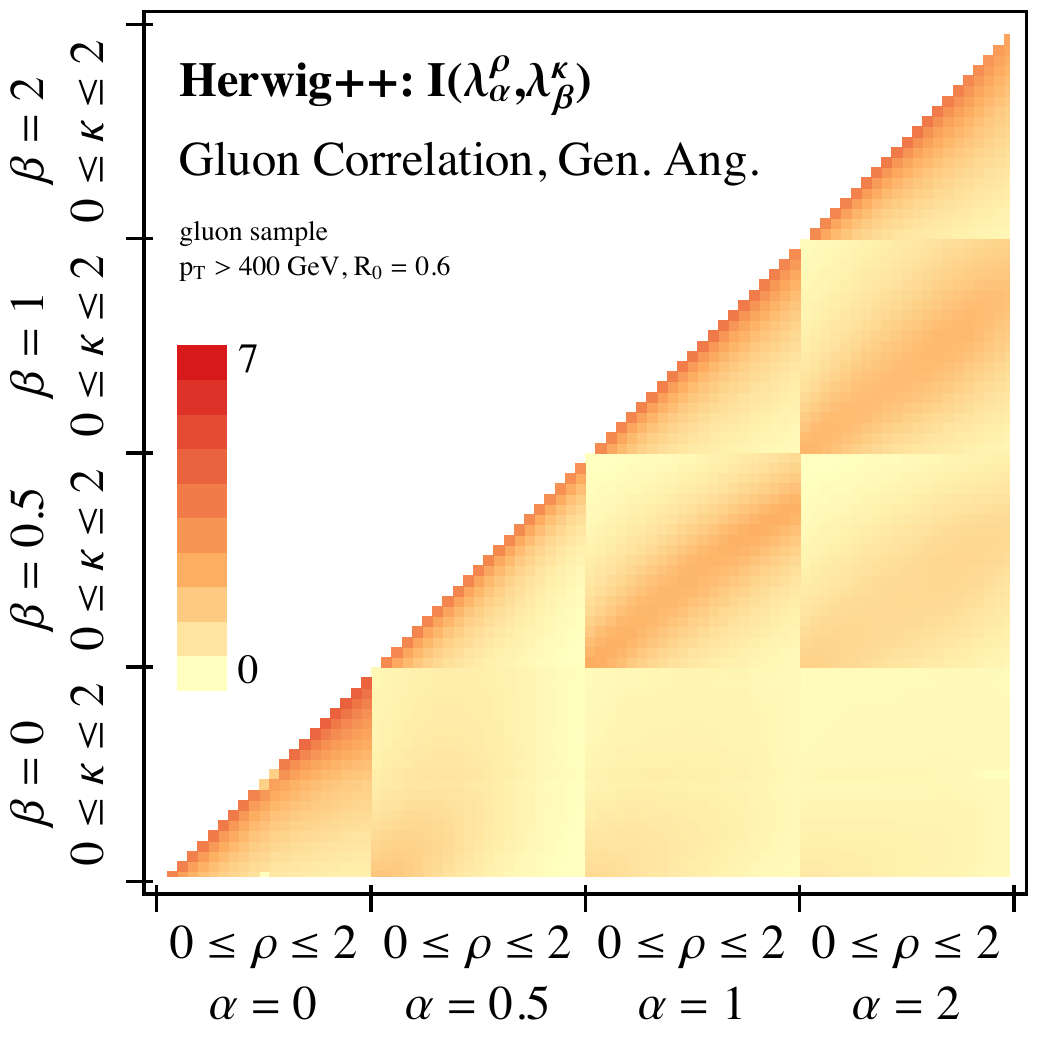}
}
\end{center}
\caption{
Same as \Fig{fig:genang_mutinf_corr_q}, but for a pure gluon sample. 
} 
\label{fig:genang_mutinf_corr_g}
\end{figure}

\begin{figure}
\begin{center}
%\subfloat[]{
%\includegraphics[width=7cm]{figures/LL_genang_corr_imp.pdf}
%}
%$\qquad$
%\subfloat[]{
%\includegraphics[width=7cm]{figures/NLL_genang_corr_imp.pdf}
%}
%\\
\subfloat[]{
\includegraphics[width=7cm]{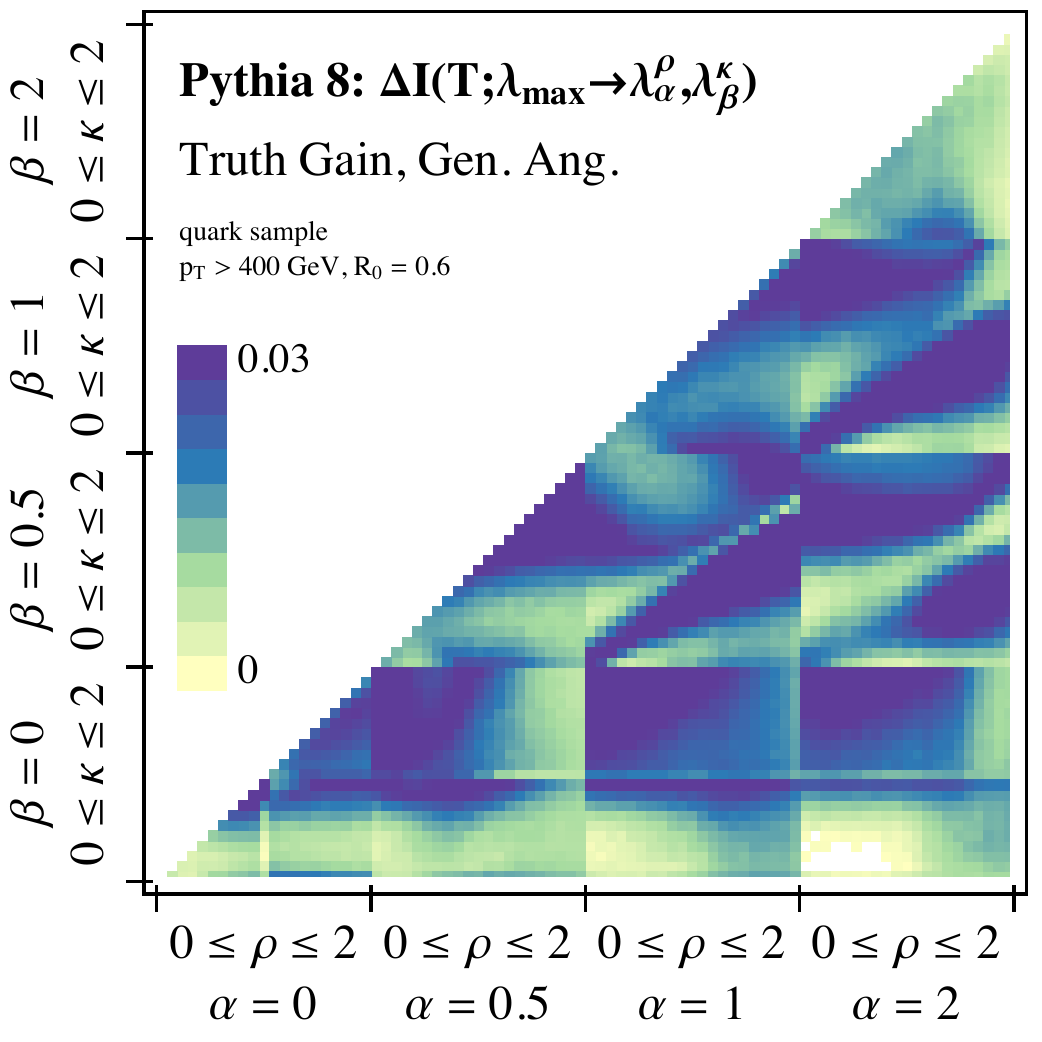}
}
$\qquad$
\subfloat[]{
\includegraphics[width=7cm]{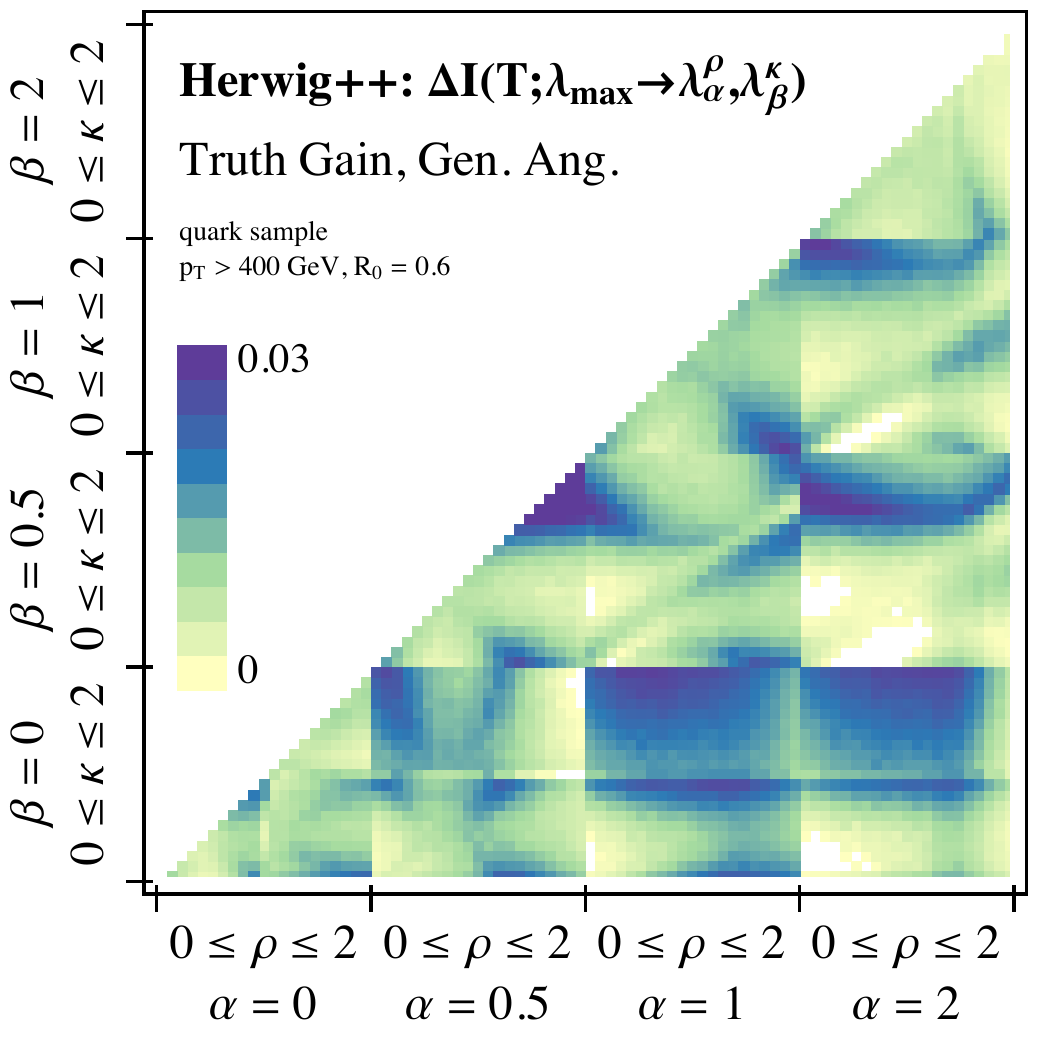}
}
\end{center}
\caption{
Improvement in the truth overlap by using two generalized angularities $(\genang{\rho}{\alpha},\genang{\kappa}{\beta})$ as compared to only one, $I(T;\genang{\rho}{\alpha},\genang{\kappa}{\beta})-\max[I(T;\genang{\rho}{\alpha}),I(T;\genang{\kappa}{\beta})]$.  We only show the \pythia{8} and \herwigpp\ parton showers since the LL and NLL analytic calculations are not sufficiently accurate to extract subtle differences in truth overlap.
}
\label{fig:genang_mutinf_imp}
\end{figure}

Turning to the generalized angularities, in \Figs{fig:genang_mutinf_corr_q}{fig:genang_mutinf_corr_g} we show the correlations on pure quark and gluon samples as measured by $I(\genang{\rho}{\alpha}, \genang{\kappa}{\beta})$.  As in the IRC unsafe case, there is broad agreement in the overall degree of correlation, though one has to be mindful of the restricted range of validity of the (N)LL calculations.  In \Fig{fig:genang_mutinf_imp}, we show the improvement of using two generalized angularities compared to one.  The LL and NLL calculations are not shown, since those calculations are not accurate enough to assess small differences.  The comparison between \pythia{8} and \herwigpp\ is similar to the IRC safe case, with \pythia{8} being more optimistic about the gains possible by combining observables.

\begin{figure}
\begin{center}
\subfloat[]{
\includegraphics[width=7cm]{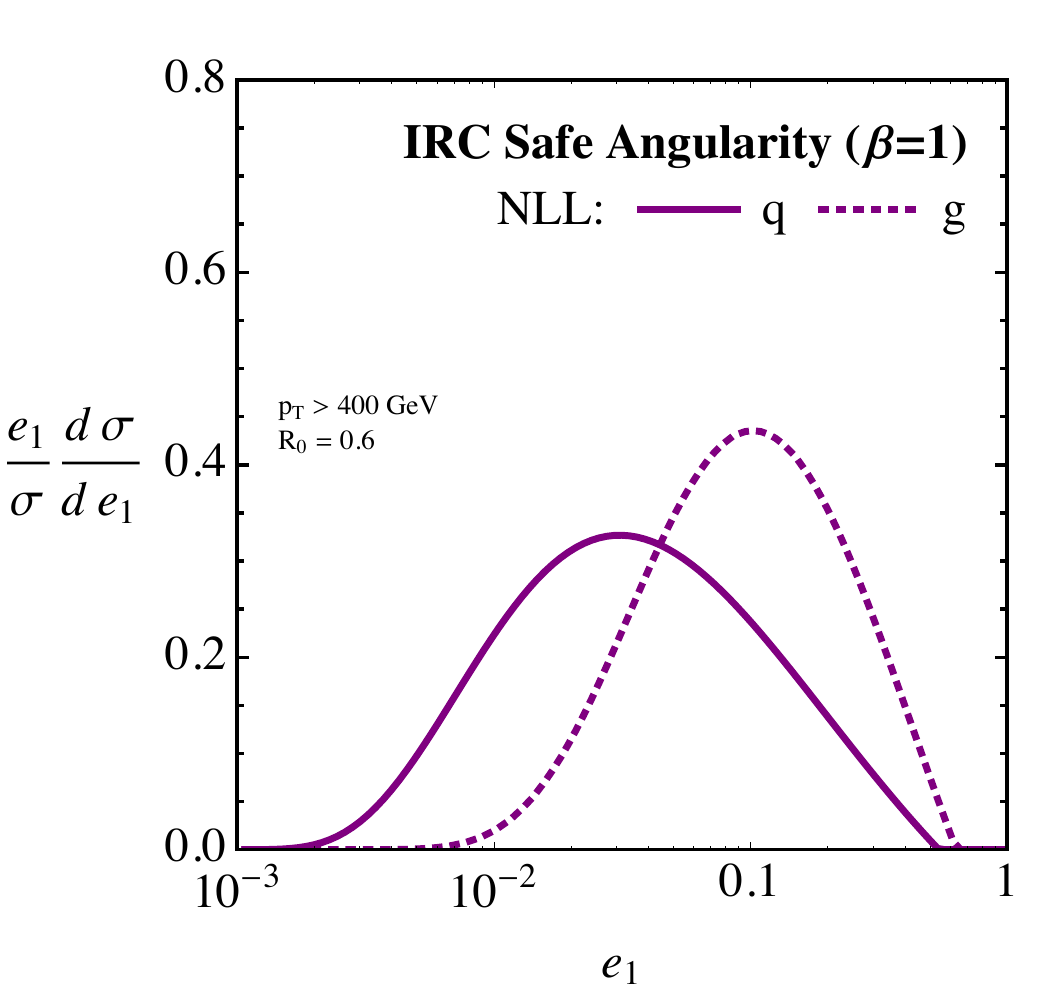}
}
$\qquad$
\subfloat[]{
\includegraphics[width=7cm]{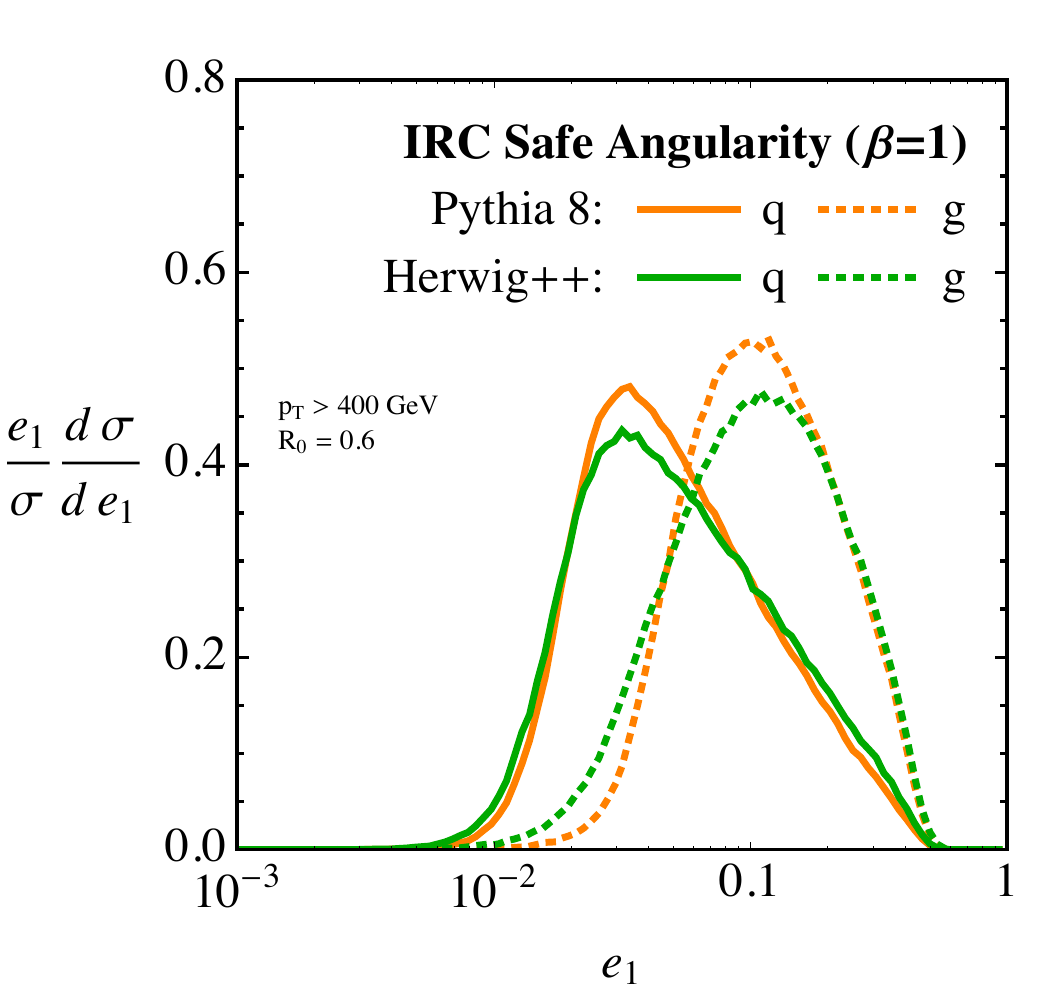}
}
\\
\subfloat[]{
\includegraphics[width=7cm]{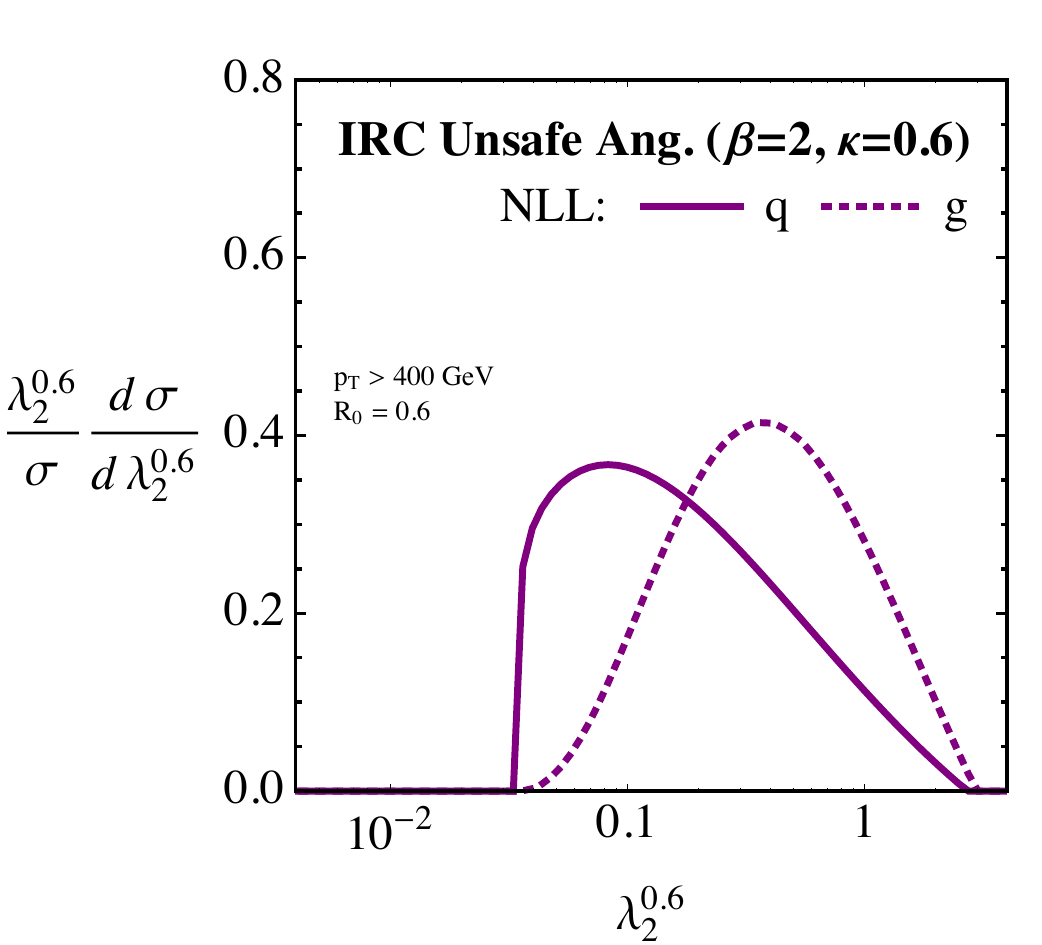}
}
$\qquad$
\subfloat[]{
\includegraphics[width=7cm]{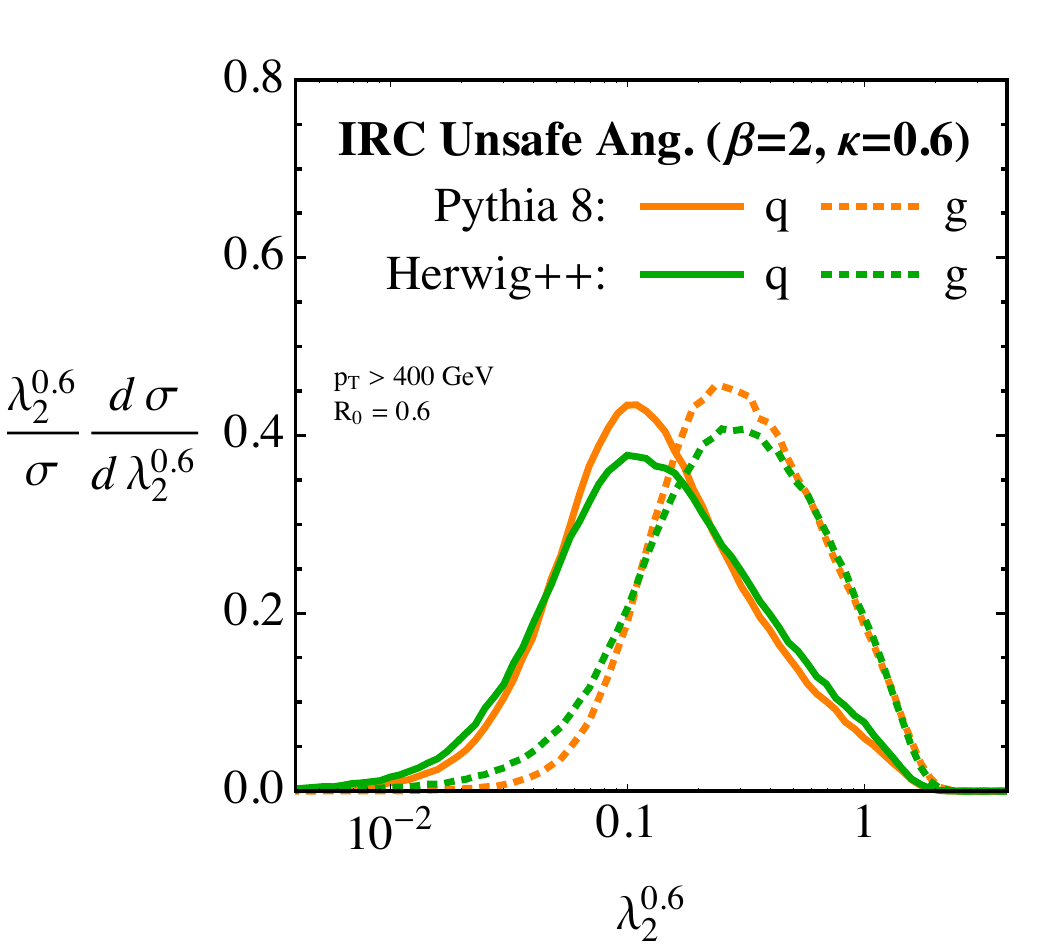}
}
\end{center}
\caption{
Raw distributions of $\safeang{1}$ (top) and $\genang{0.6}{2}$ (bottom) for the NLL calculation (left) and parton showers (right).  Note that the NLL distributions lack hadronization corrections that are present in the parton showers, which affects small values of the angularities. 
}
\label{fig:raw_single}
\end{figure}

\begin{figure}
\begin{center}
\subfloat[]{
\includegraphics[width=7cm]{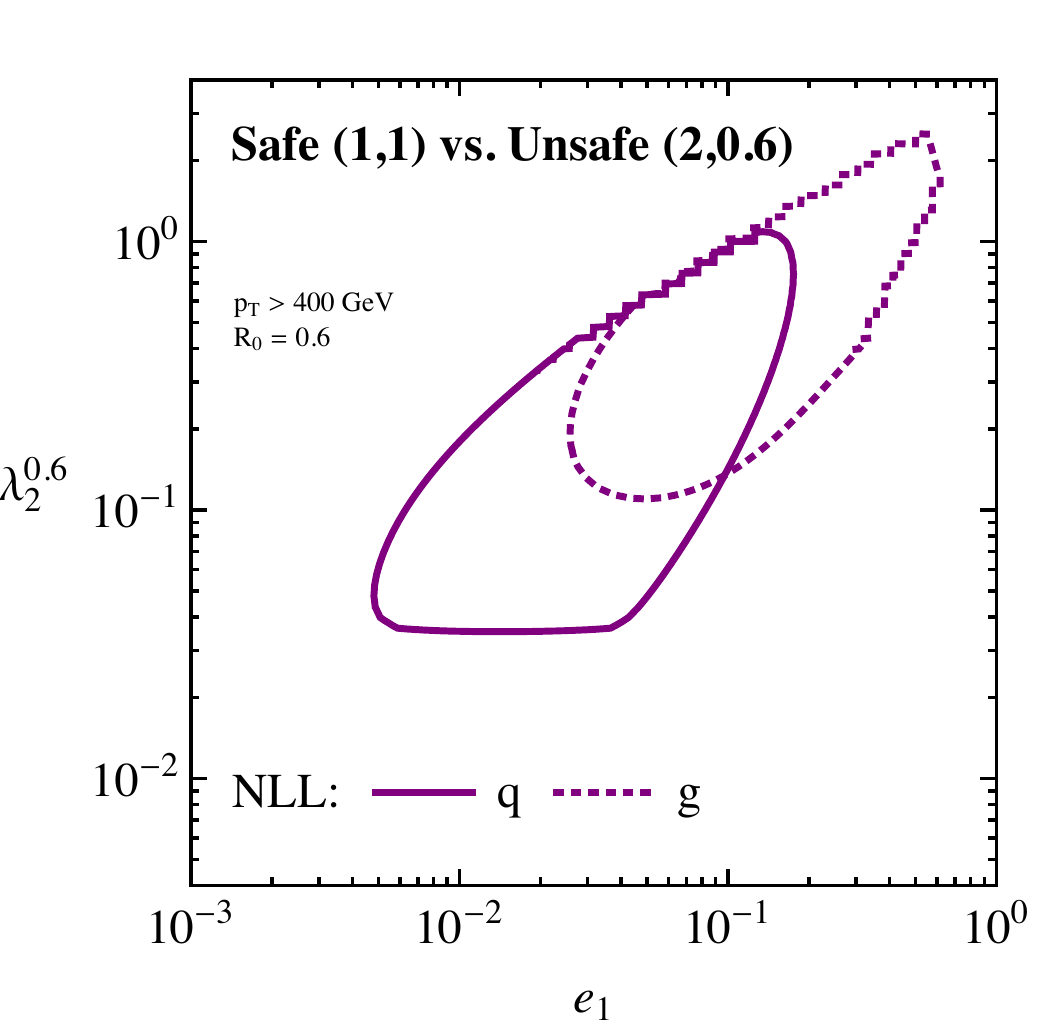}
}
$\qquad$
\subfloat[]{
\includegraphics[width=7cm]{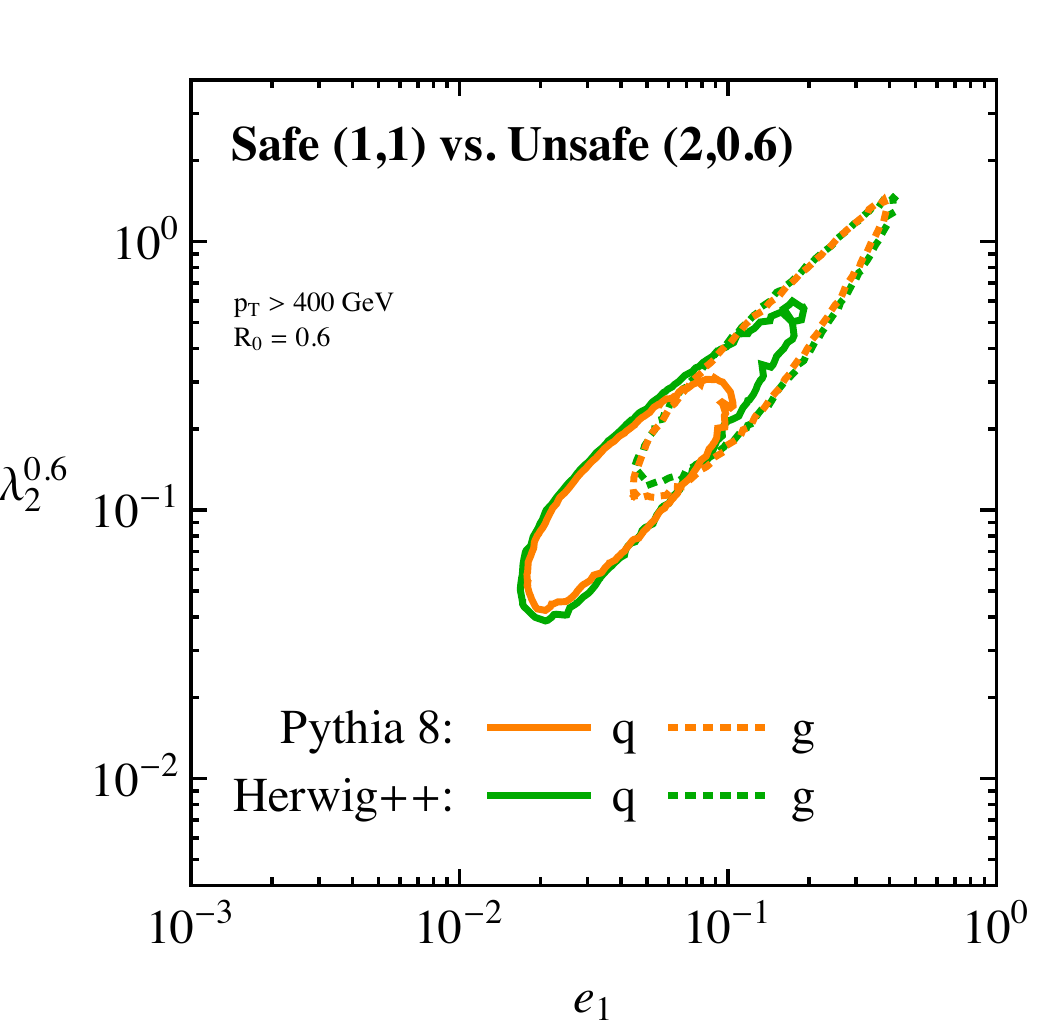}
}
\end{center}
\caption{
Double differential distributions in the $\safeang{1}$--$\genang{0.6}{2}$ plane for the NLL calculation (left) and parton showers (right).  The contours correspond to the half maximum of $\df^2 \sigma/(\df \ln \safeang{1} \, \df \ln\genang{0.6}{2})$.
}
\label{fig:raw_double}
\end{figure}

Finally, we show a few raw angularity distributions from the NLL calculation and both parton showers.  In \Fig{fig:raw_single}, we show single differential distributions for $\safeang{1}$ as an IRC safe example and $\genang{0.6}{2}$ as an IRC unsafe example.  Note that the NLL calculations lack important hadronization corrections that are modelled by the parton showers and are particularly important to correctly describe the small angularity region.  The NLL result for $\genang{0.6}{2}$ cuts off rather sharply at the low end due to our treatment of the QCD Landau pole.  The \pythia{8} distributions are more peaked than the  \herwigpp\ distributions, which is part of the reason why \pythia{8} predicts improved discrimination power compared to \herwigpp.  We then show the double differential distribution for $\safeang{1}$ and  $\genang{0.6}{2}$ in \Fig{fig:raw_double}, showing only the half-maximum contour for readability.  There is an irreducible degree of correlation between these observables due to phase space constraints (see \Eq{eq:unsafedoublephasespace}), but one can see that, while the gluon contours are similar for \pythia{8} and \herwigpp, the \pythia{8} contour for quarks is significantly smaller than \herwigpp.  This explains the enhanced discrimination power predicted by \pythia{8}.  The NLL contours are much larger in size than either \pythia{8} or \herwigpp\ because the NLL distributions do not vanish at the phase space boundaries \cite{Larkoski:2014tva}.  The analytic distribution will only vanish at the boundaries starting at NLL$'$ order, beyond the accuracy to which double differential cross sections have as-of-yet been computed.

\bibliographystyle{JHEP}
\bibliography{QvGCorr}

\end{document}